\documentclass{aa}  

\newdimen\imgwidth
\usepackage{graphicx}
\usepackage{txfonts}
\usepackage[usenames]{color}
\usepackage{longtable}

\begin{document}
%
   \title{New active galactic nuclei detected \\ in ROSAT All Sky Survey galaxies}
   \subtitle{II. The complete dataset\thanks{Based on observations collected
       at European Southern Observatory, Chile}$^{,}$\thanks{Based on
       observations taken at the German-Spanish Astronomical Center,
     Calar Alto, operated by
    the Max-Planck-Institute for Astronomy, Heidelberg jointly with the Spanish
  Commission for Astronomy.}$^{,}$\thanks{Based on observations collected at
     McDonald Observatory}
}

   \author{ W. Kollatschny \inst{1},
           R. Kotulla \inst{1},
           W. Pietsch \inst{2},
           K. Bischoff \inst{1,3},
           M. Zetzl \inst{1}
          }


   \institute{Institut f\"ur Astrophysik, Universit\"at G\"ottingen,
              Friedrich-Hund Platz 1, D-37077 G\"ottingen, Germany\\
              \email{wkollat@astro.physik.uni-goettingen.de}
          \and
             Max-Planck-Institut f\"ur Extraterrestrische Physik,
             Giessenbachstrasse, D-85740 Garching, Germany
          \and
             Halfmann Teleskoptechnik GmbH \& Co. KG, Gessertshausener Str. 8,
             D-86356 Neus\"a\ss-Vogelsang, Germany
             }

   \date{Received 2007; accepted 2008}
   \authorrunning{Kollatschny et al.}
   \titlerunning{New AGN detected in RASS galaxies. II -- The complete dataset}

  \abstract{}
{The  ROSAT ALL Sky Survey Bright Source Catalogue (RASS-BSC) has been
 correlated with the Catalogue of 
Principal Galaxies (PGC) to identify new extragalactic counterparts. 550
reliable optical counterparts have been detected. However there existed
no optical spectra for about 200 Active Galactic Nuclei (AGN) candidates
before the
ROSAT ALL Sky Survey (RASS) was been completed.}
{We took optical spectra of 176 X-ray candidates and companions at
  ESO, Calar Alto observatory and McDonald observatory. When necessary we used
a line profile decomposition to measure line fluxes, widths and centers to
classify their type of activity.}
{We discuss the redshift-,
linewidth-, as well as optical and
X-ray luminosity distribution of our ROSAT selected sample.
139 galaxies of our 166 X-ray counterparts
 have been 
identified as AGN with  93 being Seyfert 1 galaxies (61\%).
Eighteen of them (20\%) are
Narrow Line Seyfert 1 galaxies.
34 X-ray candidates (21\%) are LINERs and only eight candidates
(5\%) are  Seyfert 2. The ratio of the number of Seyfert 1 galaxies
to Seyfert 2 galaxies is about 11/1. Optical surveys result in ratios
of 1/1.4. The high fraction of detected Seyfert 1 galaxies is explained by
the sensitivity of the ROSAT to soft X-rays which are heavily
absorbed in type 2 AGN. 
Two X-ray candidates are HII-galaxies and 25 candidates
(15\%) show no signs of spectral activity. The AGN in our RASS selected sample
exhibit
slightly higher optical luminosities
 ($\rm M_B = (-20.71 \pm 1.75)$~mag) and
similar X-ray luminosities 
($\log(L_{\mathrm{X}} [\rm erg\ s^{-1}]) =42.9 \pm 1.7$)
compared to other AGN surveys. The H$\alpha$ line width distribution (FWHM)
of our newly identified ROSAT AGN sample is similar to the line widths
distribution based on SDSS AGN. 
However, our newly identified RASS AGN have rather
reddish colors explaining why they have not been detected
before in ultraviolet or blue excess surveys.}
{}

   \maketitle
%

\section{Introduction}
Canonical active galactic nuclei (AGN) are blue, UV-excess sources. They 
 are known to be strong X-ray emitters
(e.g. Fabbiano et al. \cite{Fabbiano92}) 
 in comparison to normal
galaxies. However, not all AGN detected by their optical, infrared or
radio properties have counterparts in X-ray surveys.
The X-ray satellite ROSAT detected nearly 19\,000 bright sources in
the soft X-ray band (0.1-2.4 keV). These are listed in the 
ROSAT All Sky Survey Bright Source Catalogue (RASS-BSC)
(Voges et al., \cite{Voges93}, \cite{Voges99}).
We are interested in the
optical spectral properties of ROSAT-selected AGN.
On the one hand we wanted to take optical spectra of as many ROSAT
counterparts as possible. On the other hand we wanted to test whether our newly
identified 
RASS selected AGN show spectral properties different to those AGN
that have been detected earlier with other strategies.
A correlation of the RASS with 
optical counterparts in
the Principal Catalog of Galaxies (Paturel et al., \cite{Paturel89, Paturel03})
resulted in about 900
extragalactic counterparts or candidates. These X-ray sources are mainly AGN or
nearby galaxies. However, for about 200 AGN candidates no optical spectra
existed when the ROSAT ALL Sky Survey (RASS) was completed.

This paper is the second  in a series  describing the optical spectra
and the physical properties of these AGN.
In the first paper the selection strategy was described by Pietsch et al. 
(\cite{Pietsch98} hereafter paper 1)  and the
optical spectra of 35 X-ray counterparts were presented.

In parallel to our new optical identifications we investigated X-ray and
ROSAT-related AGN papers to determine whether some of our targets had been identified
since (Moran et al. \cite{Moran96}, Simcoe et al. \cite{Simcoe97}
 Bade et al., \cite{Bade98},
Motch et al. \cite{Motch98}, Appenzeller et al. \cite{Appenzeller98},
  Fischer et al. \cite{Fischer98},  Bauer et al. \cite{Bauer00},
 Brinkmann et al. \cite{Brinkmann00}, Vaughan et al. \cite{Vaughan01}
 Xu et al., \cite{Xu01},  Grupe et al. \cite{Grupe04}).
Furthermore we checked  SDSS-related papers ( Adelman et al. \cite{Adelman06})
as well as the Veron-Cetty catalogue of AGN (\cite{Veroncat12ed}).
In our list of 166 new AGN candidates
we found some information with respect to 41 objects.
In a preprint (Bischoff et al. \cite{Bischoff99}) we presented some
preliminary data of 17 of our new AGN candidates. 
%
%
\section{Observations}
\subsection{The ROSAT AGN sample}
\label{dr_datasel}
In paper 1 we described in detail our strategy
to find new AGN in the ROSAT All Sky Survey. Further detailed information
about the galaxy identification project can be found in Zimmermann et al. 
(\cite{Zimmermann01}). 
The selection is based on the empirical finding that most active
galaxies are strong X-ray emitters in comparison to non-active galaxies.
In the RASS-BSC
(Voges \cite{Voges99}) nearly 19\,000 
X-ray sources are listed, having count rates in excess of 0.05 cts/s: 
These objects were subsequently cross-correlated with
the Principal Catalog of Galaxies (Paturel at al.
 \cite{Paturel89, Paturel03}).
904 X-ray sources were identified as having extragalactic counterparts. 

Extended X-ray sources such as early-type galaxies or clusters of
galaxies were rejected from our sample in the next step.

Finally, 547 correlations have been determined as reliable counterparts
of compact extragalactic X-ray sources. These sources were cross-checked
with the NASA Extragalactic Database (NED):
349 sources were known active galaxies. However, no spectral information
existed for 198 AGN candidates.  
Our goal was to obtain spectra of as many of these objects as possible.

\subsection{Optical observations}
Table 2
contains information on our observed sample of 166 X-ray counterparts
as well as some information on ten additional companions of theses
counterparts.
Given are the ROSAT name (1), the optical positions 
derived from the Digitized Sky Survey (DSS) (2 and 3), the object name (4),
the X-ray count rate (5), derived X-ray luminosity (6), the apparent (7) and
absolute (8) B-band magnitudes (for details see paper 1).

Figure 1
 shows the spatial distribution -- in equatorial
coordinates -- of all AGN candidates for which we have
acquired spectra. They are distributed homogeneously over the whole sky, except
for the regions blocked by the Milky Way. The sky distribution of all
904 extragalactic X-ray counterparts has been shown by Zimmermann et al.
 (\cite{Zimmermann01}) (their Fig.~3). 
\begin{figure}
  \centering
  \includegraphics[width=\columnwidth]{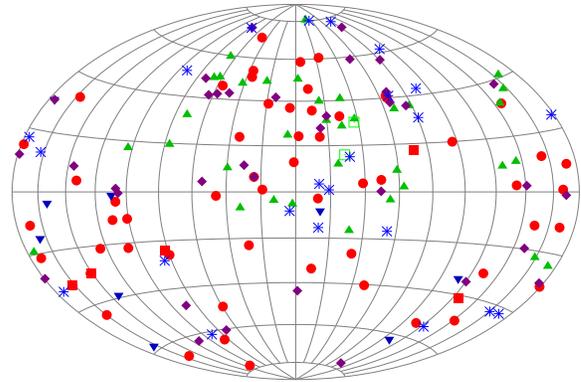}
  \caption{Aitoff projection of all selected sources in J2000 equatorial
   coordinates (Center of the plot: RA=12h, DEC=0$^{\circ}$;
   RA increases to the
   right.).There is a homogenous spatial distribution of our X-ray selected
   galaxies, except for the regions blocked
  by the Milky Way. The symbols used in this and all following plots:
  Sy~1 type objects: filled red circles, intermediate types (Sy~1.5/1.8/1.9):
  green upward triangles, Sy~2: blue downward triangles, LINERS: purple
  diamonds, H~II: green open squares, non-active: blue stars.} 
  \label{fig:aitoff}
\end{figure}

We obtained optical spectra of our X-ray candidate galaxies during five
observing campaigns. We used the 2.2\,m telescope
at Calar Alto Observatory in Spain, the 2.2\,m telescope
at La Silla/ESO in Chile, as well as the 2.7\,m telescope at
McDonald Observatory
in Texas. We obtained spectra of 176 AGN candidates and companion
galaxies.
In a few cases we observed
more than one object when there was more than one candidate in the
ROSAT error box. We identified 166 AGN and non-AGN as optical counterparts of
ROSAT RASS sources. Details of the observations, such as the observing
periods, the telescopes and spectrographs we used, as well as the wavelength
coverage of the spectra are listed in Table 1.
%
\begin{table}[h]
\footnotesize
\caption[Observing runs]{List of our observing runs giving the telescopes and
 instruments used for the
   acquisition of the spectra as well as their wavelength range.}
\begin{center}
\begin{tabular}{l@{~~~}l@{~~~}l@{~~~}l@{~~~}l}
\hline\hline
\# & Observing run & Telescope & Instrument (CCD) & $\lambda\lambda$ [\AA] \\
\hline
1 & 96/11/02-06 &  ESO 2.2m$^{a}$ & EFOSC2 (LORAL) & 3800--8000 \\
2 & 97/07/02-06 &  ESO 2.2m$^{a}$ & EFOSC2 (LORAL) & 4100--7400 \\
3 & 97/07/28-08/01 & CA 2.2m$^{b}$ & CAFOS (SITe\#1d) & 4200--8200 \\
4 & 98/03/01-05 &  CA 2.2m$^{b}$ & CAFOS (SITe\#1d) & 4200--8200 \\
5 & 00/02/08-09 & MDO 2.7m$^{c}$ & LCS (TK3/CC1) &  4700--7300\\
\hline\hline
\multicolumn{5}{l}{$^{a}$ La Silla, Chile} \\
\multicolumn{5}{l}{$^{b}$ Calar Alto, Spain} \\
\multicolumn{5}{l}{$^{c}$ McDonald Observatory, Texas, USA}
\end{tabular}
\end{center}
\label{tab:obsruns}
\end{table}

The observing dates of the individual galaxies and exposure 
times are listed in Table 3. Exposure times range from 10
to 45 minutes.
The spectrograph slits had projected widths 
of 1.5 to 2 arc sec. We had typical seeing conditions of 1 to 1.5 arc sec.
The slit was oriented in a north-south direction, in most cases,
to minimize the impact of light losses caused by differential refraction.

Our spectra typically cover a wavelength range from 3800~/~4700\,\AA\ to 
7300~/~8200\,\AA\,  (see Table~1) with a spectral
resolution of 8 to 10\,\AA. 
Additional spectra were taken after each object exposure for
wavelength calibration. Various standard stars were
observed for flux calibration. 

\subsection{Data reduction}
The reduction of the spectra (bias subtraction, cosmic ray correction,
flat-field correction, 2D-wavelength calibration, night sky subtraction,
flux calibration) was done in a homogeneous way with the IRAF reduction
packages\footnote{IRAF is distributed by the National Optical Astronomy
  Observatories, which are operated by the Association of Universities for
  Research in Astronomy, Inc., under cooperative agreement with the National
  Science Foundation.}. We extracted spectra of the central 3 arc sec. 
The wavelength calibration was done using comparison spectra
of He-Ar (runs \#1, \#2 ), Hg-Cd-He (\#3, \#4 ) and He-Ne (\#5).
The flux calibration was done by means of the standard stars HD 49798 (\#1),
NGC 7293 (\#1, Turnshek et al. \cite{Turnshek90}), BD 332642 (\#4) and
BD G191B2B (\#4,5).

Redshifts, emission line intensities and line widths were derived from
individual emission/absorption lines as well as
determined by fitting line
complexes
(H$\alpha$ narrow and broad, [N~II], [S~II]; H$\beta$ narrow and
broad). Only the narrow Balmer line components were used for the
diagnostic diagram (see Fig. \ref{fig:baldwin}) to derive their activity type.

\section{Results}

The optical spectra of the X-ray counterparts (AGN candidates)
of our first observing run are published in Paper 1. All spectra  
taken during the remaining 4 observing runs are presented in
Fig. \ref{fig:spectra}.

We discriminated between the main X-ray counterpart and the
galaxy companion when
we observed more than one galaxy in the ROSAT error box.

The results derived from our optical spectroscopy are given in Table~3.
 We list the name of the object (1), the observing run (2),
 total exposure time used (3), observed redshift (4), flux ratios of
[\ion{O}{iii}]$\lambda$5007/H$\beta$ (5) and
[\ion{N}{ii}]$\lambda$6583/H$\alpha$ (6), Balmer decrement of the
broad components (7), the width (FWHM) of the broad H$\alpha$ component
(8) and the
derived nuclear activity class (9). The galaxies already presented in
paper 1 are marked [P1]. Those galaxies that since have been classified as an
AGN are marked [V]. The detailed literature can be found  
in the Veron-Cetty Catalogue of AGN (\cite{Veroncat12ed}).
\\

\subsection{Redshift distribution}

The redshifts we derived from the object spectra are given
in Table~3.
 The typical error is less than
160~km s$^{-1}$.
 There is no redshift information for the
BL\,Lac candidates.

Figure~2 shows the redshift distribution of our newly
identified RASS AGN (dark columns). Most objects have redshifts below
$z=0.1$. 
 Two objects not shown in Fig.~2 have redshifts
of $z~=~0.55$ and $z~=~1.90$.  
Our median redshift is  $z~=~0.0327$.
The redshift distribution of the already known RASS AGN (Zimmermann et al.
 \cite{Zimmermann01}) with a median redshift of $z=0.0330$
 is shown for comparison (grey columns). 
The median redshift of all RASS AGN also amounts to $z=0.0330$. 

\begin{figure}
  \centering
  \includegraphics[width=\columnwidth]{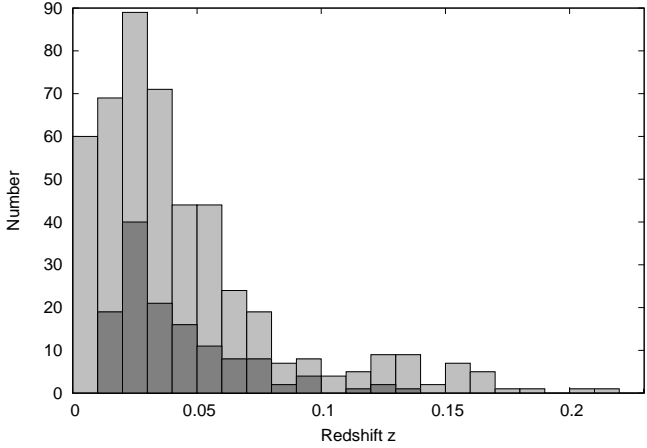}
  \caption{
Redshift distribution of our newly identified RASS AGN (dark columns).
Two objects have redshifts of $z=0.55$ and $z=1.90$.  
The redshift distribution of the already known RASS AGN (Zimmermann et al.
 \cite{Zimmermann01}) (grey columns) is shown for comparison.}
  \label{fig:hist_z}
\end{figure}
These numbers can be compared with
the SDSS redshift distribution of AGN found by Anderson et al.
 (\cite{Anderson07}). Limiting
their X-ray count rate to sources in excess of 0.05 cts/s, they find
a peak at $z \approx
0.15$. This shows that our ROSAT selected AGN
are  nearby objects compared to the Anderson sample.

\subsection{Optical and X-ray luminosity distribution}

 The B-Band flux, the ROSAT soft (0.1 -- 2.4 keV) X-ray flux as well as
 the luminosities
 of our sample galaxies are given in Table 2.
A Hubble
 constant of H$_0$~=~75~km s$^{-1}$ Mpc$^{-1}$
was used throughout this paper. For
all objects with redshift information (all objects except those classified as
BL\,Lac, see Table 3, cols 4 and 9) we
calculated distances and luminosities. Four objects have
absolute magnitudes $\leq-23$~mag classifying them as QSO. 

The optical luminosity distribution of our sources is shown in Figs. 3 and 4.
Optical magnitudes of our galaxies are
taken from the PGC.
We calculated  in a homogeneous way
the absolute M$_B$ magnitudes of our sample galaxies with the formula
given by Veron-Cetty (\cite{Veroncat12ed}) to compare our magnitudes
with the AGN magnitudes in
their Quasar and AGN catalogue.
Figure~3 shows 
the B-band magnitudes
 $M_{\rm B}$ of our newly identified RASS AGN (black dots)
as a function of redshift, the $M_{\rm B}$ distribution of
the already known RASS AGN (red dots) as well as
the Seyfert magnitude distribution in the Veron-Cetty Catalogue.
The grey points 
represent all Seyfert/LINER galaxies given in the Veron-Cetty
Catalog (\cite{Veroncat12ed}).
%
\begin{figure}
  \centering
  \includegraphics[width=\columnwidth]{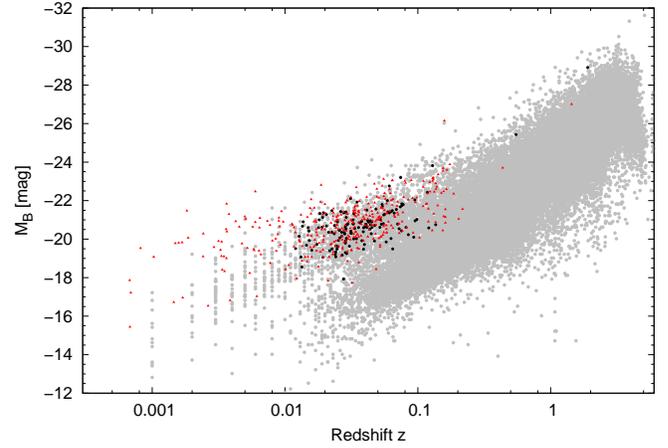}
  \caption{
$M_{\rm B}$ distribution of our newly identified RASS AGN (black dots)
as a function of redshift as well as the $M_{\rm B}$ distribution of
the already known RASS AGN (red dots).
 Grey points are
 objects from the 
12th edition Quasar catalogue of Veron-Cetty (\cite{Veroncat12ed}).
We converted all magnitudes 
 to a Hubble constant of
  H$_0$~=~75~km s$^{-1}$ Mpc$^{-1}$.} 
\label{fig:veroncetty}
\end{figure}
%
The absolute
magnitudes from the Veron-Cetty Catalogue 
have been transformed to a Hubble constant of
H$_0 = 75$ km s$^{-1}$ Mpc$^{-1}$.

The RASS selected AGN are located in optically bright galaxies compared to
the AGN in the catalogue of
Veron-Cetty (\cite{Veroncat12ed}) (see Fig.~3).   
The mean absolute magnitude of our newly identified RASS AGN is
$\rm M_B = (-20.71 \pm 1.75 )$~mag. This value is identical to the
magnitude of the
already known RASS AGN (Zimmermann et al. \cite{Zimmermann01})(see Fig.~4)
and only slightly
above the mean value $\rm M_B = -20.46\,$~mag
of Ho et al. (\cite{Ho97}).  Ho et al. (\cite{Ho97}) derived their absolute
magnitude from a sample of all nearby AGN.
%
%
\begin{figure}
  \centering
  \includegraphics[width=\columnwidth]{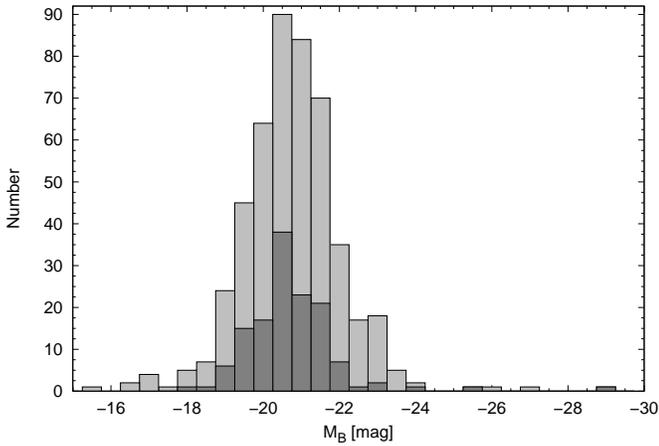}
  \caption{
$M_{\rm B}$ distribution of our newly identified RASS AGN
(dark columns) as well as the $M_{\rm B}$ distribution
 of the already known RASS AGN (Zimmermann et al.
 \cite{Zimmermann01}) (grey columns) shown for comparison.
}
  \label{fig:hist_mb}
\end{figure}
%
%
The optical brightness of our newly identified ROSAT AGN
cannot be responsible for them remaining undetected so far.

To compute the soft X-ray (0.1 -- 2.4 keV) energy fluxes from the ROSAT count
rates we assumed a power-law spectrum 
\begin{equation}
f_E\ \mathrm{d}E \propto E^{-\Gamma+1}\ \mathrm{d}E\ ,
\end{equation}
where $f_E\ \mathrm{d}E$ is the energy flux between $E$ and $E+\mathrm{d}E$. 
The spectral index was fixed to $\Gamma = 2.3$, a typical value for galaxies
observed with ROSAT (Hasinger \cite{Hasinger91}, Boller \cite{Boller98}).
We only
consider the Galactic \ion{H}{i}-column density (Dickey \cite{Dickey90})
 along the
line of sight as absorbing column density, so the given fluxes have to be
understood as lower limits for the intrinsic fluxes emitted by the active
nuclei from within its host galaxy.

\begin{figure}
  \centering
  \includegraphics[width=\columnwidth]{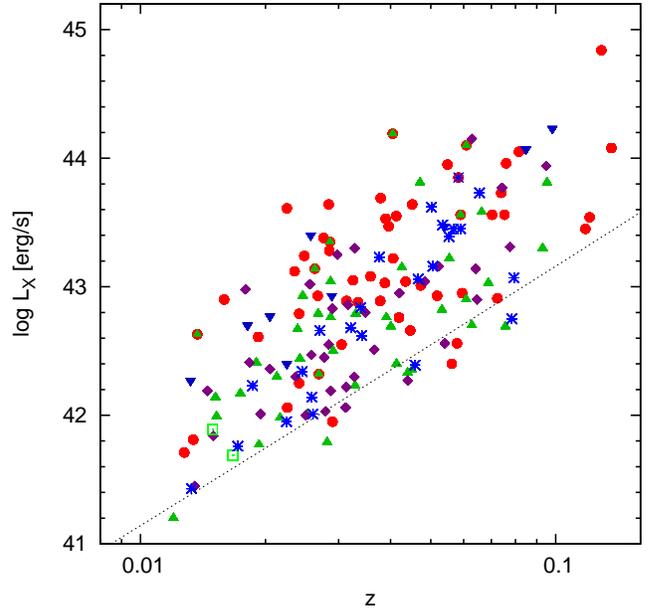}
  \caption{
X-ray luminosities as function of redshift z 
of our newly identified RASS galaxies. Symbol
    definition as in Fig. \ref{fig:aitoff}. The dashed line marks the
    detection limit of the RASS bright source catalog of $\rm 0.05 cts/s$,
    converted to luminosities using the median galactic HI column density of
    our sample ($\rm 0.258 \times 10^{21}\, cm^{-2}$).}
 \label{fig:lx_vs_z}
\end{figure}

In Fig. \ref{fig:lx_vs_z} we plot
 the X-ray luminosity  of our newly identified
RASS galaxies
 as a function of redshift
z. Most objects lie one to two orders above the
 selection-limit of 0.05 cts/s indicated by the dotted line. The points below
 the selection limit are located at positions with a lower HI column density
 than the mean and therefore have lower flux levels than expected for objects
 with an average column density. 

\begin{figure}
  \centering
  \includegraphics[width=\columnwidth]{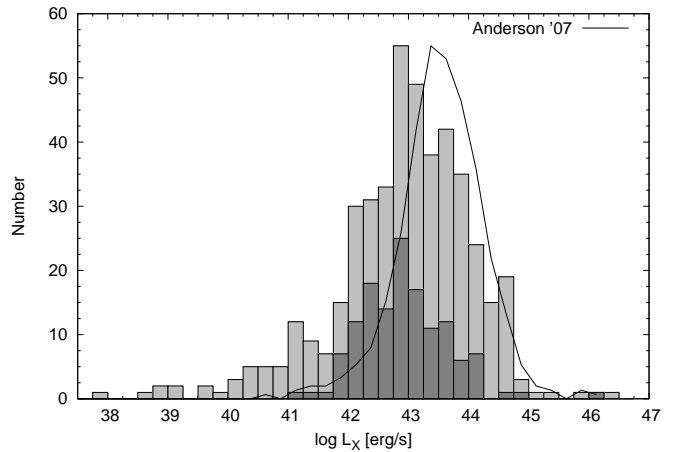}
  \caption{
Distribution of the X-ray luminosity $L_{\rm X}$
of our newly identified RASS AGN (dark columns) as well as
the distribution of the already known RASS AGN (Zimmermann et al.
 \cite{Zimmermann01}) (grey columns).
 Also shown  for comparison is the relative distribution of Anderson et al.
 (\cite{Anderson07}) (solid line) for their X-ray AGN
from the SDSS. 
}
  \label{fig:hist_lx}
\end{figure}

\begin{figure}
  \centering
  \includegraphics[width=\columnwidth]{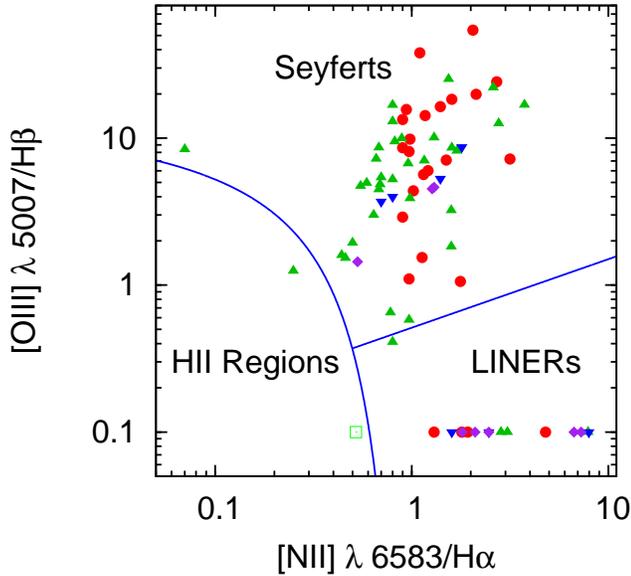}
  \caption{Diagnostic Baldwin diagram
of all our newly identified emission line galaxies with measured
narrow components in all
  four lines. Also shown are the empirically determined dividing
 lines of Kewley et al.
  (\cite{Kewley01}) (solid line). Symbols are as described in
  Fig. \ref{fig:aitoff}.
} 
 \label{fig:baldwin}
\end{figure}

The X-ray luminosity distribution $L_{\rm X}$
of our newly identified RASS AGN (dark columns)
 is shown in Fig. \ref{fig:hist_lx}
together with the luminosity distribution $L_{\rm X}$
of the already known RASS AGN (Zimmermann et al.
 \cite{Zimmermann01}) (grey columns).
The distribution of both samples is best described by a Gaussian
with FWHM $=1.66\,$dex around
$\log(L_{\mathrm{X}} [\rm erg\ s^{-1}]) =42.9$.
Zimmermann et al. (\cite{Zimmermann01}) obtained  X-ray luminosities of 
$\log(L_{\mathrm{X}}) \approx 43.5 \pm1.5$ for their sample of active
galaxies  and $\log(L_{\mathrm{X}}) \approx 42.5 \pm1.5$ for the
candidate galaxies (their Fig.~9a). They rated those X-ray sources
as candidate galaxies that were likely to possess hitherto unreported active
galactic nuclei.
The X-ray luminosity distribution of X-ray AGN selected from the SDSS
(Anderson et al. \cite{Anderson03}, \cite{Anderson07}) and having  
redshifts $\rm z\leq 0.15$ is shown for comparison. Their mean X-ray
luminosity $\log(L_{\mathrm{X}}) \approx 43.5 \pm 1.1$ is slightly
above our value. However,
they identified mostly 
quasars and Seyfert 1 galaxies in their sample.
%
%


\subsection{AGN type distribution}

To classify the activity type of our galaxies
we measured the H$\alpha$, H$\beta$,
[\ion{O}{iii}]$\lambda$5007, and [\ion{N}{ii}]$\lambda$6583 emission line
 intensities whenever it was
possible. The broad Balmer lines were fitted with one or multiple line
components.
Table~3 lists the measured line ratio of the
[\ion{O}{iii}]$\lambda$5007/H$\beta$ and
[\ion{N}{ii}]$\lambda$6583/H$\alpha$ line ratios.
We considered only the narrow Balmer line
components for these line ratios. 

The FWHM linewidth W$_{\alpha}$ of the broadest line component in the H$\alpha$
complex is given in Table~3
 (column 8). The typical error is
200~km s$^{-1}$.
Column 7 gives the Balmer decrement H$\alpha$/H$\beta$ derived from the
broad-line components only.

We classified our galaxies in the following way:
Seyfert 1 type galaxies have
single broad Balmer line components only in their spectra .
Intermediate
Seyfert types 1.5, 1.8 and 1.9 show broad as well as narrow Balmer line
components;
the intensity ratio of the broad Balmer line component decreases
with respect to the narrow Balmer line component.
We recalibrated all our galaxies in a homogeneous way - including those
of Paper 1. In a few cases we made small changes in the classification
of the intermediate type Seyferts.
Narrow Line Seyfert~1 galaxies have single component
Balmer lines with line widths of about less than 2500 km s$^{-1}$
(Osterbrock \cite{Osterbrock85}), only slightly broader than the
forbidden narrow emission lines. Furthermore, these objects have relatively
weak [\ion{O}{iii}]$\lambda$5007 emission and strong FeII emission. 
Seyfert 2 galaxies, LINERs and HII galaxies show narrow emission
lines only.
On basis of their narrow
line ratios [\ion{O}{iii}]$\lambda$5007/H$\beta$ and
[\ion{N}{ii}]$\lambda$6583/H$\alpha$
we discriminate between Seyfert 2, LINER, and HII spectra
(see Fig.~7). 
%
%
In Fig.~7 we show the dividing line for the different types
(Kewley et al. \cite{Kewley01}).
In some LINERs we measured weak [\ion{N}{ii}] and H$\alpha$ emission only,
with upper limits for the [\ion{O}{iii}]$\lambda$5007 line intensity.
Here we set a value of 0.1 for
the [\ion{O}{iii}]$\lambda$5007/H$\beta$ intensity
ratio.

BL\,Lac objects are characterized by their blue featureless
continuum.
Galaxies showing absorption line spectra only  are classified
as non-active galaxies.
%

\begin{figure}
  \centering
  \includegraphics[width=\columnwidth]{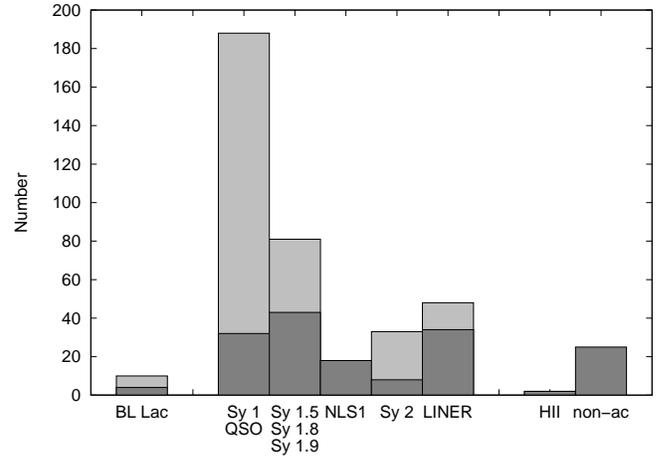}
  \caption{Distribution
of our newly identified RASS 
AGN and non-AGN types (dark columns)
as well as the distribution of the already known
RASS AGN (Zimmermann et al.
 \cite{Zimmermann01}) (grey columns).
}
  \label{fig:hist_sytypes}
\end{figure}

Figure~\ref{fig:hist_sytypes} shows the distribution of our newly identified RASS 
AGN and non-AGN types (dark columns)
as well as the distribution of the already known
RASS AGN (Zimmermann et al.
 \cite{Zimmermann01}) (grey columns).

Only 27 galaxies (16.3\%)
 of our 166 ROSAT selected galaxies show no nuclear activity:
2 HII-galaxies (1.2\%) and 25 absorption line galaxies (15.1\%).
Some of the X-ray counterparts we classified as 
absorption line galaxies have very high  X-ray luminosities (see Fig.~5).
 Seven of them show 
$\log(L_{\mathrm{X}}) \geq 43.45$. It is not clear whether we
took spectra of
the wrong counterpart or whether these objects are highly obscured AGN.
ROSAT could also have detected X-ray emission 
from an unknown galaxy group or cluster of which our candidate galaxy may be 
a member. 

Most of our galaxies (60.8\%)
 are of Seyfert type 1: 32 pure Seyfert 1 (19.3\%),
43 intermediate Seyfert 1 (25.9\%), and 18 NLS1 (10.8\%).
About one fifth of the Seyfert 1 galaxies are NLS1 galaxies.  
Among all Seyfert galaxies only 8 galaxies (4.8\%) are of Seyfert type 2.
Furthermore, we found 34 LINERs (20.5\%) and 4 BL~Lac objects (2.4\%).

The AGN type distribution in our newly identified RASS AGN 
sample is similar to the distribution
in the sample of Zimmermann et al.
 (\cite{Zimmermann01}) (see Fig.~8). 
 However, the relative number of the pure Seyfert 1 types is greater
in the sample of Zimmermann. On the other hand the relative number
of intermediate
Seyferts and LINERs is greater in our new RASS sample.
This may be caused by the 
high quality of our spectra which allowed a more detailed classification.

\subsection{Companion galaxies and galaxy pairs}
The source of the X-ray emission could not always
be identified  with one optical galaxy only.
In some cases we took more than one spectrum of objects within the
X-ray error box.

We assigned that galaxy to the X-ray source that
showed the highest activity degree in the optical spectrum.
Data for the galaxies
we identified as companion galaxies of the X-ray galaxies
 are given at the end of Table~2 and 3.  MCG ~-01-22-27 is the companion of
NGC~2617.



The X-ray quasar 1RXS~J085001.4~+~701804 is an interesting object.
This quasar  shines
through the outer disk of the spiral galaxy NGC~2650.
We see absorption lines from the foreground galaxy in the quasar spectrum
as a result of this projected superposition.

The rest of the new companion galaxies in our list are galaxy pairs.

\subsection{Line width distribution}

\begin{figure}[htbp]
        \centering
        \includegraphics[width=\columnwidth]{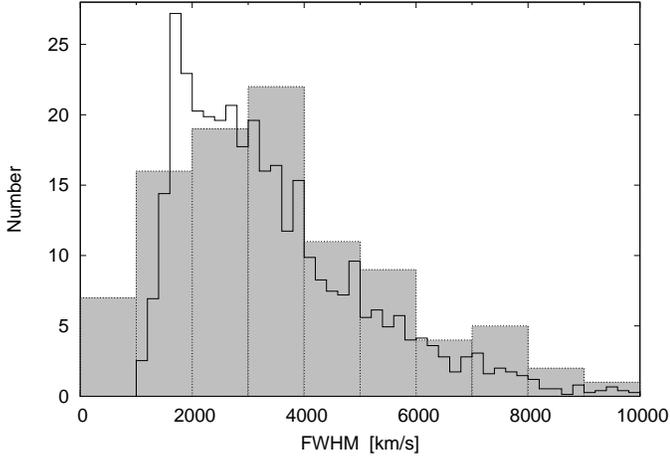}
        \caption{H$\alpha$ line width distribution (FWHM) of our
          newly identified ROSAT AGN sample in steps of 1000 km s$^{-1}$.
          The solid line gives the scaled-down distribution 
         of  H$\alpha$ line widths  of AGN derived from the 
         Sloan Digital Sky Survey (Hao et al. \cite{Hao05}).}
        \label{fig:hist_fwhm}
\end{figure}

We derived H$\alpha$, H$\beta$, and [\ion{O}{iii}]$\lambda$5007
line widths (FWHM)
of all emission line galaxies of our new sample
 from single or multi-component fits.
The spectral lines were adapted
by single-component fits for starburst HII galaxies, LINERs,
Seyfert 2 galaxies,
Narrow-Line Seyfert 1 galaxies (NLS1), and pure Seyfert 1 galaxies.
If the Balmer lines were fitted by more than one broad component and if the
broadest component had a relative line intensity of at least 30 percent
we present the line widths (FWHM) of the broadest component
in Table~3 (column 8) 
and Fig. \ref{fig:hist_fwhm}. 

We show the H$\alpha$ line width distribution (FWHM) of our
AGN sample in Fig.~\ref{fig:hist_fwhm}.
A step size of 1000 km s$^{-1}$ was used. 
For comparison we also plot (solid line) the relative distribution of the
H$\alpha$ line widths (FWHM) of Hao et al. (\cite{Hao05})
who used a sample of AGN
derived from the Sloan Digital Sky Survey (SDSS).  
The H$\alpha$ FWHM distribution of our ROSAT AGN sample 
resembles the H$\alpha$ FWHM distribution of the SDSS
AGN sample.

\subsection{Optical to blue color distribution}

Finally, we inspected the optical colors of our galaxy spectra to check
whether the hitherto unknown ROSAT selected AGN have specific
continuum  properties.
We derived B and V-band colors of our newly identified RASS-AGN with
the IRAF task 'sbands'. Not all of our spectra cover the full wavelength range
of the B-band filter. Therefore, we derived our B and V-band colors
with a bandpass width of $200\,$\AA\ only for the sake of homogeneity.
We compared these 'narrow-band' colors with 'full bandwidth' colors
whenever we had the full wavelength range at our disposal. 
Furthermore, we compared our 'narrow-band' colors with data from the
literature. They agree amongst each other within 0.1 mag.

\begin{figure}
  \centering
  \includegraphics[width=\columnwidth]{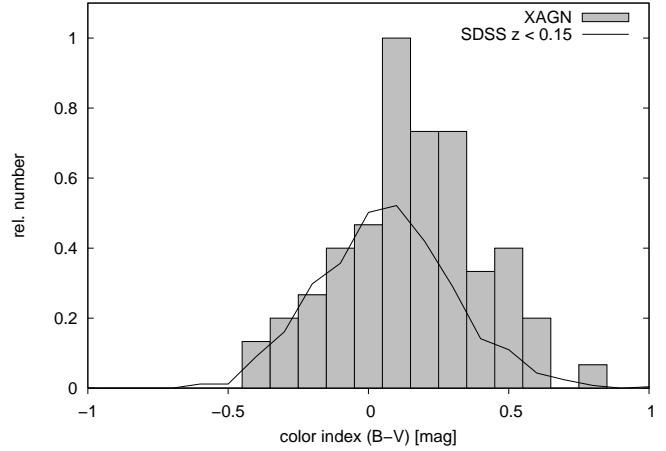}
  \caption{ Color distribution of our newly identified RASS AGN
 (shaded grey) and of a
    comparison sample consisting of all 764 z$\leq 0.15$ AGN from the
     Second Data Release of SDSS (black line).}
  \label{fig:hist_conticolor}
\end{figure}

We then compared the colors of our newly identified RASS AGN
with an independent sample of all 764 AGN spectra derived from the
Second Data Release of SDSS (Abazajian et al. \cite{Abazajian04}).
These SDSS-AGN were selected by their classification as QSO in the SDSS
database and limited to $z\leq 0.15$ to fit the redshift range of our sample.
We determined the colors of these SDSS-AGN
in the same way as explained above. The
results are shown in Fig. \ref{fig:hist_conticolor}.

The color distribution of the SDSS sample (black solid line) shows a
Gaussian-like shape with a mean value of ($\rm B-V\approx 0.05\,$mag).
The distribution of
our newly identified RASS AGN contains far more red galaxies.

The shift of the (B-V) distribution to the red by about 0.15 mag
might be caused by the fact that the newly identified RASS AGN are
systematically redder and/or that the relative number of
intermediate Seyfert galaxies and LINERs is higher compared to
optically selected AGN samples.
Intermediate Seyfert galaxies and LINERs have redder colors than pure
Sy~1 types.
We derived the following colors from the spectra of our
individual AGN types:
Sy~1: $0.04\pm0.16\,$mag,
Sy~1.5~-~1.9: $0.13\pm0.26\,$mag,
NLS1: $-0.09\pm0.21\,$mag,
Sy~2: $0.28\pm0.27\,$mag,
LINER: $0.34\pm0.16\,$mag.
%
%

\section{Discussion and conclusion}

About 550 compact X-ray sources have been identified as AGN or AGN candidates
by cross-correlating the ROSAT Bright Source Catalog with the
Principal Catalog of Galaxies. 350 of these sources have been identified
before as AGN on basis of their optical spectra. 
Cross-correlating the ROSAT Bright Source Catalog with the
Principal Catalog of Galaxies yielded a sample of 198 galaxies with as yet
unknown X-ray emission. 

We took spectra of as many  X-ray counterparts as possible in five optical
spectroscopic campaigns. At the end we obtained spectra of 166
X-ray counterparts. Since the start of our identification campaign,
 41 of these galaxies have likewise been
identified as AGN by other observers (see introduction).
Our AGN classification of the different types agrees with the
classification of  other authors - except for minor discrepancies in
the subdivision. For homogeneity we used
our spectra and classification schema for this investigation.

Most of our galaxies are nearby objects.
We determined 
redshifts up to 0.15 with a median of $z~=~ 0.033$.
Only two AGN of our new
sample are more distant
objects (up to $z=1.9$). 

The RASS selected AGN show a mean absolute magnitude of
$\rm M_B = (-20.71 \pm 1.75)$~mag in the optical. They are
hosted by brighter galaxies
compared to the AGN in the catalogue of
Veron-Cetty (\cite{Veroncat12ed}) (see Fig.~3).  Ho et al. (\cite{Ho97})
derived
a mean magnitude of $\rm M_B = -20.46\,$~mag from a sample of all nearby AGN.
Therefore it is not due to their optical faintness that our ROSAT selected AGN
have not been detected before.

Our X-ray luminosity distribution
of the RASS selected AGN peaks at $\log(L_{\mathrm{X}}
[\rm erg\ s^{-1}]) =42.9 \pm 1.7$ ( Fig. \ref{fig:hist_lx}). 
This value should be compared with the peak in the luminosity
 $\log(L_{\mathrm{X}}) \approx 43.5 \pm 1.1$
of X-ray AGN selected from the SDSS
(Anderson et al. \cite{Anderson03}, \cite{Anderson07}) with 
redshifts below $\rm z\leq 0.15$.
They identified mostly quasars and Seyfert 1 galaxies
in their sample which might have caused the  
peak at slightly higher luminosities in the  luminosity distribution.
 
We determined the activity type of our emission line galaxies
on the basis of their Balmer line widths as well as narrow
emission line ratios.
139 galaxies of our 166 ROSAT selected objects have been classified
as AGN.
 This indicates that the X-ray selection criterion is very successful
in finding new AGN.
 In the remaining 
27 galaxies, faint AGN might still be buried that could not be separated in 
our spectra from the galaxy emission.

We compare the relative numbers of our Seyfert, LINER,
H~II and absorption line galaxies with the numbers derived from the 
Palomar Survey of all nearby galaxies (Ho et al. \cite{Ho97}).
While the fraction of galaxies showing no sign 
of nuclear activity is the same in both samples (XAGN: 15.1\% vs Ho: 13.6\%) we
find far more Seyfert type nuclei (XAGN: 60.8\% vs Ho: 10.7\%) 
in our ROSAT selected AGN. Contrary to this, H~II galaxies
are found in nearly half of all Palomar galaxies (Ho: 42.4\% vs XAGN: 1.2\%).
 The relative number of LINERs is higher in the Palomar Survey
(XAGN: 20.5\% vs Ho: 32.7\%). This might be caused by the fact that it is more
difficult to detect weak LINER properties in distant galaxies because of
the lower S/N ratio. 

The fraction of our new ROSAT-selected
 Seyfert 1 galaxies (pure Sy1, intermediate Sy1, and NLS1) is very high 
compared to type 2 Seyferts (see Fig.~8).
The Seyfert 1/2 ratio is 11/1.
This ratio is strongly affected by selection effects as ROSAT was most 
sensitive in the soft X-ray band which is heavily absorbed
in type 2 AGN. Optical surveys do not suffer from this bias. Therefore, an 
 Sy1/Sy2 ratio of 1:1.4 or an Sy 1 fraction of 42\% as reported by Ho et al.
 (\cite{Ho97}) may better reflect the true ratio.

We classified 18 galaxies as Narrow Line Seyfert 1
 which is
20\% of all broad-line AGN (Sy1 to Sy1.9). Other studies of
RASS selected sources also detected many Narrow Line Seyfert 1 galaxies
(e.g. Boller et al. \cite{Boller96}, Grupe et al. \cite{Grupe99}).
 From NLS1 catalogs compiled from
the SDSS AGN Catalog, Williams et al.
 (\cite{Williams02}) found that roughly 15\% of all broad
line AGN are of NLS1-type, a number already suggested by Osterbrock
 (\cite{Osterbrock87})
and later confirmed by Zhou et al. 
 (\cite{Zhou07}), who determined an average fraction of 14\%
(up to  $\approx$20\% at M$_g \approx -22\,$mag) for their sample of $>2000$
NLS1-galaxies from SDSS-DR3.  

The linewidth distribution of our ROSAT sample matches that of 
the SDSS AGN (Hao et al. \cite{Hao05}). 

We also determined optical broad-band colors from the galaxy spectra. 
We found that our hitherto
 undetected ROSAT AGN have redder colors in comparison to
an independent sample of AGN spectra derived from the
Second Data Release of SDSS (Abazajian et al. \cite{Abazajian04})
(see Fig. \ref{fig:hist_conticolor}).


The redder color of our newly identified RASS AGN might explain why they
have not been detected in earlier AGN searches which
concentrated on blue or UV-excess galaxies or on optical identifications
based on low resolution objective prism spectra. All these surveys were not
complete and were biased towards single broad band or spectral properties.
 That our galaxies predominantly are fainter in the blue
might be caused by internal absorption of the central nonthermal
radiation within the host galaxy. Another explanation could be that the redder 
distribution of our newly identified RASS AGN might be caused by the fact
 that the relative number of
intermediate Seyfert galaxies and Liners in our sample is higher compared to 
other
optically selected AGN samples.
Intermediate Seyfert galaxies and Liners have redder colors than pure
Sy~1 types.

In a future paper
 we will supplement our optical and X-ray data with
data at radio and far-infrared wavelengths to further constrain our
classification and to gain a deeper insight into their X-ray production
mechanisms.

\begin{acknowledgements}
The ROSAT project was supported by the German Bundesministerium f\"ur Bildung,
Wissenschaft, Forschung und Technologie (BMBF/DARA) and by the Max Planck
Gesellschaft (MPG). \\
This work was partially supported by the \emph{Deut\-sche
For\-schungs\-ge\-mein\-schaft, DFG\/} project number Ko~857 and the
\emph{''Verbundforschung Astrophysik'' (Grant 05 AE5PD 1/4)}.

\end{acknowledgements}
%



\longtab{2}{\footnotesize
\label{tab:xraydata}
\begin{longtable}{lrrlr@{$\,\pm\,$}lrrr}
\caption{X-ray identification information} \\
\hline\hline 
ROSAT-Name & 
\multicolumn{2}{c}{RA, DEC (J2000.0)} & 
Name & 
\multicolumn{2}{c}{Countrate} & 
$\log(\rm L_X)$ & 
$\rm m_B$ & 
$\rm M_B$ \\ 
 & 
[h m s] & 
[d m s] & 
 & 
\multicolumn{2}{c}{[cts s$^{-1}$]} & 
$[\rm erg s^{-1}]$ & [mag] & [mag] \\
(1) & (2) & (3) & (4) & \multicolumn{2}{c}{(5)} & (6) & (7) & (8) \\
\hline \endfirsthead

\caption{(continued)} \\
\hline\hline 
ROSAT-Name & 
\multicolumn{2}{c}{RA, DEC (J2000.0)} & 
Name & 
\multicolumn{2}{c}{Countrate} & 
$\log(\rm L_X)$ & 
$\rm m_B$ & 
$\rm M_B$ \\ 
 & 
[h m s] & 
[d m s] & 
 & 
\multicolumn{2}{c}{[cts s$^{-1}$]} & 
$[\rm erg s^{-1}]$ & [mag] & [mag] \\
(1) & (2) & (3) & (4) & \multicolumn{2}{c}{(5)} & (6) & (7) & (8) \\
\hline \endhead
\hline\hline \endfoot

$\rm 1RXS~J000156.7-273748$ & 
         $00~01~55.8$ &  $-27~37~38$ &  ESO~409-003 &  
         $0.067$ &  $0.0171$ &  $42.55$ &  $14.66$ &  $-20.63$ \\

$\rm 1RXS~J000459.1+114205$ & 
         $00~04~58.4$ &  $11~42~03$ &  UGC~32 &  
         $0.148$ &  $0.0205$ &  $43.77$ &  $14.20$ &  $-23.20$ \\

$\rm 1RXS~J000805.6+145027$ & 
         $00~08~05.6$ &  $14~50~23$ &  CGCG~433-025 &  
         $0.179$ &  $0.0180$ &  $43.64$ &  $15.70$ &  $-20.60$ \\

$\rm 1RXS~J001530.2+172009$ & 
         $00~15~30.9$ &  $17~19~42$ &  NGC~57 &  
         $0.072$ &  $0.0198$ &  $42.23$ &  $13.14$ &  $-21.23$ \\

$\rm 1RXS~J001823.8+300357$ & 
         $00~17~23.6$ &  $30~08~48$ &  NGC~71 &  
         $0.006$ &  $0.0202$ &  $42.40$ &  $14.42$ &  $-20.36$ \\

$\rm 1RXS~J001823.8+300357$ & 
         $00~18~22.8$ &  $30~04~47$ &  NGC~70 &  
         $0.006$ &  $0.0202$ &  $42.30$ &  $14.18$ &  $-20.71$ \\

$\rm 1RXS~J002108.1-190950$ & 
         $00~21~13.2$ &  $-19~10~44$ &  PKS~0018-19 &  
         $0.104$ &  $0.0195$ &  $43.81$ &  $17.00$ &  $-20.95$ \\

$\rm 1RXS~J002534.9-330255$ & 
         $00~25~31.3$ &  $-33~02~48$ &  ESO~350-015 &  
         $0.249$ &  $0.0310$ &  $43.62$ &  $14.23$ &  $-22.31$ \\

$\rm 1RXS~J003413.7-212619$ & 
         $00~34~13.5$ &  $-21~26~20$ &  HCG~4a &  
         $0.330$ &  $0.0350$ &  $43.14$ &  $13.70$ &  $-21.43$ \\

$\rm 1RXS~J004236.9-104919$ & 
         $00~42~36~7$ &  $-10~49~22$ &  VIII~Zw~36 &  
         $0.244$ &  $0.0280$ &  $43.55$ &  $14.60$ &  $-21.51$ \\

$\rm 1RXS~J010013.4-151755$ & 
         $01~00~15.9$ &  $-15~17~57$ &  MCG~-03-03-017 &  
         $0.151$ &  $0.0207$ &  $43.47$ &  $14.90$ &  $-21.80$ \\

$\rm 1RXS~J010517.5-582618$ & 
         $01~05~16.5$ &  $-58~26~13$ &  ESO~113-010 &  
         $0.182$ &  $0.0390$ &  $43.40$ &  $14.60$ &  $-20.47$ \\

$\rm 1RXS~J010918.0+131011$ & 
         $01~09~18.4$ &  $13~10~08$ &  UGC~716 &  
         $0.069$ &  $0.0150$ &  $43.45$ &  $15.60$ &  $-21.30$ \\

$\rm 1RXS~J011219.5-320338$ & 
         $01~12~19.2$ &  $-32~03~44$ &  NGC~427 &  
         $0.153$ &  $0.0230$ &  $42.88$ &  $15.10$ &  $-20.55$ \\

$\rm 1RXS~J011233.3-320139$ & 
         $01~12~32.6$ &  $-32~01~44$ &  RX~J011232.8-320140 &  
         $0.033$ &  $0.0020$ &  $~$ &  $21.10$ &  $~$ \\

$\rm 1RXS~J012018.8-440748$ & 
         $01~20~19.6$ &  $-44~07~43$ &  ESO~244-017 &  
         $0.305$ &  $0.0310$ &  $43.12$ &  $14.60$ &  $-20.28$ \\

$\rm 1RXS~J013124.2+330837$ & 
         $01~31~23.8$ &  $33~08~38$ &  KUG~0128+328 &  
         $0.104$ &  $0.0187$ &  $43.56$ &  $16.50$ &  $-20.78$ \\

$\rm 1RXS~J014526.6-034945$ & 
         $01~45~25.3$ &  $-03~49~39$ &  MCG~-01-05-031 &  
         $0.175$ &  $0.0290$ &  $42.70$ &  $13.30$ &  $-21.00$ \\

$\rm 1RXS~J014739.5-660952$ & 
         $01~47~39.5$ &  $-66~09~49$ &  ESO~080-005 &  
         $0.091$ &  $0.0200$ &  $42.32$ &  $16.00$ &  $-19.17$ \\

$\rm 1RXS~J023513.9-293616$ & 
         $02~35~13.4$ &  $-29~36~18$ &  ESO~416-002 &  
         $0.356$ &  $0.0340$ &  $43.56$ &  $14.90$ &  $-22.00$ \\

$\rm 1RXS~J023536.7-293845$ & 
         $02~35~36.6$ &  $-29~38~44$ &  PHL~1389 &  
         $0.043$ &  $0.0040$ &  $~$ &  $16.10$ &  $~$ \\

$\rm 1RXS~J025552.4+091853$ & 
         $02~55~52.2$ &  $09~18~42$ &  IC~1867 &  
         $0.083$ &  $0.0190$ &  $43.02$ &  $14.60$ &  $-20.46$ \\

$\rm 1RXS~J030606.3-390212$ & 
         $03~06~05.9$ &  $-30~02~10$ &  NGC~1217 &  
         $0.193$ &  $0.0230$ &  $42.77$ &  $13.30$ &  $-21.28$ \\

$\rm 1RXS~J030825.9+040637$ & 
         $03~08~26.3$ &  $04~06~40$ &  NGC~1218 &  
         $0.175$ &  $0.0210$ &  $43.35$ &  $13.90$ &  $-21.41$ \\

$\rm 1RXS~J034203.8-211428$ & 
         $03~42~02.8$ &  $-21~14~26$ &  ESO~548-081 &  
         $0.258$ &  $0.0260$ &  $42.63$ &  $12.80$ &  $-20.90$ \\

$\rm 1RXS~J042710.2-624712$ & 
         $04~27~12.5$ &  $-62~47~09$ &  AM~0426-625 &  
         $0.097$ &  $0.0150$ &  $42.41$ &  $13.00$ &  $-21.33$ \\

$\rm 1RXS~J043520.2-780150$ & 
         $04~35~16.2$ &  $-78~01~57$ &  ESO~015-011 &  
         $0.274$ &  $0.0250$ &  $44.10$ &  $14.20$ &  $-22.76$ \\

$\rm 1RXS~J043813.8-104740$ & 
         $04~38~13.8$ &  $-10~47~45$ &  MCG~-02-12-050 &  
         $0.066$ &  $0.0160$ &  $43.08$ &  $15.20$ &  $-20.60$ \\

$\rm 1RXS~J044148.1-011806$ & 
         $04~41~48.3$ &  $-01~18~10$ &  UGC~3134 &  
         $0.083$ &  $0.0160$ &  $42.93$ &  $14.20$ &  $-21.13$ \\

$\rm 1RXS~J045142.3-034834$ & 
         $04~51~41.4$ &  $-03~48~34$ &  MCG~-01-13-025 &  
         $0.281$ &  $0.0290$ &  $42.90$ &  $14.60$ &  $-19.42$ \\

$\rm 1RXS~J045256.6+011854$ & 
         $04~52~58.9$ &  $01~19~53$ &  UGC~3194 &  
         $0.053$ &  $0.0140$ &  $42.03$ &  $14.90$ &  $-20.35$ \\

$\rm 1RXS~J045840.4-215922$ & 
         $04~58~40.3$ &  $-21~59~32$ &  ESO~552-039 &  
         $0.106$ &  $0.0180$ &  $43.47$ &  $16.30$ &  $-19.72$ \\

$\rm 1RXS~J045859.6-002904$ & 
         $04~58~54.6$ &  $-00~29~20$ &  NGC~1713 &  
         $0.120$ &  $0.0199$ &  $41.84$ &  $13.30$ &  $-20.59$ \\

$\rm 1RXS~J050820.9+172134$ & 
         $05~08~20.5$ &  $17~21~58$ &  CGCG~469-001 &  
         $0.060$ &  $0.0128$ &  $41.77$ &  $15.40$ &  $-19.05$ \\

$\rm 1RXS~J051621.5-103341$ & 
         $05~16~21.2$ &  $-10~33~40$ &  MCG~-02-14-009 &  
         $0.307$ &  $0.0290$ &  $43.64$ &  $15.50$ &  $-19.79$ \\

$\rm 1RXS~J054715.3+505209$ & 
         $05~47~14.9$ &  $50~52~15$ &  UGC~3355 &  
         $0.057$ &  $0.0144$ &  $42.01$ &  $15.60$ &  $-19.50$ \\

$\rm 1RXS~J055559.4-612438$ & 
         $05~55~52.6$ &  $-61~24~14$ &  ESO~120-023 &  
         $0.093$ &  $0.0070$ &  $43.23$ &  $15.30$ &  $-20.60$ \\

$\rm 1RXS~J060635.4-472957$ & 
         $06~06~35.8$ &  $-47~29~56$ &  ESO~254-017 &  
         $0.213$ &  $0.0160$ &  $43.25$ &  $14.60$ &  $-20.79$ \\

$\rm 1RXS~J062122.8-280712$ & 
         $06~21~26.2$ &  $-28~06~53$ &  ESO~425-019 &  
         $0.077$ &  $0.0129$ &  $41.95$ &  $13.40$ &  $-21.38$ \\

$\rm 1RXS~J062307.7-643618$ & 
         $06~23~07.7$ &  $-64~36~21$ &  PMN~J0623-6436 &  
         $0.417$ &  $0.0130$ &  $44.84$ &  $14.80$ &  $-23.82$ \\

$\rm 1RXS~J063059.7-240636$ & 
         $06~30~59.4$ &  $-24~06~46$ &  PMN~J0630-2406 &  
         $0.084$ &  $0.0140$ &  $~$ &  $15.40$ &  $~$ \\

$\rm 1RXS~J064011.5-255337$ & 
         $06~40~11.8$ &  $-25~53~43$ &  ESO~490-026 &  
         $0.273$ &  $0.0290$ &  $43.24$ &  $14.10$ &  $-20.90$ \\

$\rm 1RXS~J064710.4+741433$ & 
         $06~47~14.0$ &  $74~14~12$ &  NGC~2256 &  
         $0.078$ &  $0.0148$ &  $41.76$ &  $14.00$ &  $-20.19$ \\

$\rm 1RXS~J064748.3+742850$ & 
         $06~47~45.8$ &  $74~28~54$ &  NGC~2258 &  
         $0.062$ &  $0.0131$ &  $41.45$ &  $13.00$ &  $-20.67$ \\

$\rm 1RXS~J065214.0+193147$ & 
         $06~52~15.1$ &  $19~31~57$ &  CGCG~085-010 &  
         $0.080$ &  $0.0153$ &  $42.90$ &  $15.60$ &  $-21.36$ \\

$\rm 1RXS~J070907.6+483657$ & 
         $07~09~08.0$ &  $49~36~56$ &  NGC~2329 &  
         $0.106$ &  $0.0194$ &  $42.01$ &  $13.10$ &  $-21.37$ \\

$\rm 1RXS~J071148.0+321902$ & 
         $07~11~47.7$ &  $32~18~36$ &  B2~0708+32b &  
         $0.332$ &  $0.0293$ &  $43.58$ &  $15.80$ &  $-21.35$ \\

$\rm 1RXS~J071204.2-603005$ & 
         $07~12~03.3$ &  $-60~30~30$ &  ESO~122-016~(NE) &  
         $0.106$ &  $0.1000$ &  $43.30$ &  $14.60$ &  $-21.00$ \\

$\rm 1RXS~J073353.4+491737$ & 
         $07~33~53.1$ &  $49~17~31$ &  UGC~3901 &  
         $0.080$ &  $0.0164$ &  $41.98$ &  $15.30$ &  $-19.40$ \\

$\rm 1RXS~J073727.1+594143$ & 
         $07~37~30.1$ &  $59~41~03$ &  UGC~3927 &  
         $0.062$ &  $0.0177$ &  $42.40$ &  $15.50$ &  $-20.61$ \\

$\rm 1RXS~J074701.7+413211$ & 
         $07~47~02.0$ &  $41~32~10$ &  UGC~4018 &  
         $0.077$ &  $0.0143$ &  $42.19$ &  $14.80$ &  $-20.52$ \\

$\rm 1RXS~J075151.6+494854$ & 
         $07~51~51.9$ &  $49~48~52$ &  MCG~+08-15-09 &  
         $0.067$ &  $0.0141$ &  $42.01$ &  $15.10$ &  $-19.93$ \\

$\rm 1RXS~J080157.7-494639$ & 
         $08~01~57.9$ &  $-49~46~42$ &  ESO~209-012 &  
         $0.163$ &  $0.0160$ &  $44.19$ &  $15.30$ &  $-20.77$ \\

$\rm 1RXS~J081021.3+421657$ & 
         $08~10~23.3$ &  $42~16~26$ &  CGCG~207-040 &  
         $0.138$ &  $0.0215$ &  $43.14$ &  $15.60$ &  $-21.48$ \\

$\rm 1RXS~J081517.8+460429$ & 
         $08~15~16.9$ &  $46~04~31$ &  MGC~+08-15-56 &  
         $0.138$ &  $0.0190$ &  $42.76$ &  $15.20$ &  $-20.95$ \\

$\rm 1RXS~J082321.4+042231$ & 
         $08~23~21.7$ &  $04~22~21$ &  IC~505 &  
         $0.095$ &  $0.0231$ &  $42.30$ &  $14.80$ &  $-20.80$ \\

$\rm 1RXS~J083539.1-040508$ & 
         $08~35~38.8$ &  $-04~05~17$ &  NGC~2617 &  
         $0.193$ &  $0.0313$ &  $41.99$ &  $14.00$ &  $-19.93$ \\

$\rm 1RXS~J084456.2+425826$ & 
         $08~44~56.6$ &  $42~58~35$ &  MCG~+07-18-43 &  
         $0.064$ &  $0.0134$ &  $42.56$ &  $15.00$ &  $-21.70$ \\

$\rm 1RXS~J085001.4+701804$ & 
         $08~50~02.3$ &  $70~18~06$ &  QSO east~of NGC~2650 &  
         $0.057$ &  $0.0144$ &  $46.01$ &  $16.00$ &  $-28.91$ \\

$\rm 1RXS~J085617.7-013809$ & 
         $08~56~17.8$ &  $-01~39~08$ &  CGCG~005-037 &  
         $0.128$ &  $0.0219$ &  $42.95$ &  $15.50$ &  $-21.41$ \\

$\rm 1RXS~J091345.4+474208$ & 
         $09~13~45.4$ &  $47~42~06$ &  MCG~+08-17-60 &  
         $0.148$ &  $0.0212$ &  $42.82$ &  $15.90$ &  $-20.77$ \\

$\rm 1RXS~J092115.9+101747$ & 
         $09~21~15.5$ &  $10~17~41$ &  VIII~Zw~45 &  
         $0.141$ &  $0.0296$ &  $42.69$ &  $15.40$ &  $-20.65$ \\

$\rm 1RXS~J093308.2+534754$ & 
         $09~33~08.9$ &  $53~47~49$ &  KUG~0929+540 &  
         $0.067$ &  $0.0181$ &  $42.56$ &  $17.00$ &  $-19.85$ \\

$\rm 1RXS~J093642.6+505249$ & 
         $09~36~43.1$ &  $50~52~49$ &  KUG~0933+511 &  
         $0.054$ &  $0.0133$ &  $42.40$ &  $15.50$ &  $-21.29$ \\

$\rm 1RXS~J094204.0+234106$ & 
         $09~42~04.8$ &  $23~41~07$ &  CGCG~122-055 &  
         $0.120$ &  $0.0181$ &  $42.06$ &  $15.30$ &  $-19.49$ \\

$\rm 1RXS~J095154.9-064916$ & 
         $09~51~55.0$ &  $-06~49~23$ &  NGC~3035 &  
         $0.263$ &  $0.0274$ &  $42.14$ &  $13.50$ &  $-20.42$ \\

$\rm 1RXS~J095948.0+112926$ & 
         $09~59~46.8$ &  $11~28~20$ &  CGCG~064-024 &  
         $0.159$ &  $0.0238$ &  $43.31$ &  $15.70$ &  $-21.80$ \\

$\rm 1RXS~J100554.9-230318$ & 
         $10~05~55.4$ &  $-23~03~25$ &  ESO~499-041 &  
         $0.131$ &  $0.0188$ &  $41.71$ &  $13.40$ &  $-20.15$ \\

$\rm 1RXS~J102403.1+062954$ & 
         $10~23~59.8$ &  $06~29~08$ &  CGCG~037-022 &  
         $0.119$ &  $0.0198$ &  $42.66$ &  $16.00$ &  $-20.28$ \\

$\rm 1RXS~J102445.5+062455$ & 
         $10~24~43.5$ &  $06~26~03$ &  CGCG~037-028 &  
         $0.052$ &  $0.0141$ &  $42.27$ &  $15.50$ &  $-20.75$ \\

$\rm 1RXS~J102841.7+490425$ & 
         $10~28~42.0$ &  $49~04~19$ &  CGCG~240-047 &  
         $0.073$ &  $0.0138$ &  $42.33$ &  $15.30$ &  $-20.95$ \\

$\rm 1RXS~J104333.4+010109$ & 
         $10~43~32.9$ &  $01~01~09$ &  GNB~069 &  
         $0.070$ &  $0.0183$ &  $42.91$ &  $16.80$ &  $-20.54$ \\

$\rm 1RXS~J104439.4+384541$ & 
         $10~44~39.2$ &  $38~45~34$ &  CGCG~212-045 &  
         $0.351$ &  $0.0303$ &  $42.89$ &  $15.00$ &  $-20.92$ \\

$\rm 1RXS~J110312.1+414200$ & 
         $11~03~11.0$ &  $41~42~19$ &  MCG~+07-23-15 &  
         $0.117$ &  $0.0216$ &  $42.22$ &  $15.10$ &  $-20.41$ \\

$\rm 1RXS~J110941.7-033915$ & 
         $11~09~42.9$ &  $-03~49~03$ &  CGCG~011-012 &  
         $0.161$ &  $0.0243$ &  $42.76$ &  $15.40$ &  $-20.59$ \\

$\rm 1RXS~J114014.0+244150$ & 
         $11~40~13.9$ &  $24~41~49$ &  NGC~3798 &  
         $0.065$ &  $0.0137$ &  $41.20$ &  $13.90$ &  $-19.51$ \\

$\rm 1RXS~J114429.9+365314$ & 
         $11~44~29.9$ &  $36~53~08$ &  KUG~1141+371 &  
         $1.368$ &  $0.0740$ &  $43.53$ &  $16.50$ &  $-19.48$ \\

$\rm 1RXS~J114604.3-081550$ & 
         $11~49~03.8$ &  $-08~16~04$ &  IC~734 &  
         $0.080$ &  $0.0183$ &  $43.07$ &  $16.00$ &  $-21.55$ \\

$\rm 1RXS~J115238.2-051229$ & 
         $11~52~36.2$ &  $-05~12~25$ &  MCG~-01-30-041 &  
         $0.108$ &  $0.0190$ &  $42.41$ &  $13.80$ &  $-20.62$ \\

$\rm 1RXS~J115537.0+125251$ & 
         $11~55~35.2$ &  $12~52~19$ &  VIII~Zw~153 &  
         $0.134$ &  $0.0291$ &  $43.54$ &  $17.90$ &  $-20.58$ \\

$\rm 1RXS~J120452.8-434353$ & 
         $12~04~52.9$ &  $-43~43~54$ &  ESO~267-013 &  
         $0.110$ &  $0.0279$ &  $42.19$ &  $14.15$ &  $-19.68$ \\

$\rm 1RXS~J120655.0+501744$ & 
         $12~06~55.6$ &  $50~17~37$ &  MCG~+08-22-71 &  
         $0.074$ &  $0.0160$ &  $42.70$ &  $17.00$ &  $-20.03$ \\

$\rm 1RXS~J120719.4+241204$ & 
         $12~07~19.8$ &  $24~11~56$ &  Mrk~648 &  
         $0.196$ &  $0.0235$ &  $42.93$ &  $16.00$ &  $-20.61$ \\

$\rm 1RXS~J121754.7+583936$ & 
         $12~17~55.0$ &  $58~39~35$ &  Mrk~202 &  
         $0.213$ &  $0.0246$ &  $42.25$ &  $15.60$ &  $-19.33$ \\

$\rm 1RXS~J123651.1+453907$ & 
         $12~36~51.2$ &  $45~39~04$ &  MCG~+08-23-67 &  
         $0.526$ &  $0.0356$ &  $42.89$ &  $15.20$ &  $-20.31$ \\

$\rm 1RXS~J124046.8-333408$ & 
         $12~40~47.0$ &  $-33~34~12$ &  ESO~381-007 &  
         $0.446$ &  $0.0455$ &  $43.95$ &  $15.94$ &  $-20.80$ \\

$\rm 1RXS~J124306.5+353859$ & 
         $12~43~07.4$ &  $35~39~05$ &  KUG~1240+359 &  
         $0.088$ &  $0.0158$ &  $44.68$ &  $16.50$ &  $-25.43$ \\

$\rm 1RXS~J125042.5-024929$ & 
         $12~50~42.4$ &  $-02~49~31$ &  CGCG~015-026 &  
         $0.068$ &  $0.0164$ &  $43.01$ &  $15.70$ &  $-20.71$ \\

$\rm 1RXS~J125253.5-152445$ & 
         $12~52~52.6$ &  $-15~24~48$ &  NGC~4756 &  
         $0.068$ &  $0.0185$ &  $41.43$ &  $13.00$ &  $-20.63$ \\

$\rm 1RXS~J125347.3+032620$ & 
         $12~53~47.0$ &  $03~26~30$ &  CGCG~043-056 &  
         $0.186$ &  $0.0422$ &  $43.73$ &  $15.70$ &  $-21.42$ \\

$\rm 1RXS~J125609.5-080906$ & 
         $12~56~10.1$ &  $-08~09~05$ &  MCG~-01-33-054 &  
         $0.164$ &  $0.0259$ &  $42.27$ &  $14.54$ &  $-19.07$ \\

$\rm 1RXS~J125851.4+235532$ & 
         $12~58~51.5$ &  $23~55~26$ &  KUG~1256+241 &  
         $0.430$ &  $0.0319$ &  $43.56$ &  $17.00$ &  $-20.43$ \\

$\rm 1RXS~J130158.9+274708$ & 
         $13~02~00.1$ &  $27~46~58$ &  CGCG~160-104 &  
         $0.124$ &  $0.0195$ &  $42.00$ &  $15.40$ &  $-19.61$ \\

$\rm 1RXS~J130456.7+395537$ & 
         $13~04~57.0$ &  $39~55~30$ &  IC~4165 &  
         $0.054$ &  $0.0119$ &  $41.79$ &  $15.00$ &  $-20.27$ \\

$\rm 1RXS~J131633.1+005249$ & 
         $13~16~27.0$ &  $00~51~12$ &  VIII~Zw~272 &  
         $0.053$ &  $0.0159$ &  $42.75$ &  $17.30$ &  $-20.18$ \\

$\rm 1RXS~J131905.8+310854$ & 
         $13~19~06.0$ &  $31~08~53$ &  CGCG~160-193 &  
         $0.113$ &  $0.0180$ &  $42.23$ &  $15.60$ &  $-20.01$ \\

$\rm 1RXS~J132016.3+330828$ & 
         $13~20~14.6$ &  $33~08~40$ &  NGC~5098~(W) &  
         $0.180$ &  $0.0214$ &  $42.51$ &  $15.00$ &  $-20.85$ \\

$\rm 1RXS~J133151.1+603451$ & 
         $13~31~50.7$ &  $60~34~48$ &  MCG~+10-19-84 &  
         $0.125$ &  $0.0156$ &  $43.45$ &  $16.00$ &  $-22.43$ \\

$\rm 1RXS~J133908.5+115855$ & 
         $13~39~08.5$ &  $11~58~54$ &  VIII~Zw~327 &  
         $0.133$ &  $0.0227$ &  $43.30$ &  $17.80$ &  $-20.10$ \\

$\rm 1RXS~J134352.7+803544$ & 
         $13~43~54.5$ &  $80~35~58$ &  CGCG~365-016 &  
         $0.050$ &  $0.0098$ &  $42.35$ &  $15.40$ &  $-20.90$ \\

$\rm 1RXS~J135420.2+325547$ & 
         $13~53~20.0$ &  $32~55~48$ &  UGC~8829 &  
         $0.835$ &  $0.0459$ &  $42.93$ &  $14.80$ &  $-20.36$ \\

$\rm 1RXS~J135536.3+155755$ & 
         $13~55~37.0$ &  $15~58~14$ &  CGCG~103-009 &  
         $0.112$ &  $0.0206$ &  $41.69$ &  $15.70$ &  $-18.43$ \\

$\rm 1RXS~J135625.3+283138$ & 
         $13~56~25.3$ &  $28~31~35$ &  MCG~+05-33-24a &  
         $0.059$ &  $0.0147$ &  $42.69$ &  $15.60$ &  $-21.84$ \\

$\rm 1RXS~J140543.6+405115$ & 
         $14~05~44.4$ &  $40~51~16$ &  CGCG~219-050 &  
         $0.164$ &  $0.0207$ &  $43.03$ &  $15.70$ &  $-21.54$ \\

$\rm 1RXS~J140619.7-163721$ & 
         $14~06~20.7$ &  $-16~37~00$ &  NPM1G~-16.0443 &  
         $0.057$ &  $0.0155$ &  $42.50$ &  $16.09$ &  $-19.25$ \\

$\rm 1RXS~J140723.3+150447$ & 
         $14~07~22.3$ &  $15~04~39$ &  CGCG~103-073 &  
         $0.139$ &  $0.0232$ &  $42.14$ &  $14.60$ &  $-20.49$ \\

$\rm 1RXS~J141802.6+800710$ & 
         $14~17~49.0$ &  $80~06~54$ &  CGCG~353-034 &  
         $0.057$ &  $0.0107$ &  $42.39$ &  $15.30$ &  $-21.04$ \\

$\rm 1RXS~J141922.5-263842$ & 
         $14~19~22.4$ &  $-26~38~41$ &  ESO~511-030 &  
         $1.221$ &  $0.0858$ &  $43.61$ &  $13.30$ &  $-21.48$ \\

$\rm 1RXS~J142916.8+300520$ & 
         $14~29~11.7$ &  $30~04~38$ &  MCG~+05-34-053 &  
         $0.052$ &  $0.0132$ &  $41.89$ &  $15.50$ &  $-18.38$ \\

$\rm 1RXS~J143207.9+313504$ & 
         $14~32~09.0$ &  $31~35~05$ &  CGCG~163-074 &  
         $0.079$ &  $0.0154$ &  $43.22$ &  $15.50$ &  $-21.26$ \\

$\rm 1RXS~J143450.7+033834$ & 
         $14~34~50.6$ &  $03~38~43$ &  CGCG~047-107 &  
         $0.056$ &  $0.0164$ &  $41.95$ &  $15.30$ &  $-20.04$ \\

$\rm 1RXS~J151447.8-402157$ & 
         $15~14~47.2$ &  $-40~21~35$ &  ESO~328-036-S &  
         $0.162$ &  $0.0236$ &  $42.79$ &  $15.85$ &  $-19.08$ \\

$\rm 1RXS~J151639.8+001454$ & 
         $15~16~40.2$ &  $00~15~02$ &  CGCG~021-063 &  
         $0.078$ &  $0.0190$ &  $43.16$ &  $15.60$ &  $-21.02$ \\

$\rm 1RXS~J151750.8+050615$ & 
         $15~17~51.7$ &  $05~06~28$ &  CGCG~049-106 &  
         $0.107$ &  $0.0194$ &  $43.03$ &  $15.60$ &  $-20.38$ \\

$\rm 1RXS~J153210.0+585434$ & 
         $15~32~16.1$ &  $58~54~04$ &  VII~Zw~608 &  
         $0.028$ &  $0.0086$ &  $42.90$ &  $17.60$ &  $-19.49$ \\

$\rm 1RXS~J153723.0-163551$ & 
         $15~37~22.4$ &  $-16~35~45$ &  NGC~5959 &  
         $0.054$ &  $0.0139$ &  $42.34$ &  $14.50$ &  $-20.47$ \\

$\rm 1RXS~J153845.4-032309$ & 
         $15~38~44.7$ &  $-03~22~48$ &  CGCG~022-021 &  
         $0.070$ &  $0.0149$ &  $42.44$ &  $15.29$ &  $-19.65$ \\

$\rm 1RXS~J155625.4+090311$ & 
         $15~56~25.9$ &  $09~03~19$ &  Mrk~863 &  
         $0.115$ &  $0.0169$ &  $43.15$ &  $15.30$ &  $-20.88$ \\

$\rm 1RXS~J161040.5+020626$ & 
         $16~10~40.6$ &  $02~06~39$ &  CGCG~023-021 &  
         $0.105$ &  $0.0169$ &  $42.76$ &  $15.40$ &  $-19.91$ \\

$\rm 1RXS~J161951.7+405834$ & 
         $16~19~51.3$ &  $40~58~47$ &  KUG~1618+410 &  
         $0.510$ &  $0.0289$ &  $43.69$ &  $16.00$ &  $-19.92$ \\

$\rm 1RXS~J162013.1+400858$ & 
         $16~20~12.8$ &  $40~09~06$ &  KUG~1618+402 &  
         $0.356$ &  $0.0239$ &  $43.28$ &  $16.00$ &  $-19.30$ \\

$\rm 1RXS~J162302.0+375506$ & 
         $16~23~03.1$ &  $37~55~21$ &  NGC~6137 &  
         $0.087$ &  $0.0142$ &  $42.06$ &  $13.40$ &  $-22.10$ \\

$\rm 1RXS~J162637.1+350239$ & 
         $16~26~36.4$ &  $35~02~42$ &  CGCG~196-064 &  
         $0.081$ &  $0.0128$ &  $42.81$ &  $15.70$ &  $-20.02$ \\

$\rm 1RXS~J162724.3+424041$ & 
         $16~27~25.2$ &  $42~40~47$ &  NGC~6159 &  
         $0.108$ &  $0.0135$ &  $42.86$ &  $15.20$ &  $-20.32$ \\

$\rm 1RXS~J162741.8+405536$ & 
         $16~27~41.1$ &  $40~55~37$ &  NGC~6160 &  
         $0.069$ &  $0.0122$ &  $42.68$ &  $14.20$ &  $-21.36$ \\

$\rm 1RXS~J170328.3+360400$ & 
         $17~03~27.8$ &  $36~04~20$ &  MCG~+06-37-023 &  
         $0.520$ &  $0.0095$ &  $44.15$ &  $15.60$ &  $-21.43$ \\

$\rm 1RXS~J171227.2+355256$ & 
         $17~12~28.4$ &  $35~53~03$ &  MCG~+06-38-005 &  
         $0.125$ &  $0.0139$ &  $42.79$ &  $15.00$ &  $-20.16$ \\

$\rm 1RXS~J171821.4+780118$ & 
         $17~18~16.6$ &  $78~01~05$ &  MCG~+13-12-022 &  
         $0.126$ &  $0.0102$ &  $43.45$ &  $15.30$ &  $-21.51$ \\

$\rm 1RXS~J172215.7+304250$ & 
         $17~22~15.4$ &  $30~42~40$ &  MCG~+05-41-010 &  
         $0.077$ &  $0.0137$ &  $43.06$ &  $15.60$ &  $-20.78$ \\

$\rm 1RXS~J172324.0+565840$ & 
         $17~23~25.2$ &  $56~58~28$ &  NGC~6370 &  
         $0.054$ &  $0.0079$ &  $42.45$ &  $13.88$ &  $-21.36$ \\

$\rm 1RXS~J173030.1+742244$ & 
         $17~30~36.9$ &  $74~22~34$ &  NGC~6414 &  
         $0.072$ &  $0.0070$ &  $42.95$ &  $15.40$ &  $-20.75$ \\

$\rm 1RXS~J180242.4+424734$ & 
         $18~02~39.8$ &  $42~47~46$ &  MCG~+07-37-018~(NE) &  
         $0.082$ &  $0.0114$ &  $43.16$ &  $14.40$ &  $-22.16$ \\

$\rm 1RXS~J180634.6+613554$ & 
         $18~06~35.6$ &  $61~35~38$ &  MCG~+10-26-015 &  
         $0.091$ &  $0.0054$ &  $42.66$ &  $16.20$ &  $-18.98$ \\

$\rm 1RXS~J182715.1+195624$ & 
         $18~27~14.8$ &  $19~56~19$ &  MCG~+03-47-002 &  
         $0.129$ &  $0.0179$ &  $43.22$ &  $15.30$ &  $-20.77$ \\

$\rm 1RXS~J184038.1-770930$ & 
         $18~40~38.5$ &  $-77~09~29$ &  ESO~045-011 &  
         $0.449$ &  $0.0508$ &  $42.98$ &  $13.70$ &  $-20.59$ \\

$\rm 1RXS~J190937.0-622853$ & 
         $19~09~33.3$ &  $-62~28~40$ &  ESO~104-041 &  
         $0.241$ &  $0.0467$ &  $44.07$ &  $17.46$ &  $-20.23$ \\

$\rm 1RXS~J192700.9-534241$ & 
         $19~27~01.6$ &  $-53~42~52$ &  ESO~184-068 &  
         $0.309$ &  $0.0811$ &  $43.85$ &  $15.99$ &  $-20.88$ \\

$\rm 1RXS~J193139.0-335426$ & 
         $19~31~38.6$ &  $-33~54~43$ &  PKS~1928-34 &  
         $0.254$ &  $0.0290$ &  $44.23$ &  $17.00$ &  $-21.02$ \\

$\rm 1RXS~J200052.6-343808$ & 
         $20~00~46.6$ &  $-34~38~00$ &  ESO~399-015 &  
         $0.066$ &  $0.1920$ &  $42.47$ &  $15.02$ &  $-20.06$ \\

$\rm 1RXS~J200606.1-542212$ & 
         $20~09~06.4$ &  $-54~22~48$ &  SGC~2005.13-5431.6 &  
         $0.157$ &  $0.0313$ &  $43.48$ &  $14.20$ &  $-22.47$ \\

$\rm 1RXS~J201731.2-411452$ & 
         $20~17~31.2$ &  $-41~14~52$ &  RXS~J201731.2-411452 &  
         $0.145$ &  $0.0390$ &  $~$ &  $17.00$ &  $~$ \\

$\rm 1RXS~J202304.1+093233$ & 
         $20~23~04.3$ &  $09~32~38$ &  CGCG~399-005 &  
         $0.123$ &  $0.0167$ &  $42.67$ &  $15.70$ &  $-19.21$ \\

$\rm 1RXS~J203433.1-303731$ & 
         $20~34~31.4$ &  $-30~37~29$ &  IRAS~F20315-3047 &  
         $0.167$ &  $0.0351$ &  $42.61$ &  $13.30$ &  $-21.13$ \\

$\rm 1RXS~J205522.8+022118$ & 
         $20~55~22.3$ &  $02~21~16$ &  CGCG~374-029 &  
         $0.055$ &  $0.0118$ &  $41.81$ &  $15.10$ &  $-18.55$ \\

$\rm 1RXS~J210221.7+105812$ & 
         $21~02~21.6$ &  $10~58~16$ &  CGCG~425-034 &  
         $0.203$ &  $0.0211$ &  $43.04$ &  $15.40$ &  $-19.91$ \\

$\rm 1RXS~J212324.3+021137$ & 
         $21~23~24.3$ &  $02~11~32$ &  CGCG~375-033 &  
         $0.071$ &  $0.0156$ &  $43.04$ &  $14.90$ &  $-21.56$ \\

$\rm 1RXS~J212541.2-490935$ & 
         $21~25~40.7$ &  $-49~09~38$ &  Fairall~969 &  
         $0.145$ &  $0.0237$ &  $43.73$ &  $15.71$ &  $-21.68$ \\

$\rm 1RXS~J213833.0+320507$ & 
         $21~39~33.4$ &  $32~05~06$ &  CGCG~493-002 &  
         $0.185$ &  $0.0217$ &  $42.93$ &  $15.50$ &  $-19.48$ \\

$\rm 1RXS~J214153.6+315135$ & 
         $21~41~53.5$ &  $31~51~28$ &  CGCG~493-004 &  
         $0.074$ &  $0.0149$ &  $43.04$ &  $15.70$ &  $-20.52$ \\

$\rm 1RXS~J215213.9-194256$ & 
         $21~52~09.6$ &  $-19~43~24$ &  ESO~600-014 &  
         $0.141$ &  $0.0359$ &  $43.94$ &  $15.90$ &  $-22.04$ \\

$\rm 1RXS~J215656.8-113920$ & 
         $21~56~56.5$ &  $-11~39~32$ &  NGC~7158 &  
         $0.285$ &  $0.0310$ &  $43.38$ &  $17.30$ &  $-17.93$ \\

$\rm 1RXS~J221231.8+384049$ & 
         $22~12~31.6$ &  $38~40~58$ &  UGC~11950 &  
         $0.072$ &  $0.0123$ &  $42.36$ &  $13.90$ &  $-20.68$ \\

$\rm 1RXS~J221918.8+120757$ & 
         $22~19~18.5$ &  $12~07~53$ &  II~Zw~177 &  
         $0.247$ &  $0.0316$ &  $44.05$ &  $17.00$ &  $-20.61$ \\

$\rm 1RXS~J222706.1+362140$ & 
         $22~27~05.8$ &  $36~21~42$ &  UGC~12040 &  
         $0.058$ &  $0.0113$ &  $42.30$ &  $14.30$ &  $-20.36$ \\

$\rm 1RXS~J223656.5-221321$ & 
         $22~36~55.9$ &  $-22~13~12$ &  ESO~602-031 &  
         $0.422$ &  $0.0430$ &  $42.79$ &  $14.30$ &  $-21.32$ \\

$\rm 1RXS~J224156.7-423550$ & 
         $22~41~52.8$ &  $-42~35~35$ &  ESO~290-003~(NE) &  
         $0.090$ &  $0.0225$ &  $42.84$ &  $15.64$ &  $-20.04$ \\

$\rm 1RXS~J225850.4+405610$ & 
         $22~58~55.5$ &  $40~55~53$ &  UGC~12282 &  
         $0.065$ &  $0.0128$ &  $42.17$ &  $14.49$ &  $-19.73$ \\

$\rm 1RXS~J230921.5+004540$ & 
         $23~09~20.3$ &  $00~45~23$ &  IC~5287 &  
         $0.091$ &  $0.0180$ &  $43.05$ &  $14.80$ &  $-20.79$ \\

$\rm 1RXS~J231357.7-113027$ & 
         $23~13~57.0$ &  $-11~30~19$ &  MCG~-02-59-006 &  
         $0.177$ &  $0.0260$ &  $43.96$ &  $15.70$ &  $-21.75$ \\

$\rm 1RXS~J231853.1-010338$ & 
         $23~18~53.7$ &  $-01~03~38$ &  UGC~12492 &  
         $0.069$ &  $0.0170$ &  $42.83$ &  $14.60$ &  $-20.74$ \\

$\rm 1RXS~J231906.5-420653$ & 
         $23~19~06.7$ &  $-42~05~37$ &  MCG~-07-47-032 &  
         $0.120$ &  $0.0329$ &  $43.39$ &  $15.00$ &  $-21.75$ \\

$\rm 1RXS~J232856.8+085346$ & 
         $23~28~58.5$ &  $08~54~39$ &  MCG~+01-59-085 &  
         $0.092$ &  $0.0155$ &  $44.08$ &  $18.00$ &  $-20.75$ \\

$\rm 1RXS~J233355.5-234336$ & 
         $23~33~55.2$ &  $-23~43~41$ &  PKS~2331-240 &  
         $0.437$ &  $0.0387$ &  $43.81$ &  $16.50$ &  $-19.90$ \\

$\rm 1RXS~J234547.5-293053$ & 
         $23~45~47.6$ &  $-29~31~04$ &  NGC~7749 &  
         $0.079$ &  $0.0187$ &  $42.80$ &  $13.82$ &  $-21.92$ \\

$\rm 1RXS~J235038.6+243321$ & 
         $23~50~36.6$ &  $24~33~22$ &  UGC~12804 &  
         $0.051$ &  $0.0127$ &  $42.62$ &  $15.11$ &  $-20.58$ \\

$\rm 1RXS~J235728.4-302739$ & 
         $23~57~27.3$ &  $-30~27~37$ &  AM~2354-304-E &  
         $0.057$ &  $0.0447$ &  $42.55$ &  $15.82$ &  $-19.63$ \\

\hline \hline
\\ \multicolumn{5}{l}{Companion galaxies, galaxy pairs}\\ \hline
$\rm 1RXS~J023513.9-293616$ & 
     $02~34~54.5$ & $-29~34~28$ &  RX~J023454.8-293425 &  
     $0.006$ &  $0.0020$ &  $45.00$ &  $17.90$ &  $-24.53$ \\

$\rm 1RXS~J043520.2-780150$ & 
     $04~35~16.2$ & $-78~01~57$ &  ESO~015-011~(b) &  
     $0.077$ &  $0.004$  &  $~$  &  $14.20$ &  $-22.78$ \\

$\rm 1RXS~J071204.2-603005$ & 
     $07~12~03.2$ & $-60~30~30$ &  ESO~122-016~(SW) &  
     $~$  &  $~$  &  $~$  &  $14.60$ &  $-20.99$ \\

$\rm 1RXS~J083539.1-040508$ & 
     $08~35~48.6$ & $-04~05~33$ & MCG~-01-22-27  &  
     $~$  &  $~$  &  $~$  &  $15.00$ &  $-18.93$ \\

$\rm 1RXS~J085001.4+701804$ & 
     $08~49~58.4$ & $70~17~58$ &  NGC~2650 &  
     $~$  &  $~$  &  $~$ &  $14.30$ &  $-19.53$ \\

$\rm 1RXS~J091345.4+474208$ & 
     $09~13~44.7$ & $47~42~18$ &  MCG~+08-17-60~(N) &  
     $~$  &  $~$  &  $~$ &  $15.70$ &  $-20.97$ \\

$\rm 1RXS~J102403.1+062954$ & 
     $10~23~59.8$ & $06~29~08$ &  CGCG~037-023 &  
     $~$  &  $~$  &  $~$ &  $15.60$ &  $-20.84$ \\

$\rm 1RXS~J132016.3+330828$ & 
     $13~20~17.5$ & $33~08~45$ &  NGC~5098~(E) &  
     $~$  &  $~$  &  $~$ &  $15.50$ &  $-20.35$ \\

$\rm 1RXS~J153210.0+585434$ & 
     $15~32~09.4$ & $58~54~20$ &  VII~Zw~608~(W) &  
     $~$  &  $~$  &  $~$ &  $16.10$ &  $-17.13$ \\

$\rm 1RXS~J180242.4+424734$ & 
     $18~02~39.4$ & $42~47~18$ &  MCG~+07-37-018~(SW) &  
     $~$  &  $~$  &  $~$ & $15.40$ &  $-20.99$ \\

\end{longtable}}

\longtab{3}{
\label{tab:optdata}
\begin{longtable}{lrrrrrrrr}
\caption{Optical line properties}\\
\hline\hline 
Name & 
Obs.Run & 
Exp. & 
cz & 
[\ion{O}{iii}]/H$\beta$ & 
[\ion{N}{ii}]/H$\alpha$ & 
H$\alpha$/H$\beta$ &
FWHM$_{\mathrm{H\alpha,br}}$ &
AGN type \\ 
& & [s] & [km~s$^{-1}$] &  &  &  & [km~s$^{-1}$] & \\ 
(1) & (2) & (3) & (4) & (5) & (6) & (7) & (8) & (9) \\
\hline \endfirsthead
\caption{(continued)}\\
\hline\hline 
Name & 
Obs.Run & 
Exp. & 
cz & 
[\ion{O}{iii}]/H$\beta$ & 
[\ion{N}{ii}]/H$\alpha$ &
H$\alpha$/H$\beta$ &
FWHM$_{\mathrm{H\alpha,br}}$  &
AGN type \\ 
 & 
 & 
[s] & 
[km~s$^{-1}$] & 
 & 
 & 
 & 
[km~s$^{-1}$] &  
\\ 
(1) & (2) & (3) & (4) & (5) & (6) & (7) & (8) & (9)\\
\hline \endhead 
\hline\hline \endfoot

ESO~409-003 & 97/07 & 1200 &  8520 &  4.52 &  1.27
&  ~ & ~ & LINER \\

UGC~32 & 97/08 & 1200 &  22260 &  0.10 &  2.47
&  ~ & ~ & LINER \\

CGCG~433-025~~~[P1] & 96/11 & 1800 &  13520 &  38.00 &  ~
&  2.29 & 8250 & Sy~1 \\

NGC~57 & 97/08 & 600 &  5580 &  ~ &  ~
&  ~ & ~ & non-ac \\

NGC~71 & 97/08 & 960 &  6750 &  0.10 &  7.97
&  ~ & ~ & Sy~2 \\

NGC~70 & 97/08 & 960 &  7080 &  0.10 &  7.28
&  ~ & ~ & LINER \\

PKS~0018-19 & 97/07 & 2100 &  28590 &  25.22 &  1.54
&  ~ & 8160 & Sy~1.9 \\

ESO~350-015 & 97/07 & 600 &  15060 &  ~ &  ~
&  ~ & ~ & non-ac \\

HCG~4a~~~[P1] & 96/11 & 900 &  7900 &  8.20 &  1.70
&  ~ & 2100 & Sy~1.9 \\

VIII~Zw~36~~~[P1] & 96/11 & 900 &  12390 &  13.40 &  0.90
&  3.62 & 1950 & Sy~1 \\

MCG~-03-03-017 & 97/07 & 600 &  16170 &  0.10 &  2.47
&  ~ & ~ & Sy~2 \\

ESO~113-010~~~[P1] & 96/11 & 900 &  7705 &  5.30 &  1.40
&  ~ & 2000 & Sy~2 \\

UGC~716~~~[P1] & 96/11 & 900 &  17710 &  ~ &  ~
&  ~ & ~ & non-ac \\

NGC~427~~~[P1] & 96/11 & 2700 &  10035 &  0.10 &  1.80
&  ~ & 5500 & Sy~1 \\

RX~J011232.8-320140~~~[P1] & 96/11 & 900 &  ~ &  ~ &  ~
&  ~ & ~ & BL~Lac \\

ESO~244-017~~~[P1] & 96/11 & 900 &  7045 &  7.10 &  1.50
&  2.20 & 3000 & Sy~1 \\

KUG~0128+328 & 97/08 & 900 &  21090 &  54.28 &  2.05
&  4.96 & 2710 & Sy~1 \\

MCG~-01-05-031~~~[P1] & 96/11 & 900 &  5420 &  3.70 &  0.70
&  ~ & ~ & Sy~2 \\

ESO~080-005~~~[P1] & 96/11 & 1800 &  8080 &  13.00 &  0.80
&  ~ & 3450 & Sy~1.8 \\

ESO~416-002~~~[P1] & 96/11 & 1500 &  17710 &  22.10 &  2.60
&  ~ & 16000 & Sy~1.9 \\

PHL~1389~~~[P1] & 96/11 & 1200 &  ~ &  ~ &  ~
&  ~ & ~ & BL~Lac \\

IC~1867~~~[P1] & 96/11 & 600 &  7680 &  0.10 &  6.70
&  ~ & ~ & LINER \\

NGC~1217~~~[P1] & 96/11 & 600 &  6155 &  ~ &  ~
&  ~ & ~ & Sy~2 \\

NGC~1218~~~[P1] & 96/11 & 1500 &  8590 &  0.10 &  21.50
&  ~ & 2950 & Sy~1.9 \\

ESO~548-081~~~[P1] & 96/11 & 1500 &  4110 &  0.10 &  7.90
&  ~ & 4450 & Sy~1.9 \\

AM~0426-625~~~[P1] & 96/11 & 600 &  5490 &  ~ &  ~
&  ~ & ~ & LINER \\

ESO~015-011~~~[P1] & 96/11 & 900 &  18280 &  5.40 &  0.70
&  1.99 & 2750 & Sy~1.8 \\

MCG~-02-12-050~~~[P1] & 96/11 & 1800 &  10730 &  38.10 &  1.10
&  2.52 & 5400 & Sy~1 \\

UGC~3134~~~[P1] & 96/11 & 900 &  8665 &  4.00 &  0.80
&  ~ & ~ & Sy~2 \\

MCG~-01-13-025~~~[P1] & 96/11 & 1200 &  4765 &  2.90 &  0.90
&  3.92 & 4500 & Sy~1 \\

UGC~3194 & 00/02 & 1800 &  8364 &  ~ &  0.42
&  ~ & 650 & LINER \\

ESO~552-039~~~[P1] & 96/11 & 1500 &  11870 &  18.40 &  1.60
&  2.22 & 4000 & Sy~1 \\

NGC~1713 & 98/03 & 1200 &  4491 &  ~ &  ~
&  ~ & ~ & LINER \\

CGCG~469-001~~~[V] & 98/03 & 1200 &  5787 &  9.48 &  0.82
&  36.92 & 6460 & Sy~1.9 \\

MCG~-02-14-009~~~[P1] & 96/11 & 900 &  8530 &  8.60 &  0.90
&  2.86 & 2850 & Sy~1 \\

UGC~3355 & 98/03 & 600 &  7810 &  ~ &  ~
&  ~ & ~ & non-ac \\

ESO~120-023~~~[P1] & 96/11 & 600 &  11280 &  ~ &  ~
&  ~ & ~ & non-ac \\

ESO~254-017~~~[P1] & 96/11 & 1500 &  8925 &  0.10 &  2.10
&  ~ & ~ & LINER \\

ESO~425-019 & 00/02 & 1200 &  6740 &  ~ &  ~
&  ~ & ~ & non-ac \\

PMN~J0623-6436~~~[P1] & 96/11 & 1200 &  38640 &  ~ &  ~
&  2.72 & 2500 & Sy~1 \\

PMN~J0630-2406~~~[P1] & 96/11 & 600 &  ~ &  ~ &  ~
&  ~ & ~ & BL~Lac \\

ESO~490-026~~~[P1] & 96/11 & 900 &  7450 &  16.40 &  1.40
&  2.14 & 4100 & Sy~1 \\

NGC~2256 & 98/03 & 1200 &  5145 &  ~ &  ~
&  ~ & ~ & non-ac \\

NGC~2258 & 98/03 & 1200 &  4060 &  ~ &  ~
&  ~ & ~ & LINER \\

CGCG~085-010 & 98/03 & 1200 &  18256 &  4.73 &  0.55
&  2.68 & 2060 & Sy~1.5 \\

NGC~2329 & 98/03 & 1200 &  5838 &  ~ &  ~
&  ~ & ~ & LINER \\

B2~0708+32b~~~[V] & 98/03 & 1500 &  19890 &  1.83 &  1.59
&  9.28 & 3880 & Sy~1.9 \\

ESO~122-016~(NE)~~~[P1] & 96/11 & 1200 &  9770 &  ~ &  ~
&  ~ & ~ & LINER \\

UGC~3901 & 98/03 & 1200 &  6508 &  0.65 &  0.78
&  6.43 & 9150 & Sy~1.9 \\

UGC~3927~~~[V] & 98/03 & 1200 &  12405 &  ~ &  4.98
&  ~ & 3850 & Sy~1.9 \\

UGC~4018 & 00/02 & 1200 &  8616 &  ~ &  ~
&  ~ & ~ & LINER \\

MCG~+08-15-09~~~[V] & 98/03 & 1200 &  7556 &  7.23 &  0.66
&  19.37 & 4970 & Sy~1.9 \\

ESO~209-012~~~[P1] & 96/11 & 1200 &  12140 &  8.60 &  1.60
&  2.93 & 3400 & Sy~1.5 \\

CGCG~207-040 & 00/02 & 1200 &  19249 &  ~ &  ~
&  ~ & ~ & LINER \\

MGC~+08-15-56~~~[V] & 98/03 & 1200 &  12583 &  3.97 &  0.94
&  6.07 & 1570 & NLS1 \\

IC~505 & 98/03 & 900 &  9821 &  ~ &  ~
&  ~ & 300: & LINER \\

NGC~2617~~~[V] & 98/03 & 1200 &  4575 &  7.03 &  1.16
&  21.86 & 7408 & Sy~1.5 \\

MCG~+07-18-43 & 98/03 & 1200 &  16202 &  ~ &  2.16
&  ~ & 300: & LINER \\

QSO east~of NGC~2650 & 98/03 & 420 & (z=1.90) &  ~ &  ~
&  ~ & ~ & QSO \\

CGCG~005-037~~~[V] & 00/02 & 1500 &  17843 &  1.76 &  0.27
&  2.66 & 1440 & NLS1 \\

MCG~+08-17-60~~~[V] & 98/03 & 1200 &  15953 &  8.40 &  0.07
&  10.18 & 3350 & Sy~1.5 \\

VIII~Zw~45~~~[V] & 98/03 & 1200 &  12029 &  4.95 &  0.59
&  4.33 & 3480 & Sy~1.5 \\

KUG~0929+540 & 98/03 & 1200 &  17340 &  0.94 &  0.06
&  2.07 & 1760 & NLS1 \\

KUG~0933+511~~~[V] & 98/03 & 600 &  16854 &  0.72 &  0.33
&  3.88 & 1820 & NLS1 \\

CGCG~122-055 & 98/03 & 1200 &  6769 &  14.25 &  1.17
&  3.05 & 3120 & Sy~1 \\

NGC~3035~~~[V] & 98/03 & 1200 &  4546 &  ~ &  1.92
&  ~ & 7460 & Sy~1.8 \\

CGCG~064-024 & 98/03 & 1200 &  23276 &  ~ &  4.64
&  ~ & 400: & LINER \\

ESO~499-041 & 00/02 & 1200 &  3826 &  1.80 &  0.58
&  5.26 & 1510 & NLS1 \\

CGCG~037-022 & 98/03 & 1200 &  13388 &  17.50 &  0.23
&  3.06 & 1940 & NLS1 \\

CGCG~037-028 & 98/03 & 1200 &  13204 &  ~ &  0.82
&  ~ & 500: & LINER \\

CGCG~240-047 & 98/03 & 1200 &  13191 &  16.83 &  0.80
&  7.52 & 2720 & Sy~1.5 \\

GNB~069~~~[V] & 98/03 & 1200 &  21690 &  1.54 &  1.13
&  1.99 & 3670 & Sy~1 \\

CGCG~212-045~~~[V] & 98/03 & 1200 &  11339 &  1.10 &  0.97
&  ~ & 3250 & Sy~1  \\

MCG~+07-23-15 & 98/03 & 1200 &  9394 &  ~ &  ~
&  ~ & ~ & LINER \\

CGCG~011-012 & 98/03 & 1200 &  11724 &  9.96 &  0.89
&  6.80 & 3710 & Sy~1.5 \\

NGC~3798~~~[V] & 98/03 & 1200 &  3603 &  0.41 &  0.80
&  6.36 & 7990 & Sy~1.5 \\

KUG~1141+371~~~[V] & 98/03 & 1200 &  11686 &  15.65 &  0.94
&  6.29 & 15310 & Sy~1 \\

IC~734 & 00/02 & 1800 &  23801 &  ~ &  ~
&  ~ & ~ & non-ac \\

MCG~-01-30-041~~~[V] & 97/07 & 1200 &  5700 &  3.00 &  0.64
&  2.66 & 5790 & Sy~1.8 \\

VIII~Zw~153~~~[V] & 98/03 & 1200 &  36201 &  ~ &  0.19
&  2.23 & 4750 & Sy~1 \\

ESO~267-013 & 97/07 & 1200 &  4350 &  1.44 &  0.53
&  ~ & ~ & LINER \\

MCG~+08-22-71 & 98/03 & 1200 &  18841 &  4.48 &  0.68
&  4.37 & 2370 & Sy~1.8 \\

Mrk~648 & 98/03 & 1200 &  15534 &  0.69 &  0.00
&  3.84 & 2690 & NLS1 \\

Mrk~202~~~[V] & 98/03 & 1200 &  7234 &  3.15 &  0.37
&  2.87 & 2360 & NLS1 \\

MCG~+08-23-67~~~[V] & 98/03 & 1200 &  9398 &  10.34 &  0.85
&  5.61 & 1510 & NLS1 \\

ESO~381-007~~~[V] & 97/07 & 1200 &  16470 &  9.85 &  0.98
&  3.50 & 4710 & Sy~1 \\

KUG~1240+359~~~[V] & 98/03 & 1800 &  164591 &  0.40 &  ~
&  ~ & ~ & QSO \\

CGCG~015-026~~~[V] & 97/07 & 1800 &  14190 &  19.86 &  2.13
&  2.80 & 5620 & Sy~1 \\

NGC~4756 & 00/02 & 1200 &  3981 &  ~ &  ~
&  ~ & ~ & non-ac \\

CGCG~043-056 & 97/07 & 600 &  19650 &  ~ &  ~
&  ~ & ~ & non-ac \\

MCG~-01-33-054 & 97/07 & 1200 &  3960 &  0.10 &  1.60
&  ~ & ~ & Sy~2 \\

KUG~1256+241~~~[V] & 00/02 & 1800 &  22586 &  8.12 &  0.97
&  1.87 & 5130 & Sy~1 \\

CGCG~160-104~~~[V] & 98/03 & 1200 &  7506 &  -0.54 &  -1.58
&  ~ & 690 & LINER \\

IC~4165~~~[V] & 98/03 & 2400 &  8447 &  ~ &  1.02
&  ~ & 3810 & Sy~1.9 \\

VIII~Zw~272 & 98/03 & 600 &  23513 &  ~ &  ~
&  ~ & ~ & non-ac \\

CGCG~160-193~~~[V] & 98/03 & 1200 &  9870 &  0.58 &  0.97
&  1.91 & 3270 & Sy~1.5 \\

NGC~5098 & 98/03 & 600 &  10970 &  ~ &  3.76
&  ~ & 500: & LINER \\

MCG~+10-19-84~~~[V] & 98/03 & 2400 &  35364 &  1.06 &  1.77
&  5.01 & 5610 & Sy~1 \\

VIII~Zw~327 & 98/03 & 1200 &  27898 &  3.51 &  ~
&  ~ & ~ & Sy~1.5 \\

CGCG~365-016 & 00/02 & 1200 &  13519 &  3.23 &  1.59
&  3.32 & 1100 & Sy~1.9 \\

UGC~8829~~~[V] & 98/03 & 1200 &  8015 &  5.66 &  1.15
&  9.35 & 19710 & Sy~1 \\

CGCG~103-009 & 98/03 & 1200 &  5000 &  ~ &  0.35
&  ~ & 230 & H~II \\

MCG~+05-33-24a~~~[V] & 98/03 & 1200 &  22705 &  1.53 &  0.46
&  3.89 & 1770 & Sy~1.8 \\

CGCG~219-050 & 98/03 & 1500 &  20656 &  ~ &  2.92
&  ~ & 7570 & Sy~1.5 \\

NPM1G~-16.0443 & 97/07 & 1200 &  8730 &  5.24 &  0.80
&  6.16 & 2400 & Sy~1.5 \\

CGCG~103-073 & 98/03 & 600 &  7758 &  ~ &  ~
&  ~ & ~ & non-ac \\

CGCG~353-034 & 98/03 & 600 &  13747 &  ~ &  ~
&  ~ & ~ & non-ac \\

ESO~511-030~~~[V] & 97/07 & 600 &  6750 &  24.22 &  2.71
&  3.10 & 3620 & Sy~1 \\

MCG~+05-34-053 & 97/08 & 600 &  4470 &  0.10 &  0.52
&  ~ & ~ & H~II \\

CGCG~163-074 & 97/08 & 1200 &  16620 &  8.64 &  0.68
&  2.80 & 3590 & Sy~1.5 \\

CGCG~047-107~~~[V] & 98/03 & 2700 &  8696 &  9.06 &  0.41
&  5.32 & 1320 & NLS1 \\

ESO~328-036-S & 97/07 & 1200 &  7230 &  0.10 &  1.93
&  2.86 & 3750 & Sy~1 \\

CGCG~021-063~~~[V] & 97/07 & 1800 &  15660 &  4.64 &  1.30
&  ~ & 3640 & LINER \\

CGCG~049-106 & 97/07 & 1200 &  11640 &  14.43 &  1.64
&  4.65 & 2190 & NLS1 \\

VII~Zw~608 & 97/08 & 1200 &  19380 &  0.10 &  1.81
&  ~ & ~ & LINER \\

NGC~5959 & 97/07 & 600 &  7350 &  ~ &  ~
&  ~ & ~ & non-ac \\

CGCG~022-021~~~[V] & 97/07 & 1200 &  7260 &  3.90 &  0.98
&  ~ & 4310 & Sy~1.8 \\

Mrk~863~~~[V] & 97/07 & 600 &  12780 &  1.94 &  0.50
&  2.80 & 2760 & Sy~1.5 \\

CGCG~023-021 & 97/07 & 1800 &  8610 &  0.10 &  3.07
&  ~ & 4090 & Sy~1.9 \\

KUG~1618+410~~~[V] & 97/08 & 1200 &  11370 &  14.64 &  0.27
&  2.78 & 2150 & NLS1 \\

KUG~1618+402~~~[V] & 97/08 & 1200 &  8550 &  16.16 &  0.27
&  3.76 & 2160 & NLS1 \\

NGC~6137 & 98/03 & 600 &  9356 &  ~ &  1.03
&  ~ & 1440 & LINER \\

CGCG~196-064 & 97/08 & 1200 &  10350 &  1.25 &  0.25
&  4.06 & 1650 & Sy~1.9 \\

NGC~6159 & 97/08 & 600 &  9480 &  ~ &  ~
&  ~ & ~ & LINER \\

NGC~6160 & 97/08 & 600 &  9630 &  ~ &  ~
&  ~ & ~ & non-ac \\

MCG~+06-37-023 & 97/08 & 600 &  18840 &  ~ &  ~
&  ~ & ~ & LINER \\

MCG~+06-38-005 & 97/08 & 1200 &  8040 &  6.72 &  0.96
&  2.97 & 7450 & Sy~1.5 \\

MCG~+13-12-022 & 97/08 & 900 &  17010 &  ~ &  ~
&  ~ & ~ & non-ac \\

MCG~+05-41-010 & 97/08 & 600 &  13980 &  ~ &  ~
&  ~ & ~ & non-ac \\

NGC~6370 & 97/08 & 600 &  8310 &  ~ &  ~
&  ~ & ~ & LINER \\

NGC~6414 & 97/08 & 1800 &  12600 &  ~ &  ~
&  ~ & ~ & non-ac \\

MCG~+07-37-018~(NE) & 97/08 & 600 &  15180 &  ~ &  ~
&  ~ & ~ & non-ac \\

MCG~+10-26-015 & 97/08 & 600 &  8100 &  ~ &  ~
&  ~ & ~ & non-ac \\

MCG~+03-47-002 & 97/08 & 1200 &  12180 &  4.38 &  1.02
&  2.80 & 3930 & Sy~1 \\

ESO~045-011 & 97/07 & 1800 &  5370 &  ~ &  ~
&  ~ & ~ & LINER \\

ESO~104-041~(A) & 97/07 & 1200 &  25440 &  ~ &  ~
&  2.86 & 6980 & Sy~1 \\

ESO~184-068 & 97/07 & 600 &  17490 &  ~ &  ~
&  ~ & ~ & non-ac \\

PKS~1928-34 & 97/07 & 1800 &  29460 &  8.70 &  1.80
&  ~ & ~ & Sy~2 \\

ESO~399-015 & 97/07 & 1200 &  7740 &  ~ &  ~
&  ~ & ~ & LINER \\

SGC~2005.13-5431.6 & 97/07 & 600 &  16020 &  ~ &  ~
&  ~ & ~ & non-ac \\

RXS~J201731.2-411452 & 97/07 & 900 &  ~ &  ~ &  ~
&  ~ & ~ & BL~Lac \\

CGCG~399-005 & 97/07 & 1200 &  7170 &  12.61 &  2.77
&  ~ & 6790 & Sy~1.9 \\

IRAS~F20315-3047~~~[V] & 97/07 & 1800 &  5760 &  6.02 &  1.21
&  2.92 & 3320 & Sy~1 \\

CGCG~374-029 & 97/07 & 1200 &  4020 &  36.07 &  1.62
&  5.37 & 2040 & NLS1 \\

CGCG~425-034~~~[V] & 97/07 & 2400 &  8610 &  1.60 &  0.44
&  2.80 & 6460 & Sy~1.5 \\

CGCG~375-033 & 97/07 & 600 &  14520 &  ~ &  ~
&  ~ & ~ & LINER \\

Fairall~969 & 97/07 & 3600 &  22170 &  7.22 &  3.16
&  2.80 & 3660 & Sy~1 \\

CGCG~493-002 & 97/08 & 1200 &  7380 &  4.84 &  0.69
&  ~ & 5220 & Sy~1.8 \\

CGCG~493-004 & 97/08 & 1200 &  13050 &  0.41 &  0.04
&  2.84 & 1410 & NLS1 \\

ESO~600-014 & 97/07 & 600 &  28470 &  ~ &  ~
&  ~ & ~ & LINER \\

NGC~7158~~~[P1] & 96/11 & 900 &  8275 &  10.90 &  0.90
&  ~ & 2100 & NLS1 \\

UGC~11950 & 97/08 & 600 &  6150 &  ~ &  ~
&  ~ & ~ & LINER \\

II~Zw~177~~~[V] & 97/07 & 1800 &  24450 &  10.55 &  1.28
&  2.66 & 1180 & NLS1 \\

UGC~12040 & 97/08 & 1800 &  6390 &  0.10 &  2.85
&  ~ & 4020 & Sy~1.9 \\

ESO~602-031~~~[P1] & 96/11 & 1800 &  9895 &  10.10 &  1.30
&  ~ & 5900 & Sy~1.8 \\

ESO~290-003~(NE) & 97/07 & 1200 &  10170 &  ~ &  ~
&  ~ & ~ & non-ac \\

UGC~12282 & 97/08 & 1200 &  5220 &  16.85 &  3.75
&  ~ & 5660 & Sy~1.9 \\

IC~5287~~~[P1] & 96/11 & 1800 &  9750 &  0.10 &  4.80
&  3.30 & 4200 & Sy~1 \\

MCG~-02-59-006~~~[P1] & 96/11 & 900 &  22790 &  0.10 &  1.30
&  ~ & 3700 & Sy~1 \\

UGC~12492~~~[P1] & 96/11 & 600 &  8710 &  ~ &  ~
&  ~ & ~ & LINER \\

MCG~-07-47-032 & 97/07 & 600 &  16590 &  ~ &  ~
&  ~ & ~ & non-ac \\

MCG~+01-59-085 & 97/07 & 1500 &  40860 &  ~ &  ~
&  ~ & ~ & Sy~1 \\

PKS~2331-240~~~[V] & 97/07 & 1800 &  14130 &  6.86 &  ~
&  ~ & 3700 & Sy~1.9 \\

NGC~7749 & 97/07 & 600 &  10440 &  ~ &  ~
&  ~ & ~ & LINER \\

UGC~12804 & 97/08 & 600 &  10230 &  ~ &  ~
&  ~ & ~ & non-ac \\

AM~2354-304-E~~~[V] & 97/07 & 1200 &  9150 &  0.10 &  ~
&  2.80 & 1970 & NLS1 \\

\hline \hline \\
\multicolumn{5}{l}{Companion galaxies, galaxy pairs}\\ \hline
RX~J023454.8-293425~~~[P1] & 96/11 & 3000 &  203550 &  ~ &  ~
&  ~ & ~ & QSO \\

ESO~015-011~(b)~~~[P1] & 96/11 & 900 &  18430 &  1.40 &  0.90
&  ~ & ~ & LINER \\

ESO~122-016~(SW)~~~[P1] & 96/11 & 1200 &  9780 &  ~ &  ~
&  ~ & ~ & non-ac \\

MCG~-01-22-27 & 98/03 & 1200 &  4575 &  7.03 &  1.16
&  21.86 & 7410 & H~II \\

NGC~2650 & 98/03 & 1200 &  4356 &  ~ &  0.08
&  ~ & 8610 & Sy~1.8 \\

MCG~+08-17-60~(N) & 98/03 & 1200 &  15950 & ~ & ~
&  ~ & ~ & non-ac \\

CGCG~037-023 & 98/03 & 1200 &  13158 &  ~ &  0.29
&  ~ & 410 & H~II \\

NGC~5098~(E) & 98/03 & 600 &  10970 &  ~ &  ~
&  ~ & ~ & non-ac \\

VII~Zw~608~(W) & 97/08 & 1200 &  20640 &  ~ &  ~
&  ~ & ~ & non-ac \\

MCG~+07-37-018~(SW) & 97/08 & 900 &  14070 &  0.89 &  0.29
&  ~ & ~ & H~II \\

\end{longtable}}


\onecolumn
\begin{center}
Main Targets: \\
\includegraphics[width=5cm]{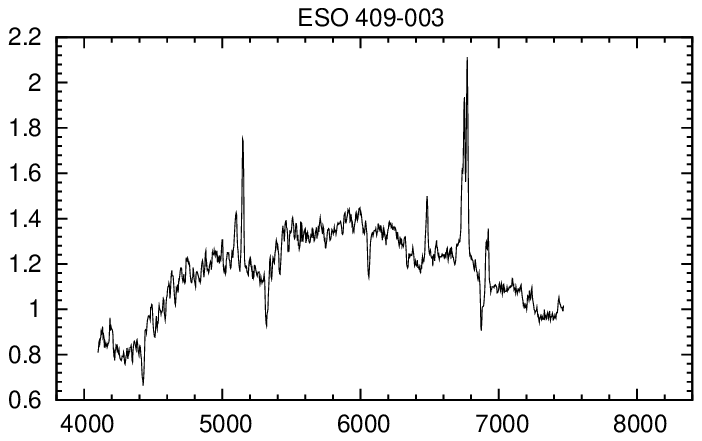}
\includegraphics[width=5cm]{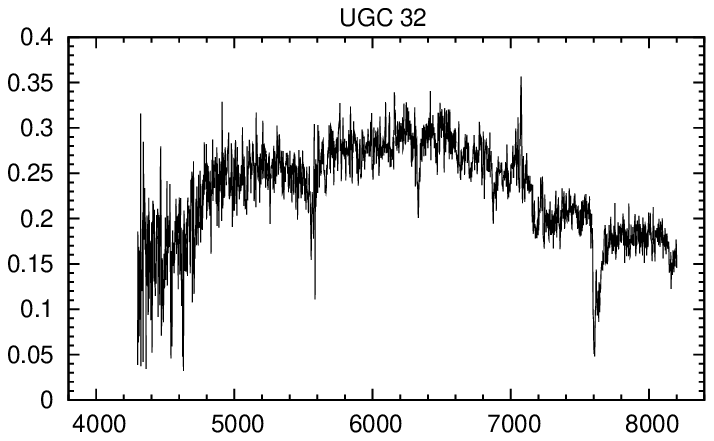}
\includegraphics[width=5cm]{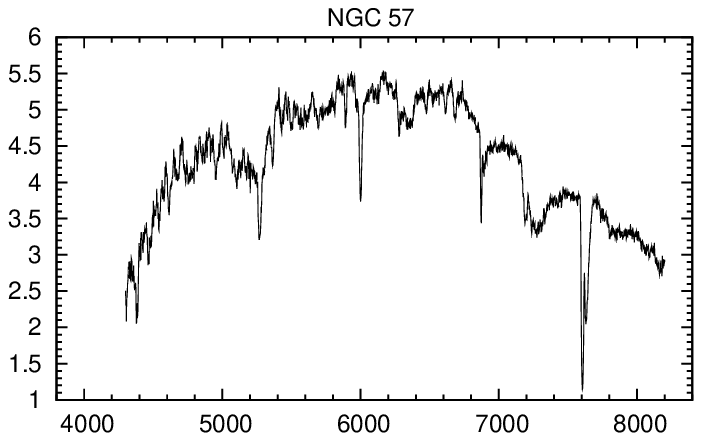}
\includegraphics[width=5cm]{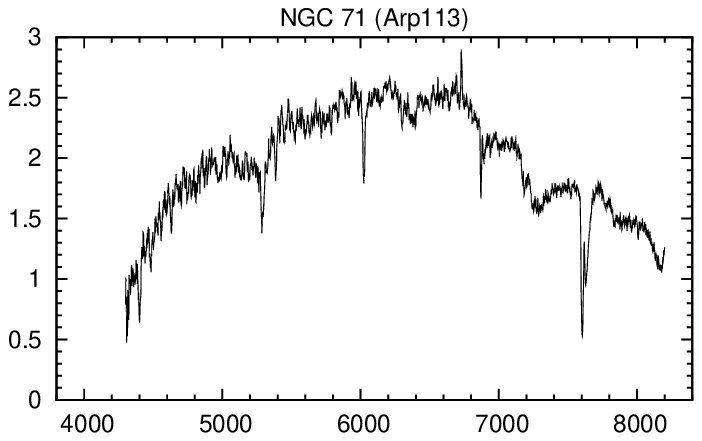}
\includegraphics[width=5cm]{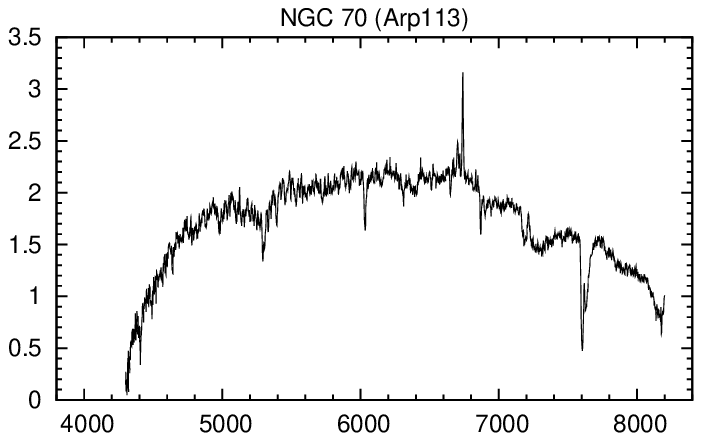}
\includegraphics[width=5cm]{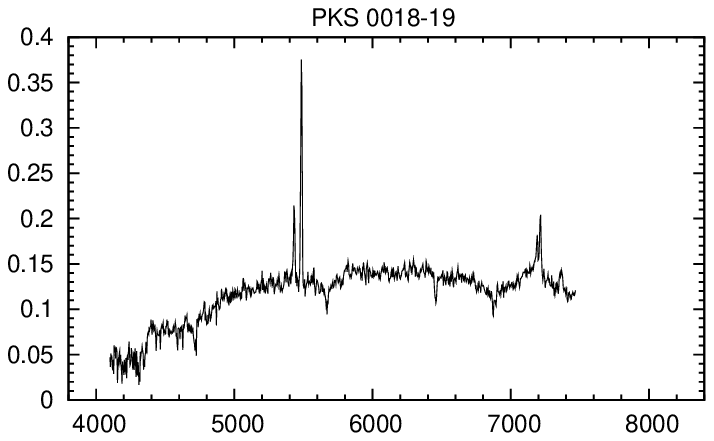}
\includegraphics[width=5cm]{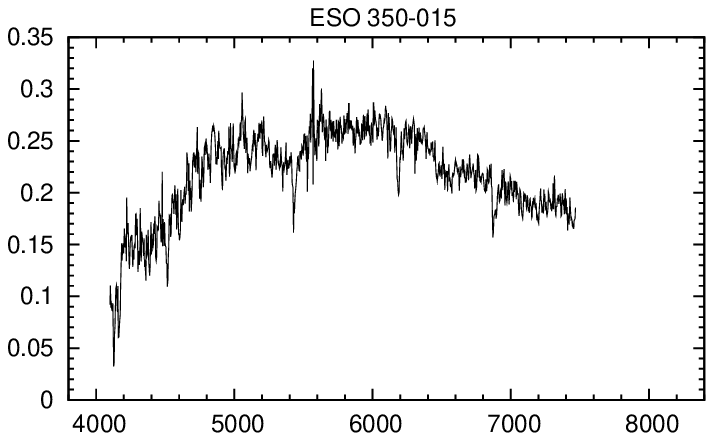}
\includegraphics[width=5cm]{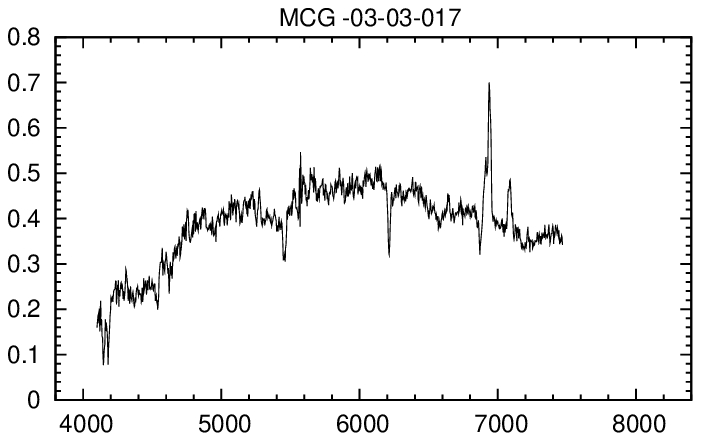}
\includegraphics[width=5cm]{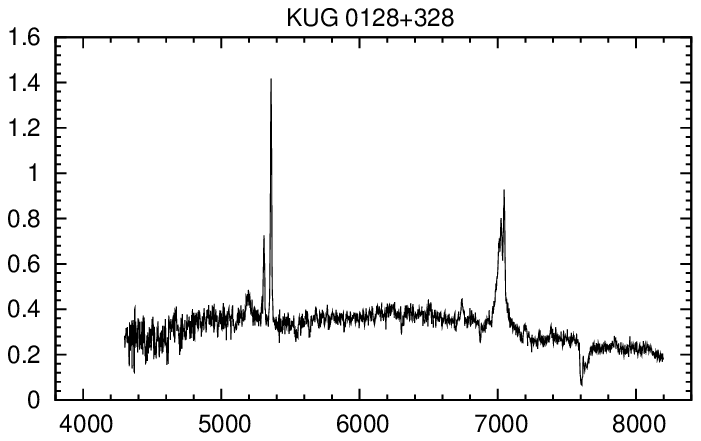}
\includegraphics[width=5cm]{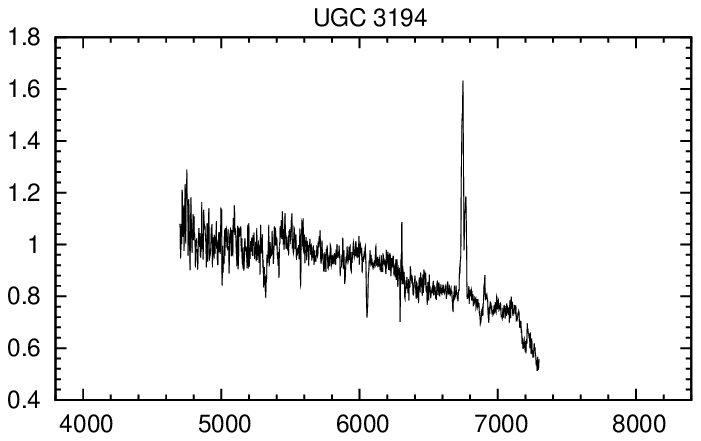}
\includegraphics[width=5cm]{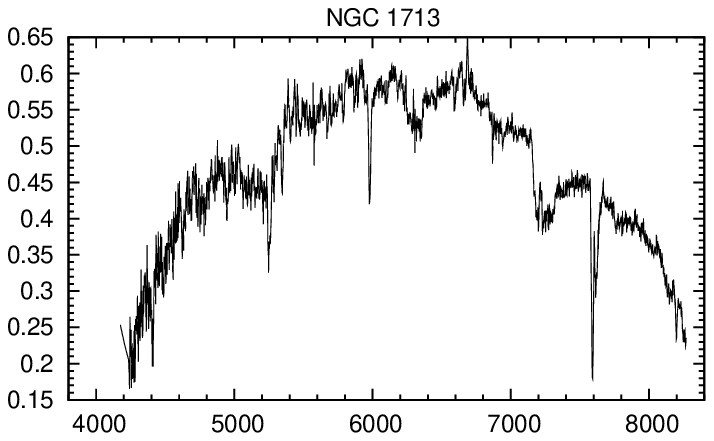}
\includegraphics[width=5cm]{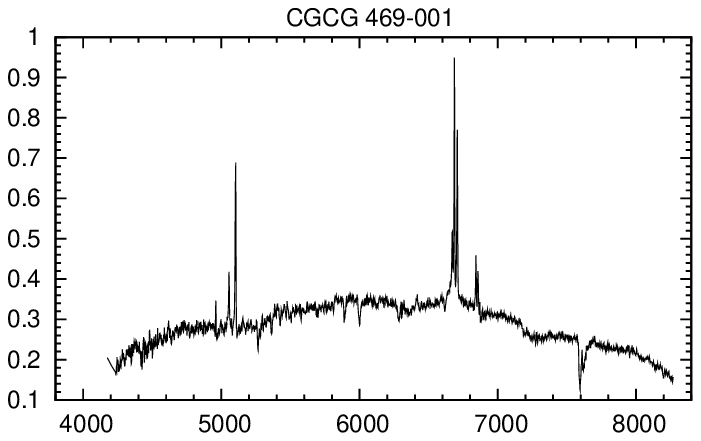}
\includegraphics[width=5cm]{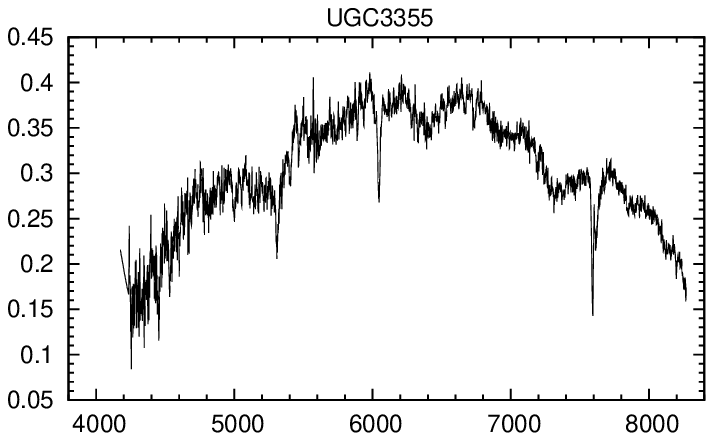}
\includegraphics[width=5cm]{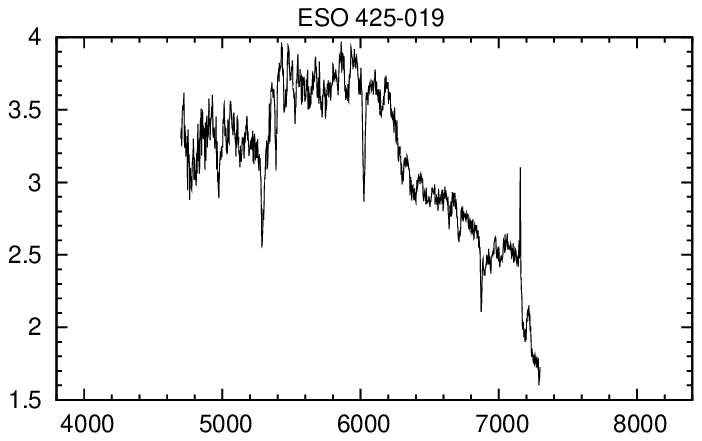}
\includegraphics[width=5cm]{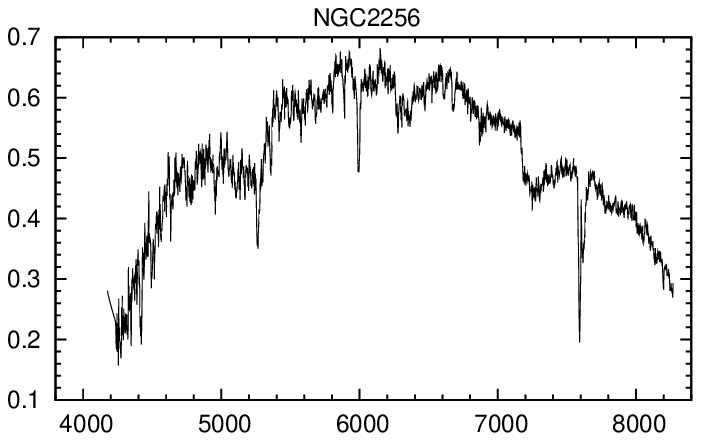}
\includegraphics[width=5cm]{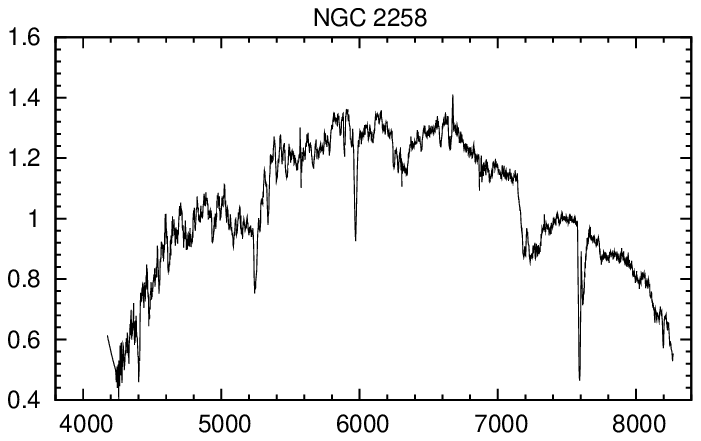}
\includegraphics[width=5cm]{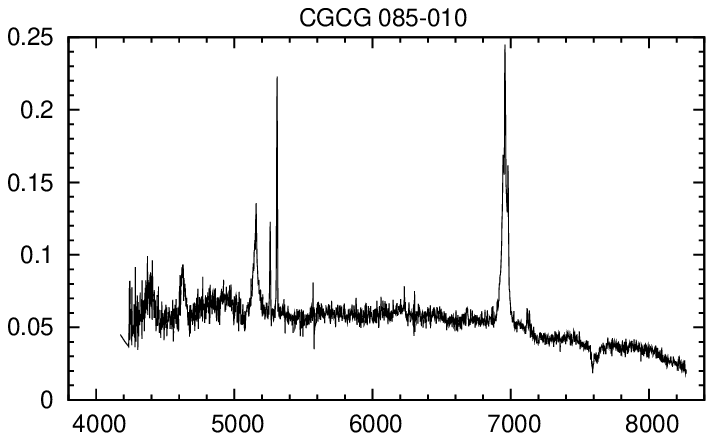}
\includegraphics[width=5cm]{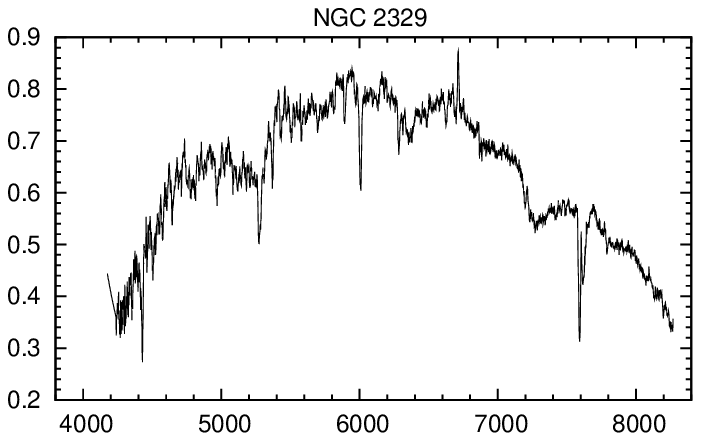}
\includegraphics[width=5cm]{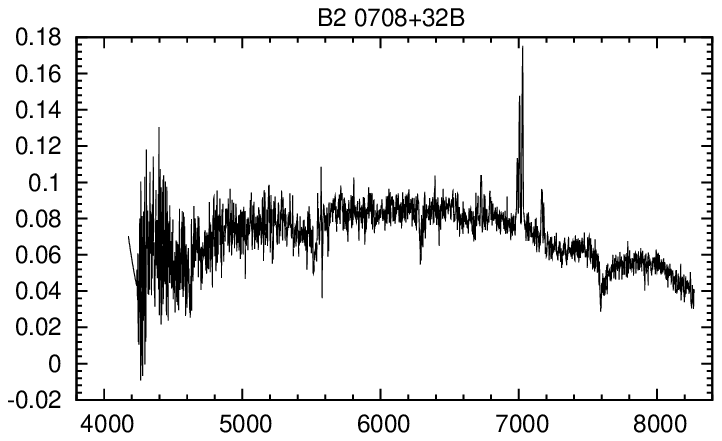}
\includegraphics[width=5cm]{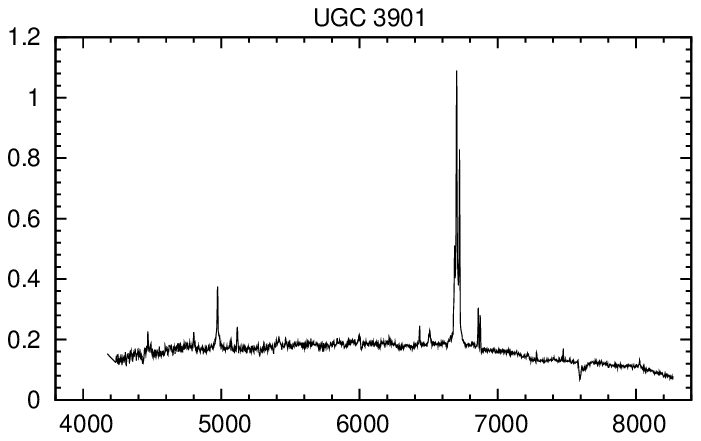}
\includegraphics[width=5cm]{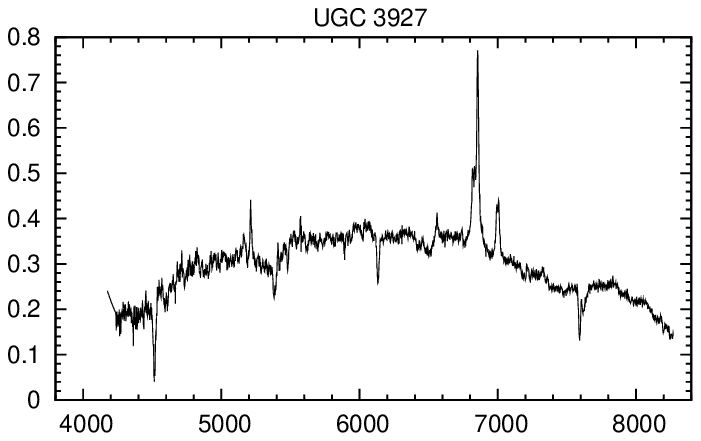}
\includegraphics[width=5cm]{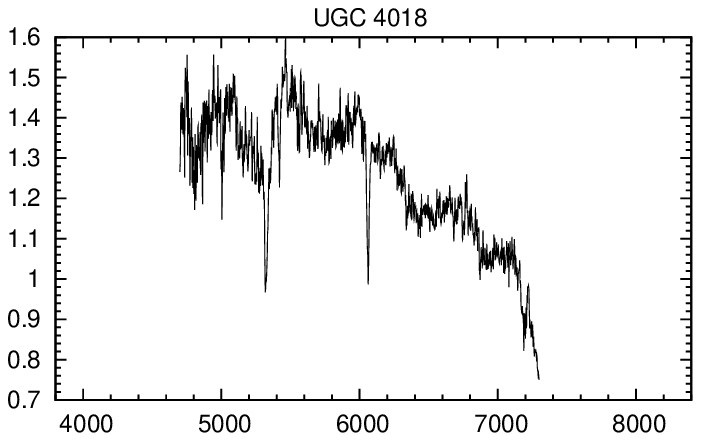}
\includegraphics[width=5cm]{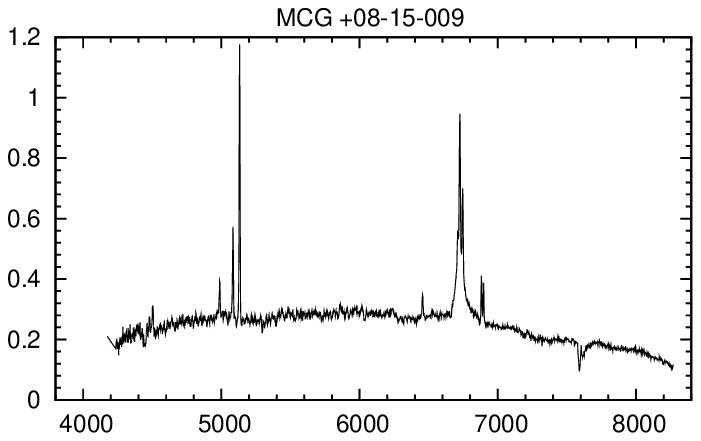}
\includegraphics[width=5cm]{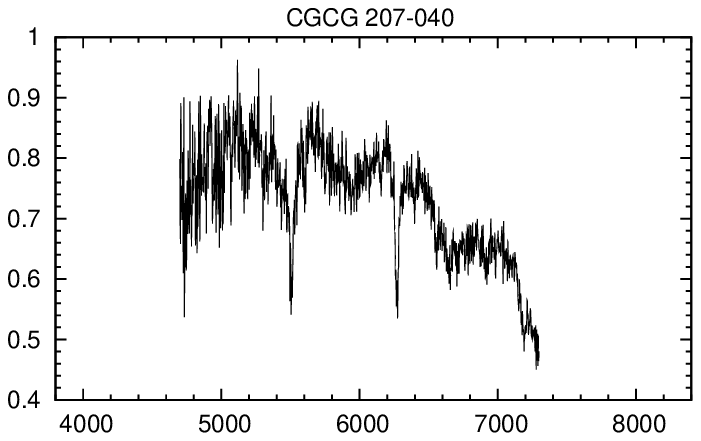}
\includegraphics[width=5cm]{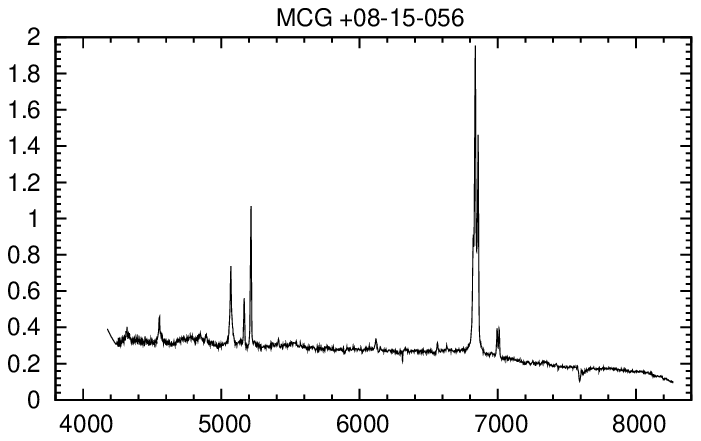}
\includegraphics[width=5cm]{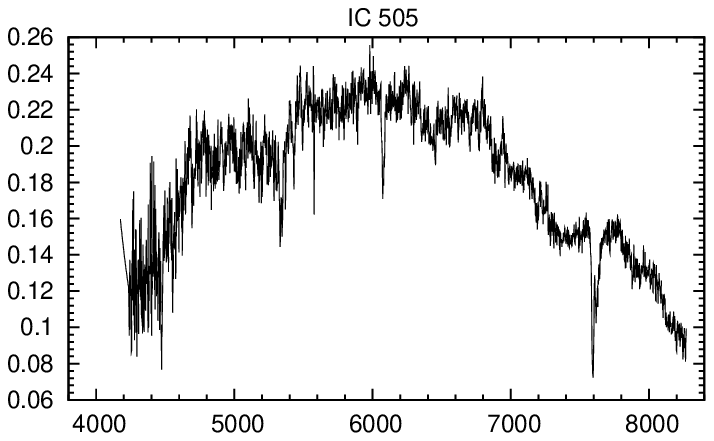}
\includegraphics[width=5cm]{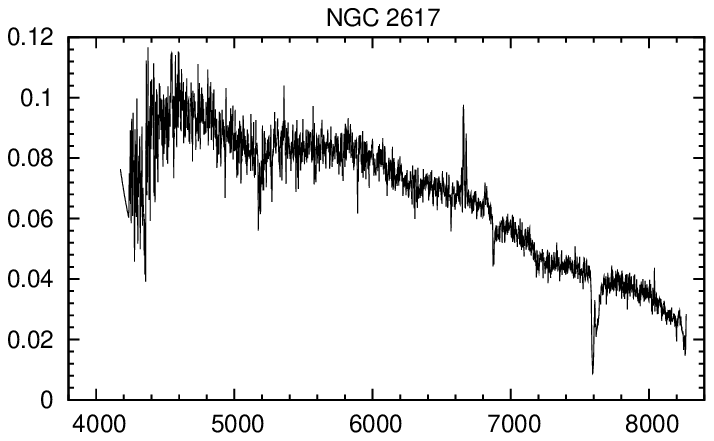}
\includegraphics[width=5cm]{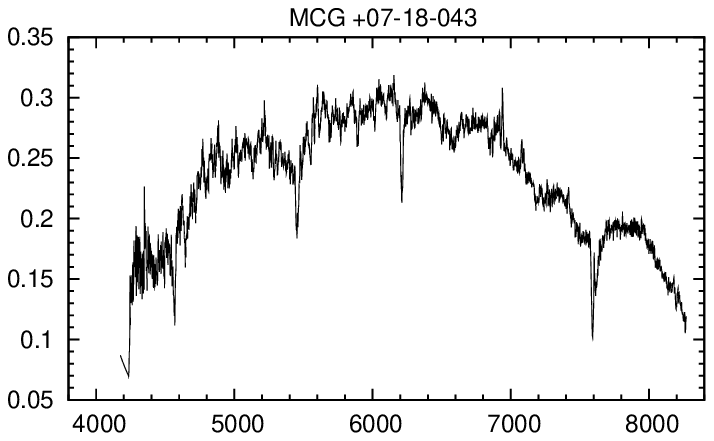}
\includegraphics[width=5cm]{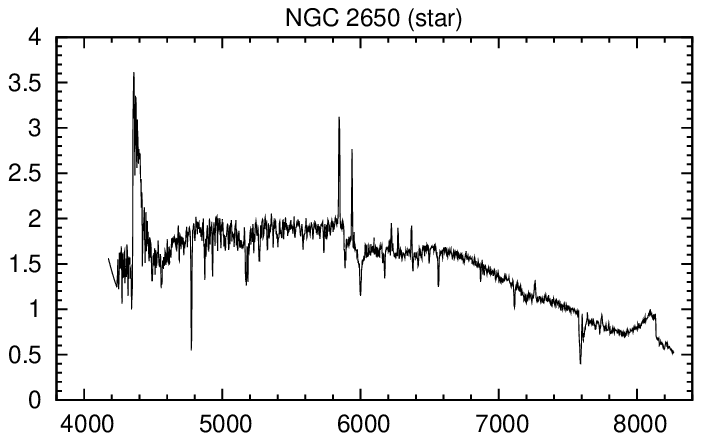}
\includegraphics[width=5cm]{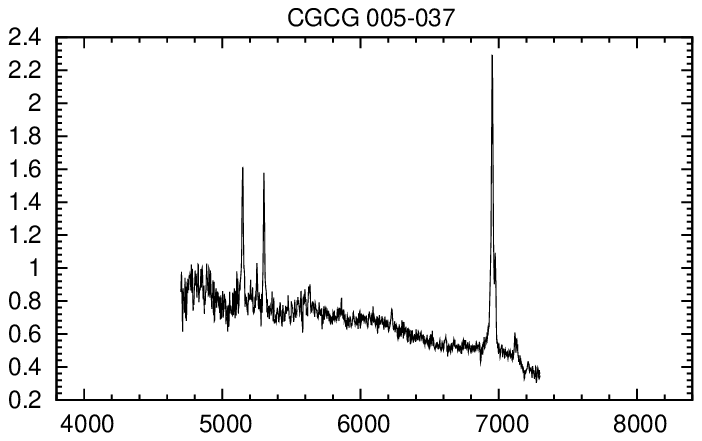}
\includegraphics[width=5cm]{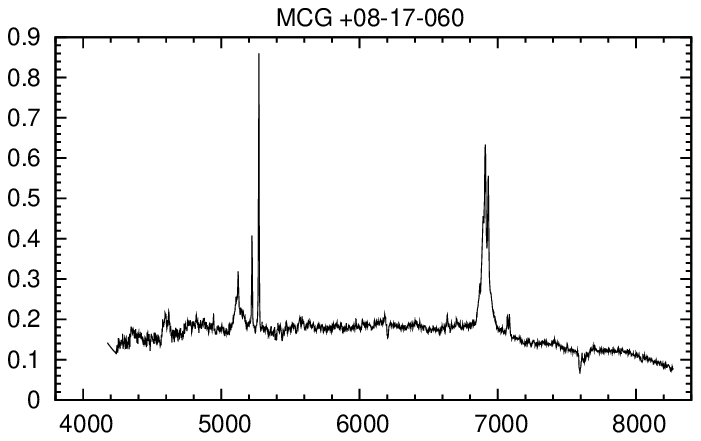}
\includegraphics[width=5cm]{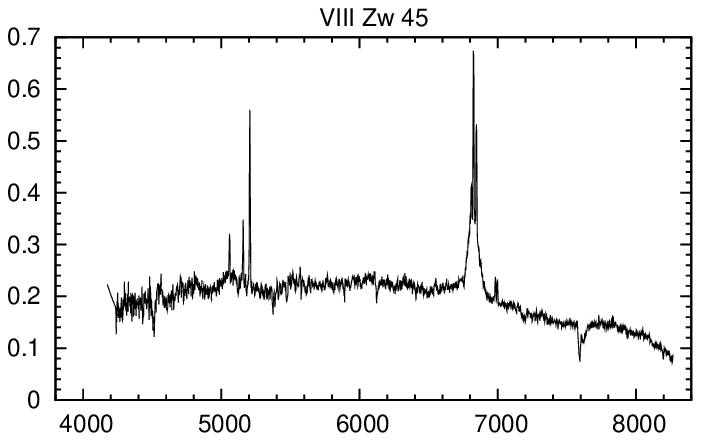}
\includegraphics[width=5cm]{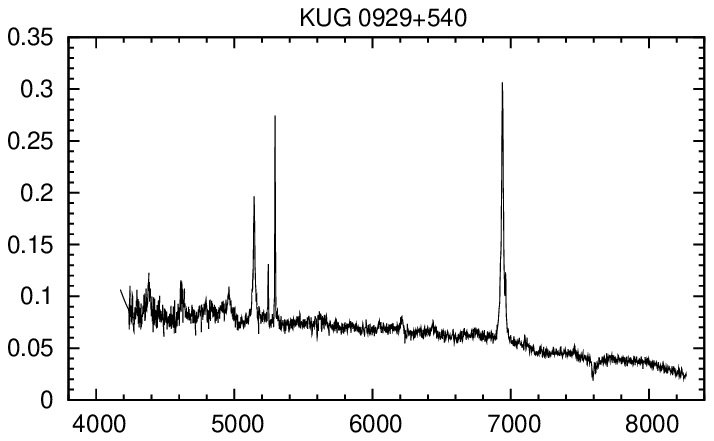}
\includegraphics[width=5cm]{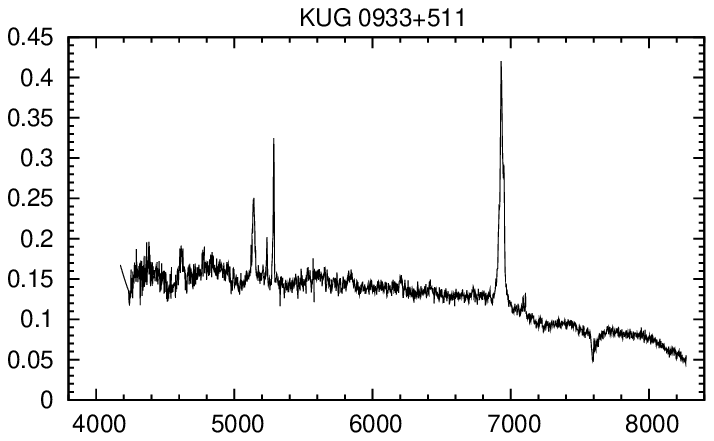}
\includegraphics[width=5cm]{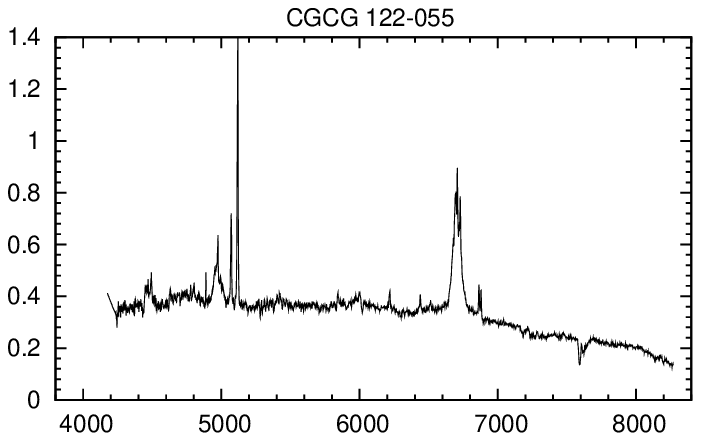}
\includegraphics[width=5cm]{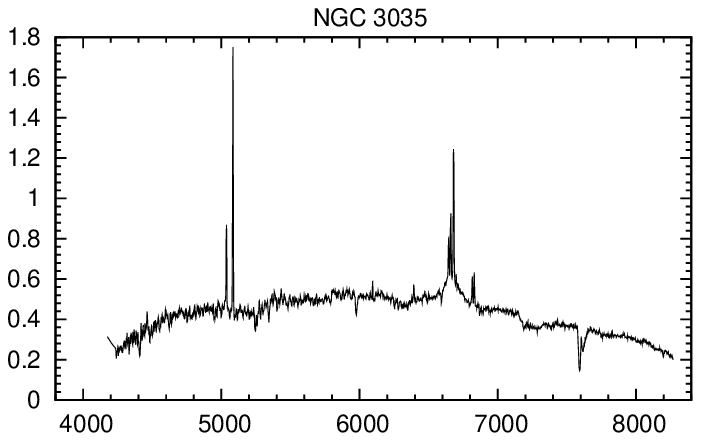}
\includegraphics[width=5cm]{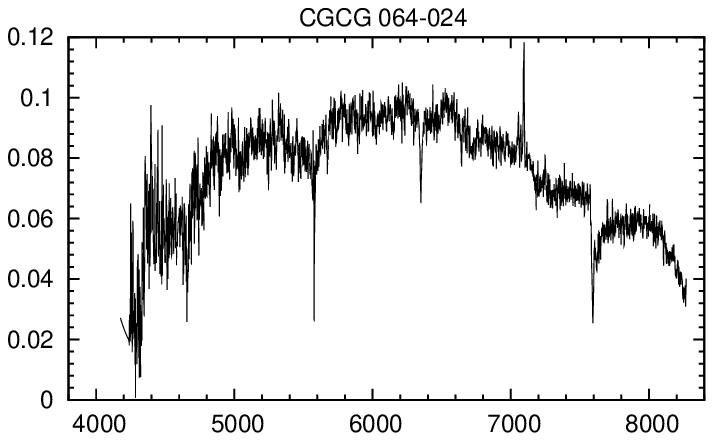}
\includegraphics[width=5cm]{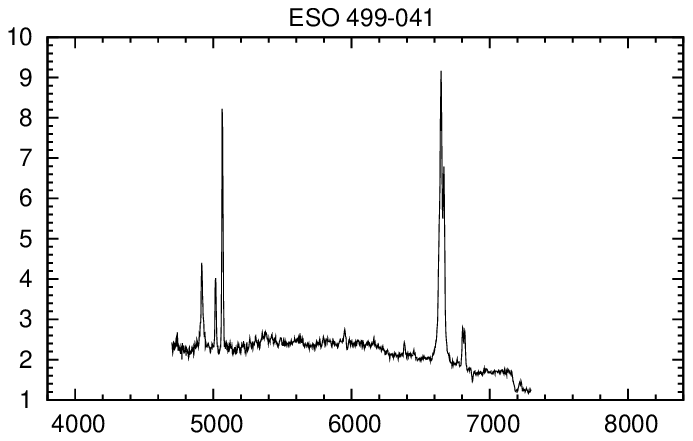}
\includegraphics[width=5cm]{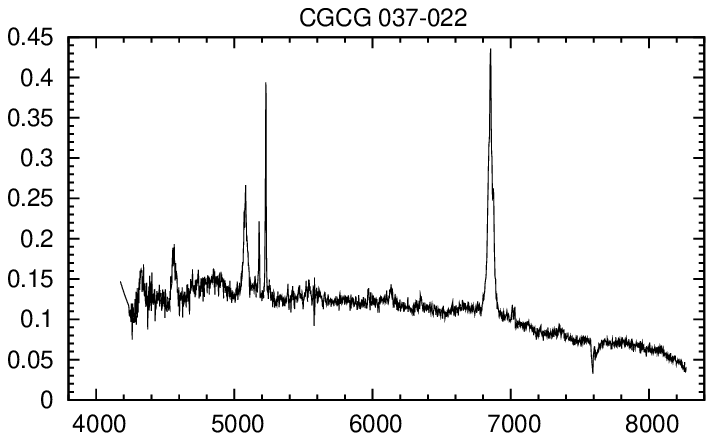}
\includegraphics[width=5cm]{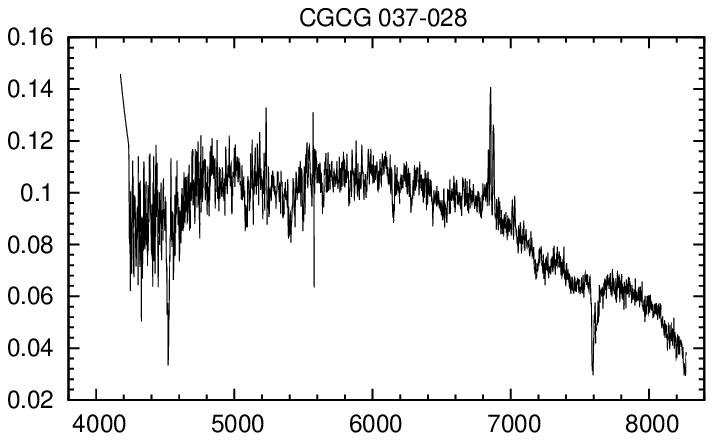}
\includegraphics[width=5cm]{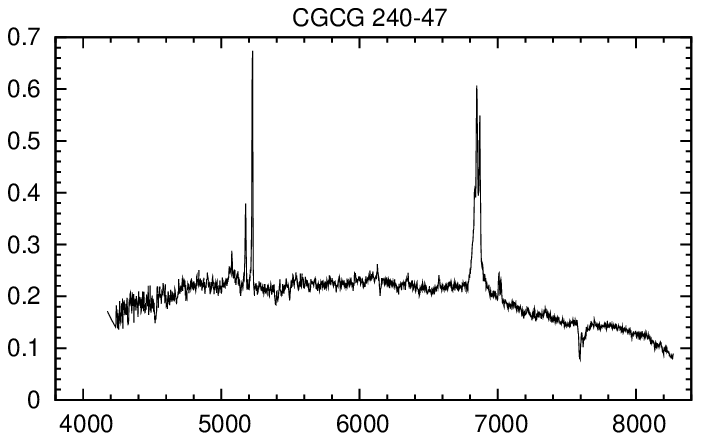}
\includegraphics[width=5cm]{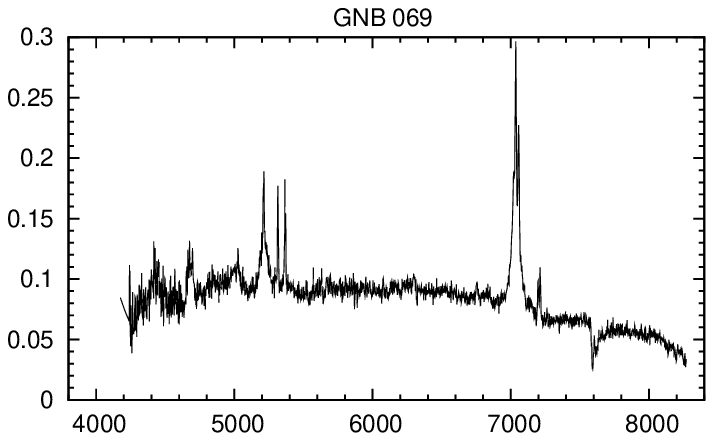}
\includegraphics[width=5cm]{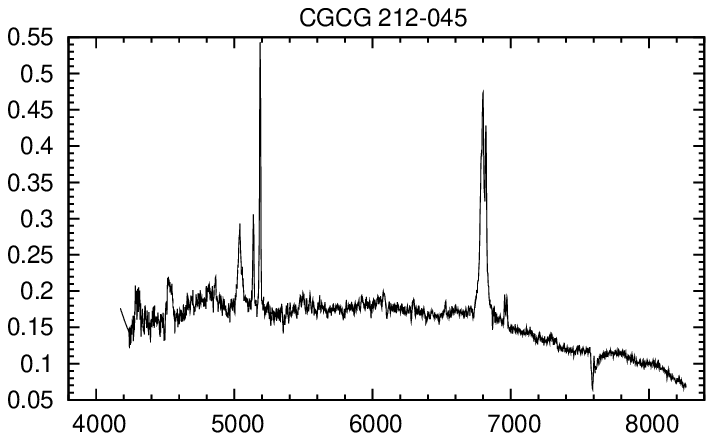}
\includegraphics[width=5cm]{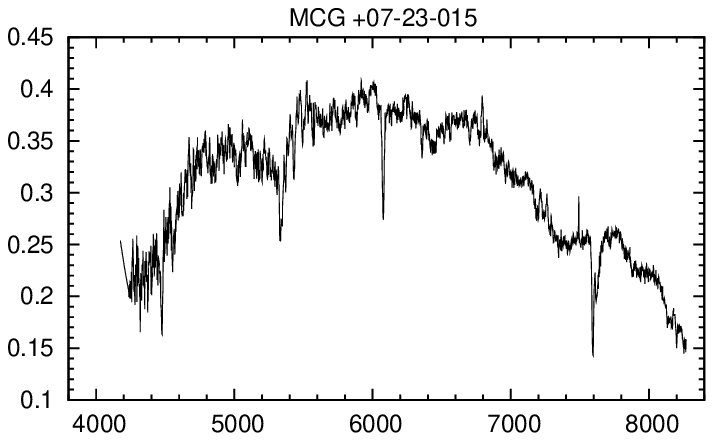}
\includegraphics[width=5cm]{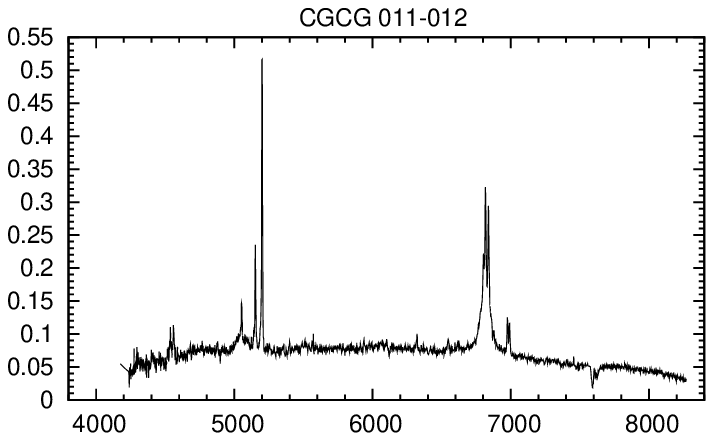}
\includegraphics[width=5cm]{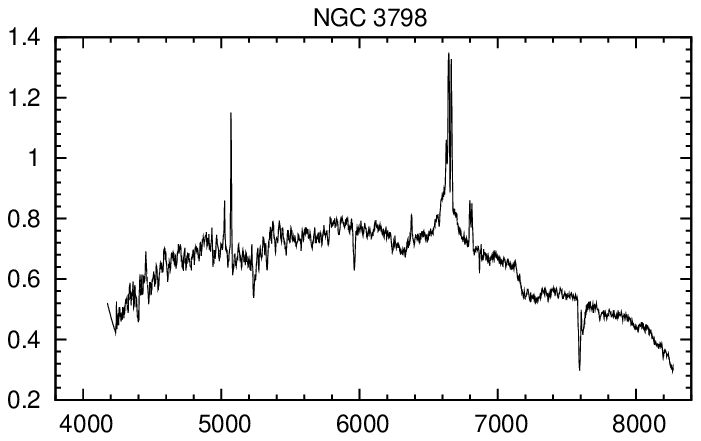}
\includegraphics[width=5cm]{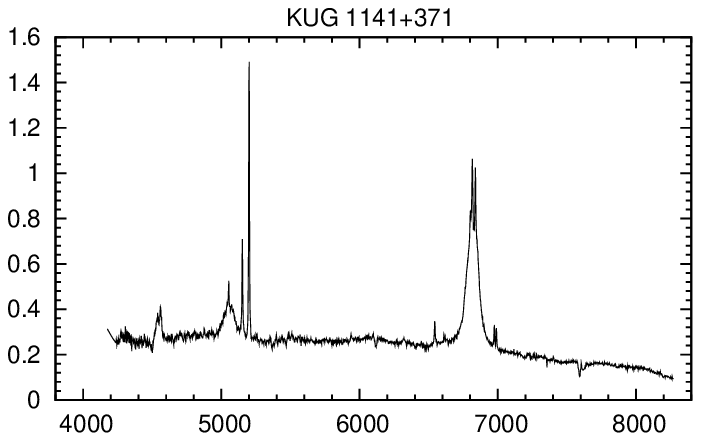}
\includegraphics[width=5cm]{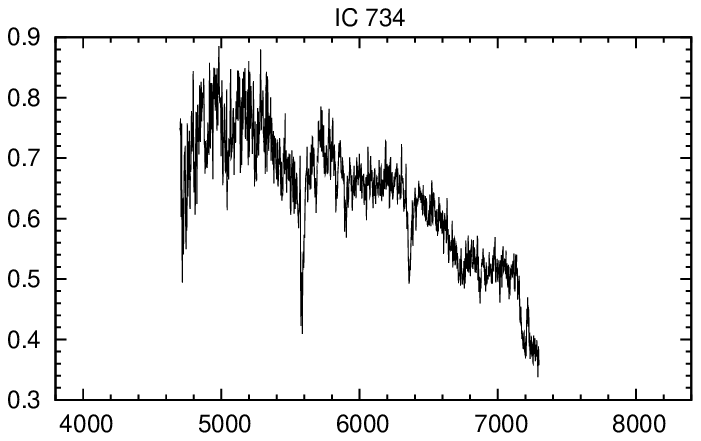}
\includegraphics[width=5cm]{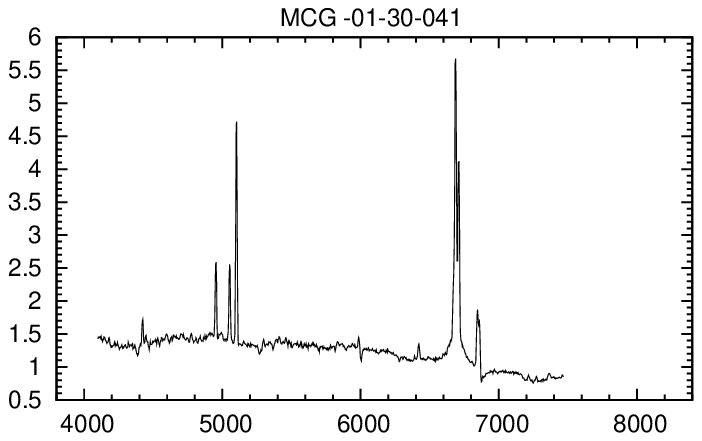}
\includegraphics[width=5cm]{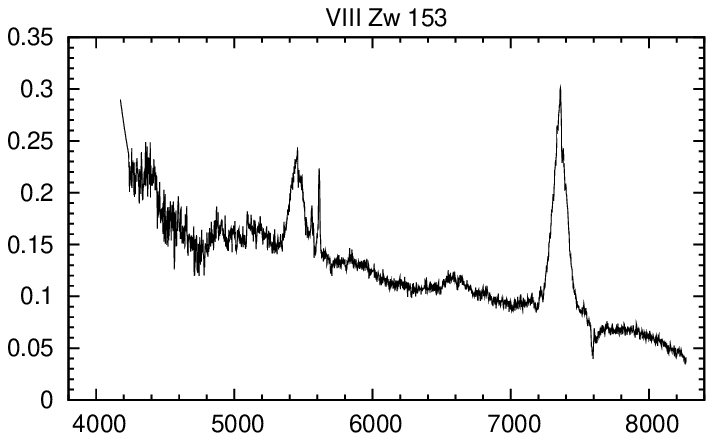}
\includegraphics[width=5cm]{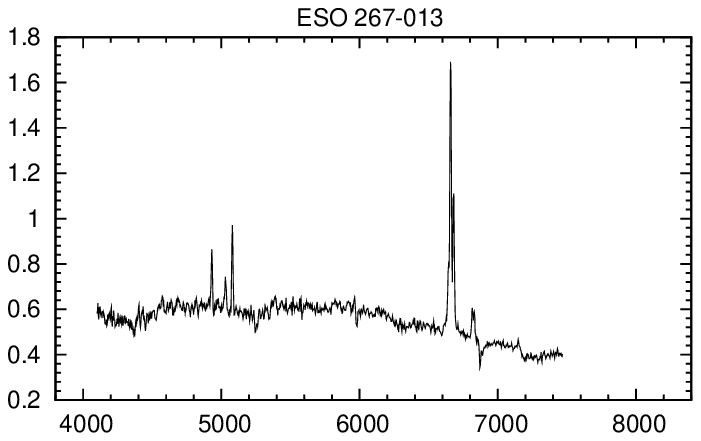}
\includegraphics[width=5cm]{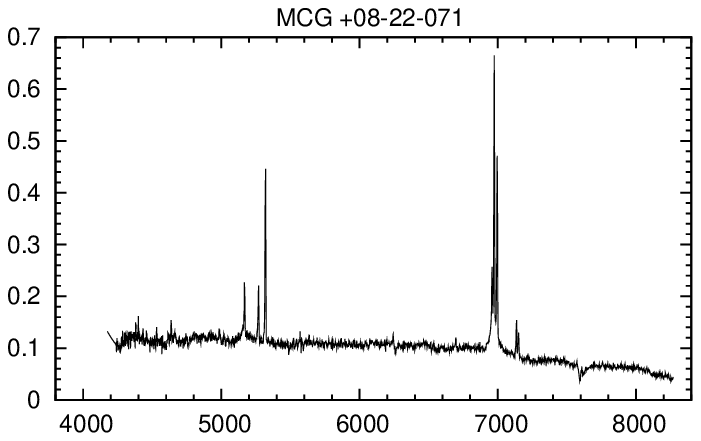}
\includegraphics[width=5cm]{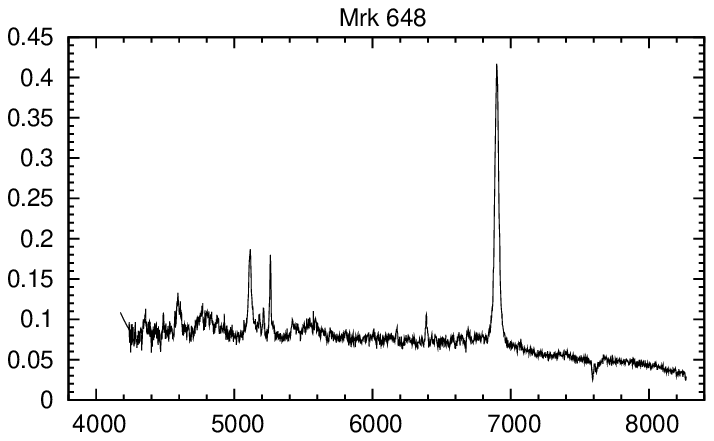}
\includegraphics[width=5cm]{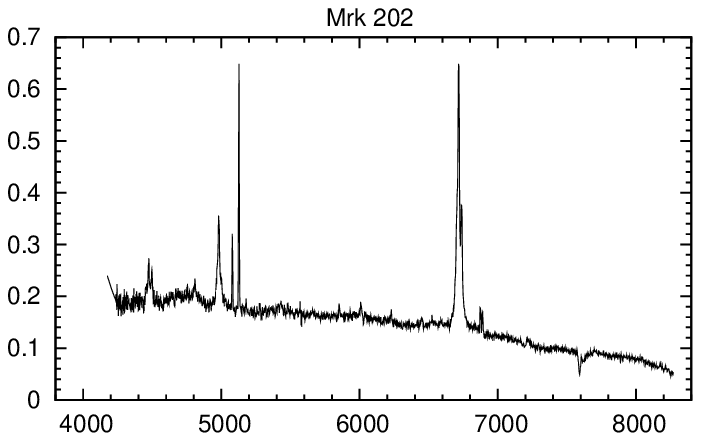}
\includegraphics[width=5cm]{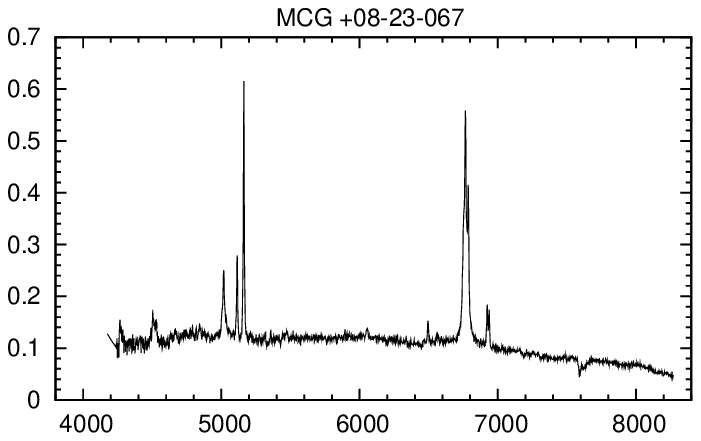}
\includegraphics[width=5cm]{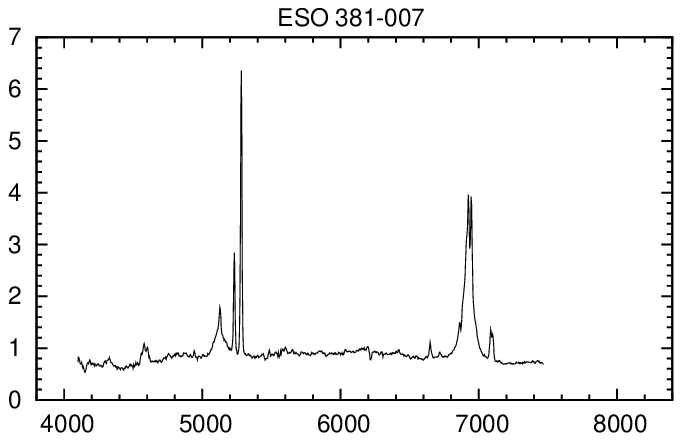}
\includegraphics[width=5cm]{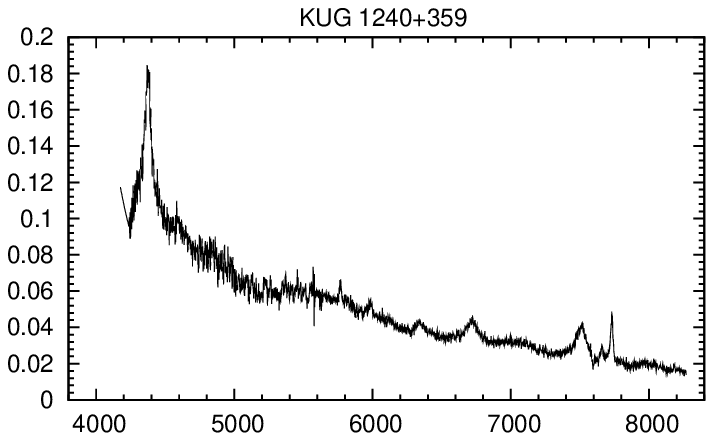}
\includegraphics[width=5cm]{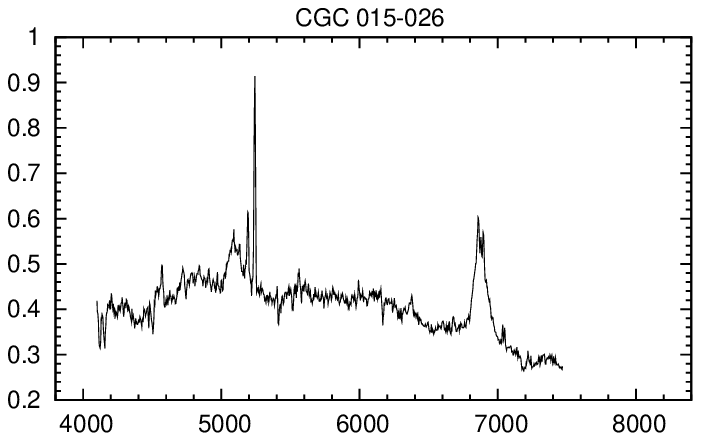}
\includegraphics[width=5cm]{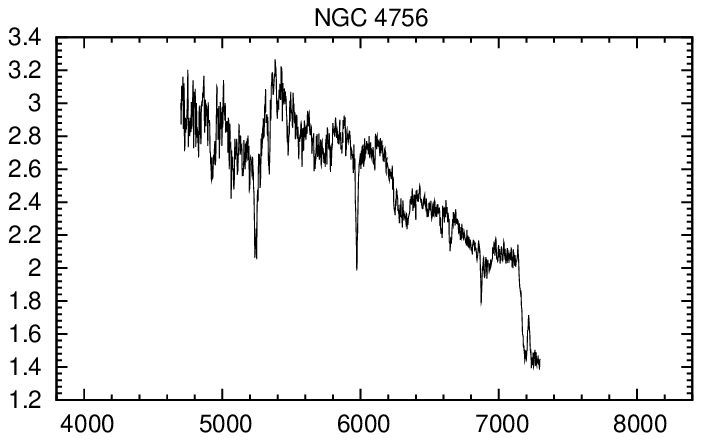}
\includegraphics[width=5cm]{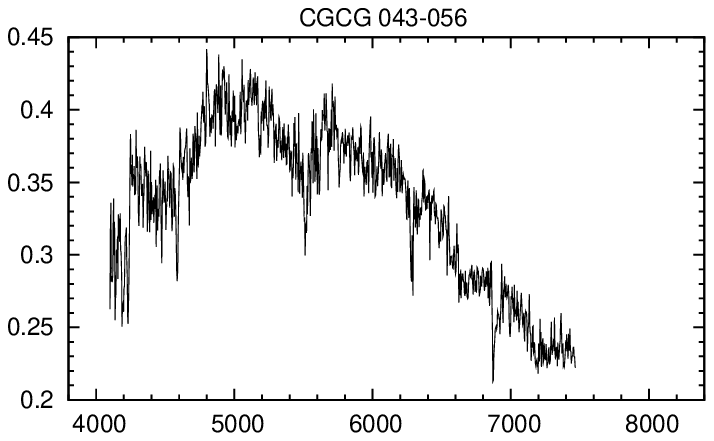}
\includegraphics[width=5cm]{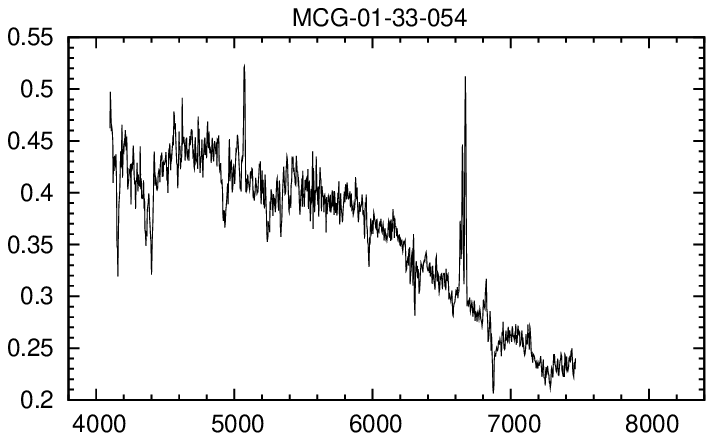}
\includegraphics[width=5cm]{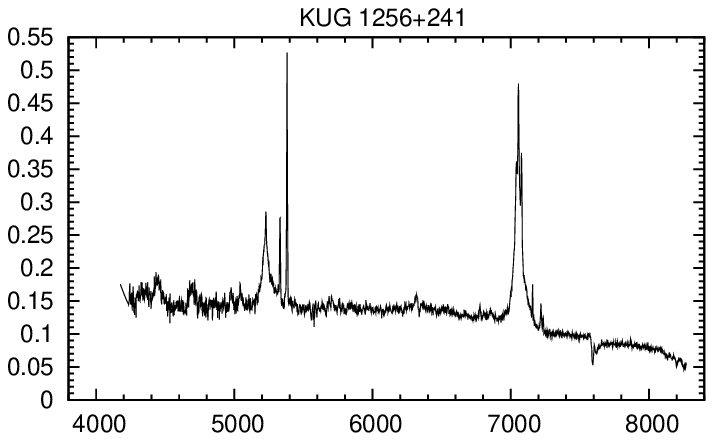}
\includegraphics[width=5cm]{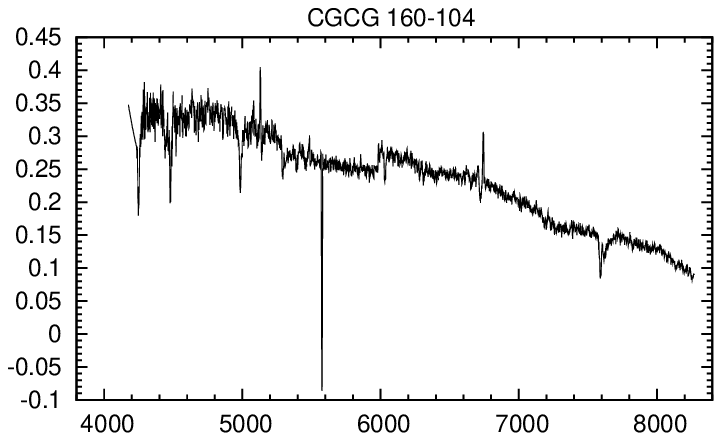}
\includegraphics[width=5cm]{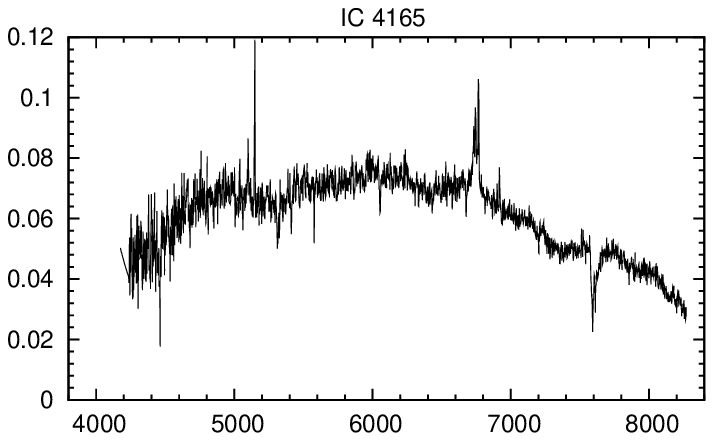}
\includegraphics[width=5cm]{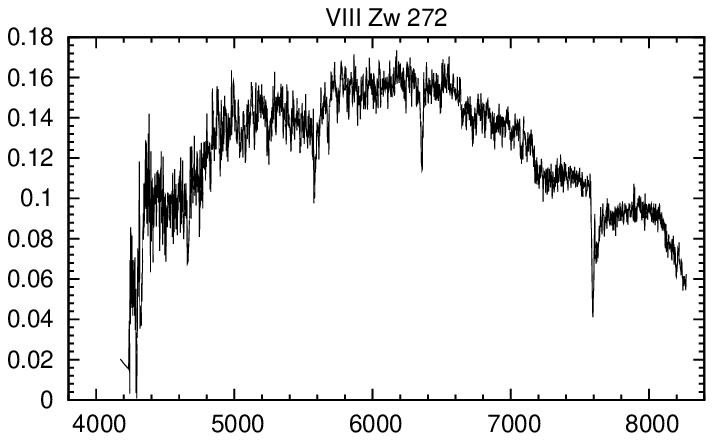}
\includegraphics[width=5cm]{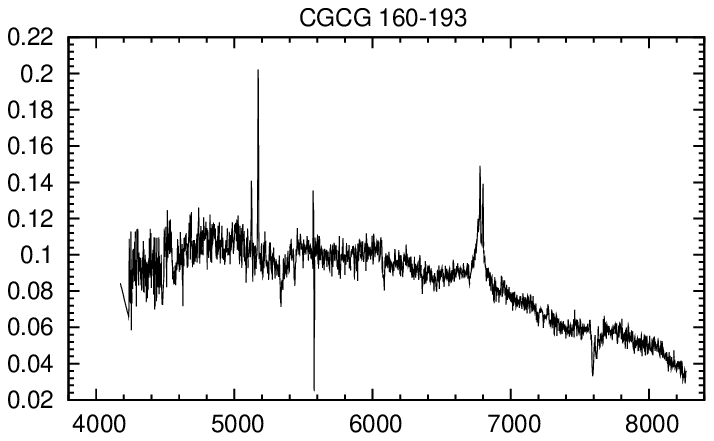}
\includegraphics[width=5cm]{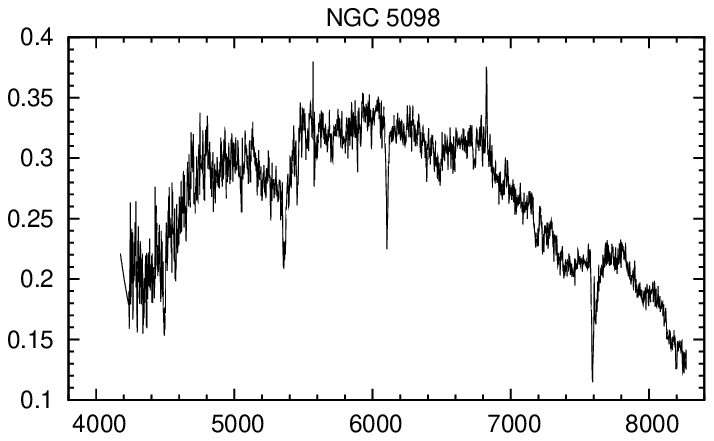}
\includegraphics[width=5cm]{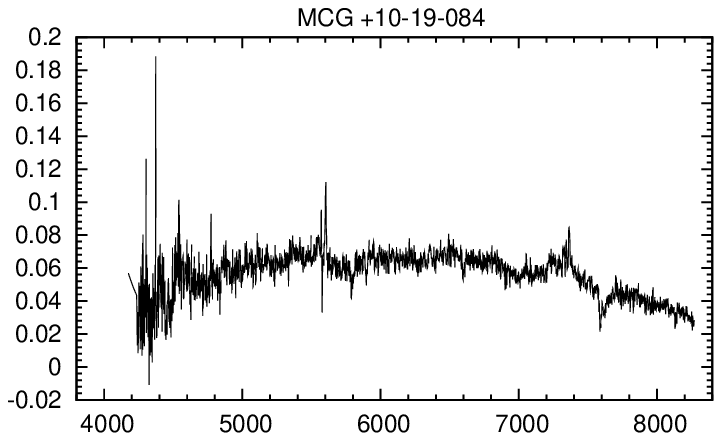}
\includegraphics[width=5cm]{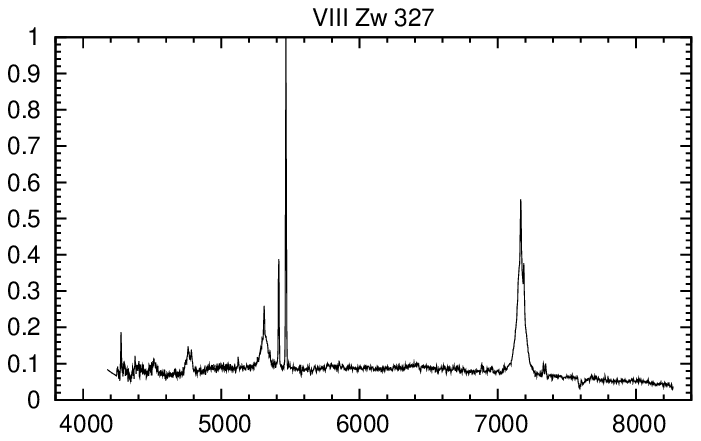}
\includegraphics[width=5cm]{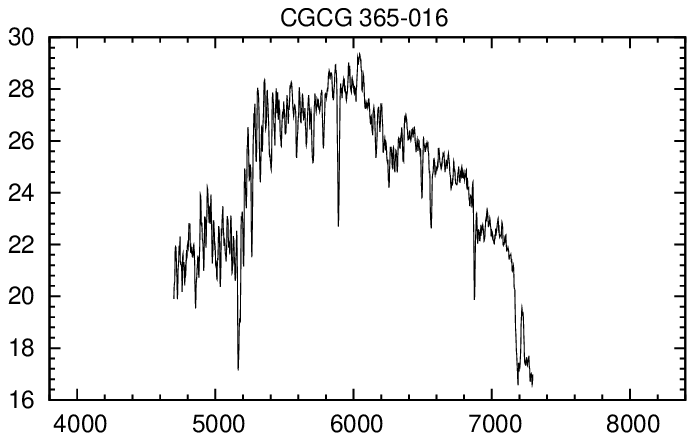}
\includegraphics[width=5cm]{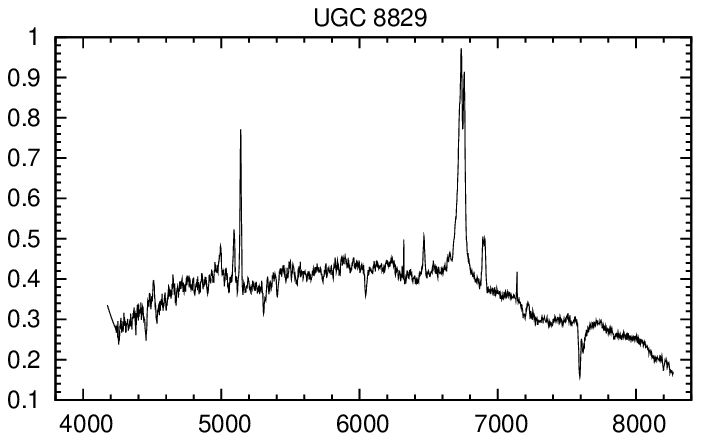}
\includegraphics[width=5cm]{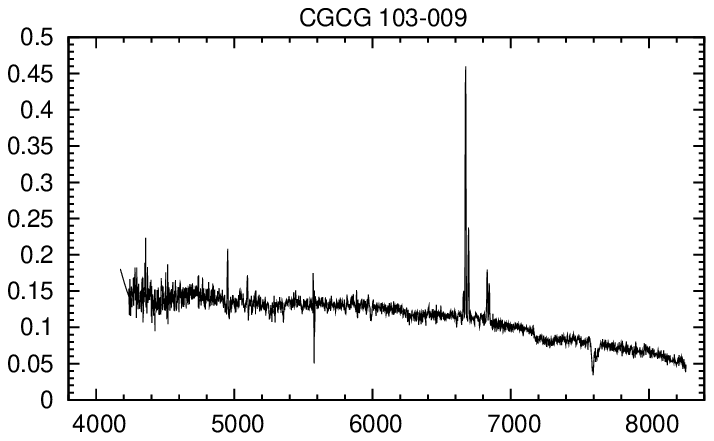}
\includegraphics[width=5cm]{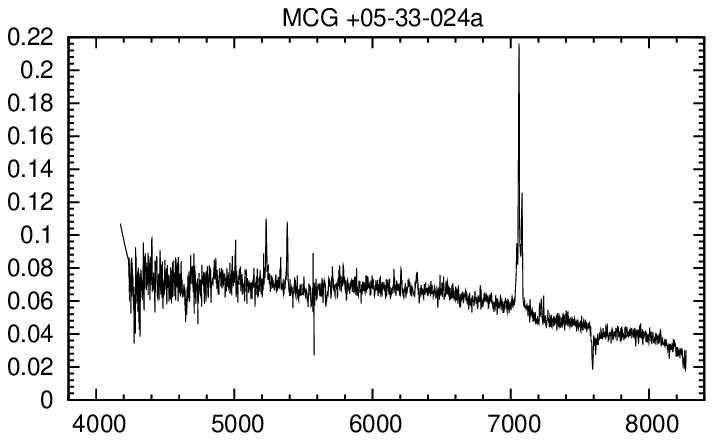}
\includegraphics[width=5cm]{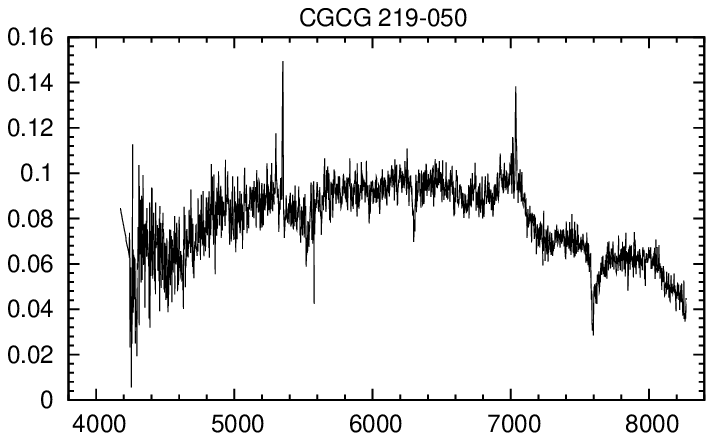}
\includegraphics[width=5cm]{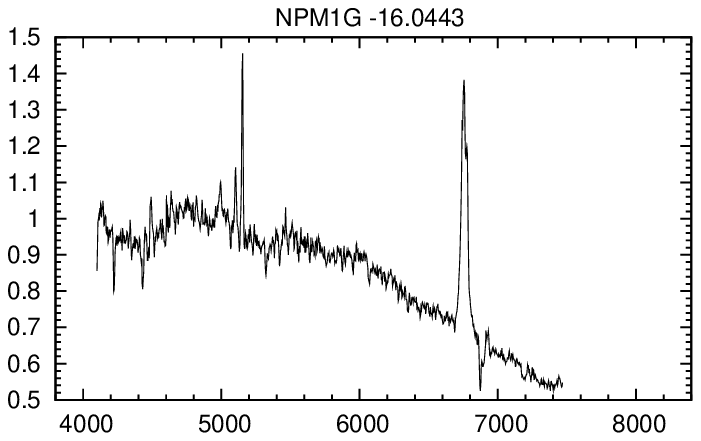}
\includegraphics[width=5cm]{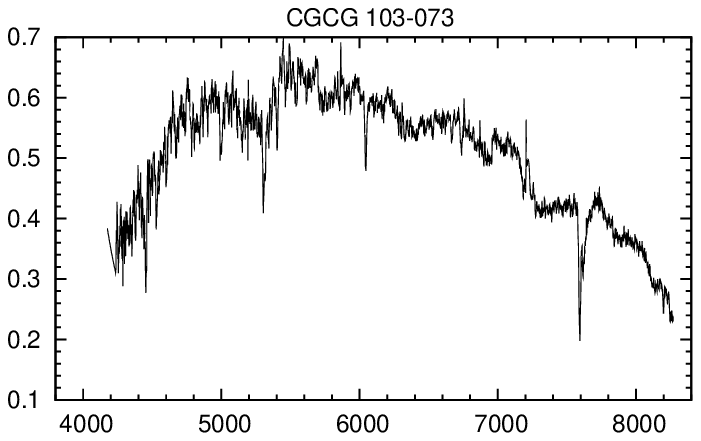}
\includegraphics[width=5cm]{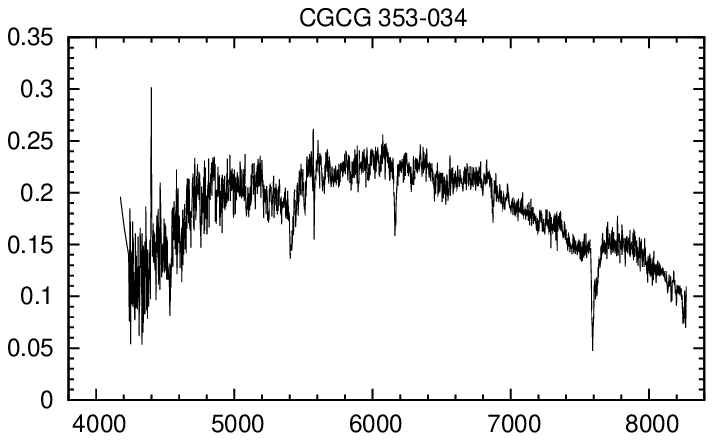}
\includegraphics[width=5cm]{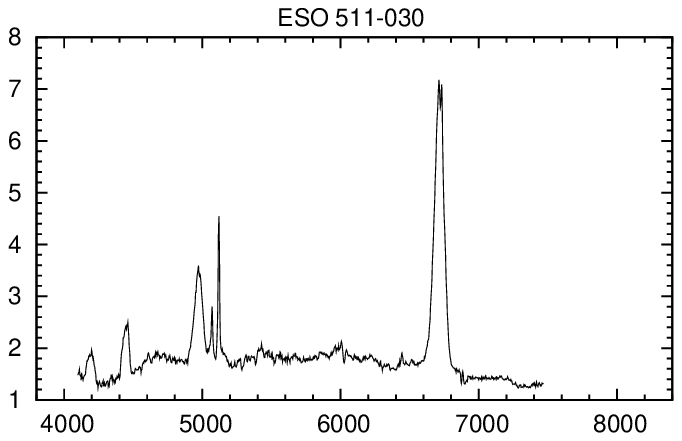}
\includegraphics[width=5cm]{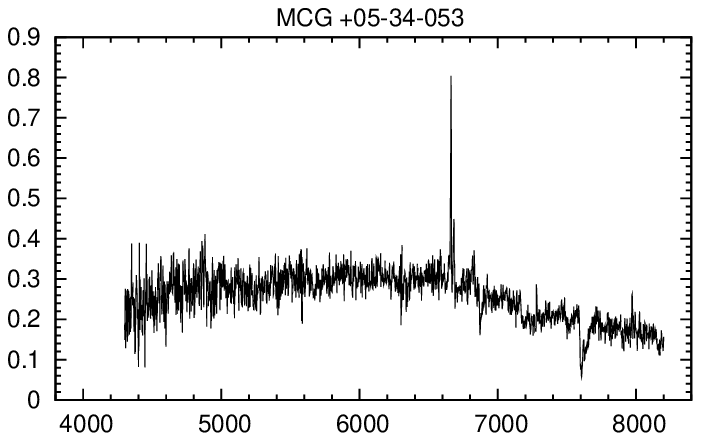}
\includegraphics[width=5cm]{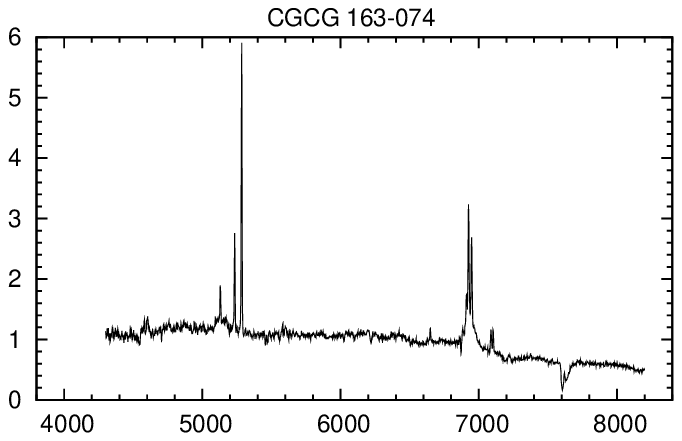}
\includegraphics[width=5cm]{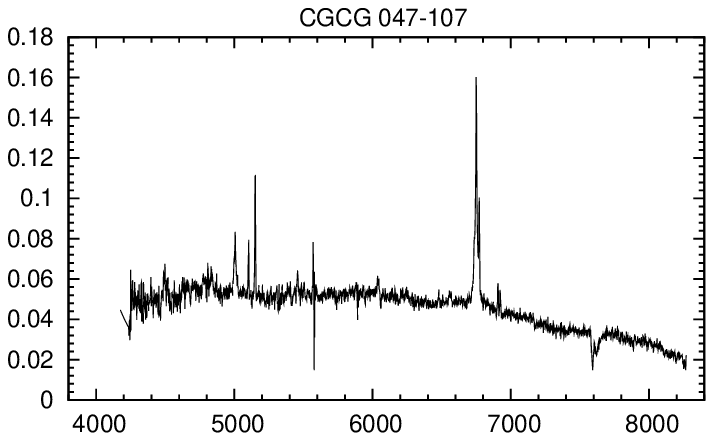}
\includegraphics[width=5cm]{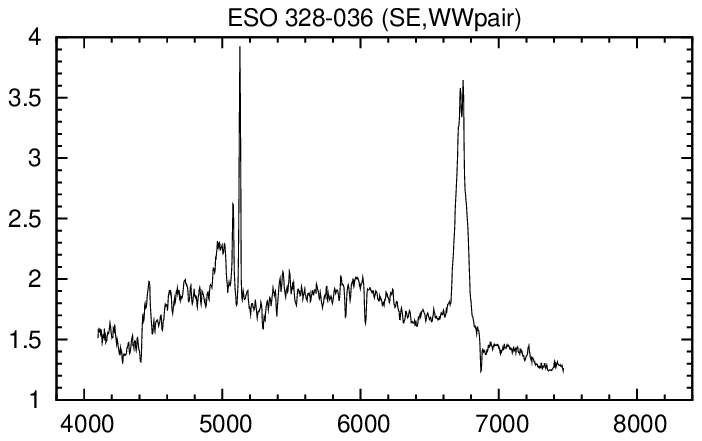}
\includegraphics[width=5cm]{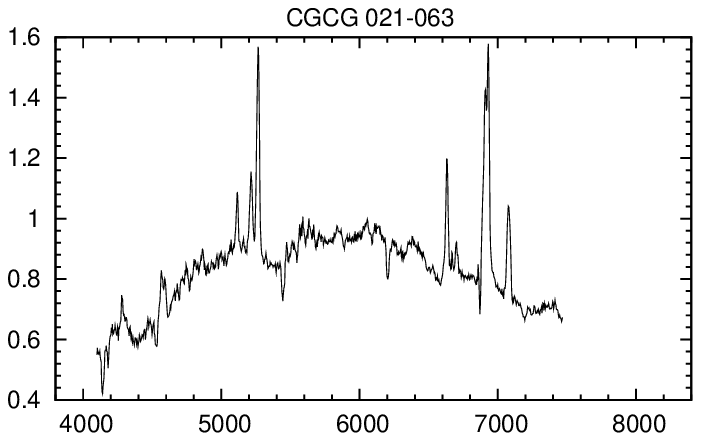}
\includegraphics[width=5cm]{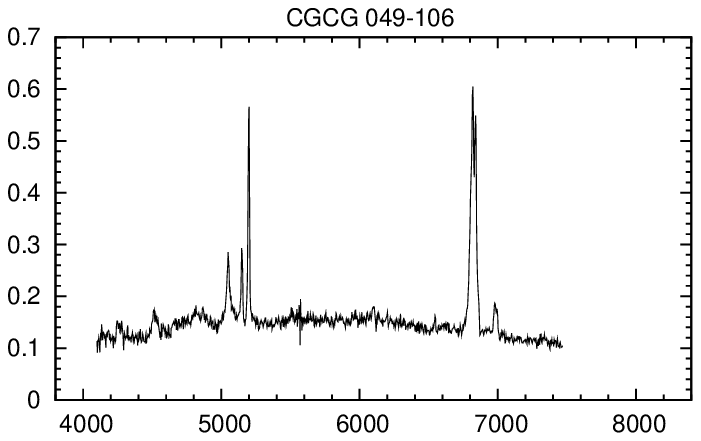}
\includegraphics[width=5cm]{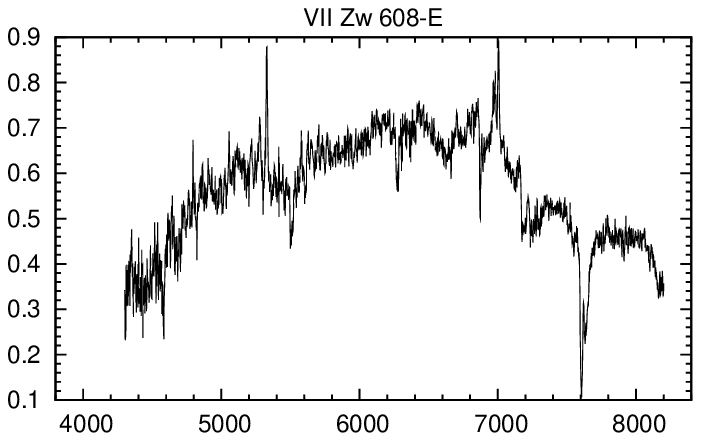}
\includegraphics[width=5cm]{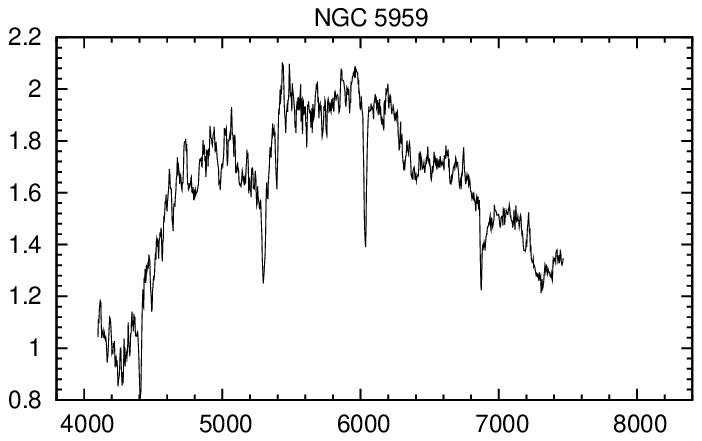}
\includegraphics[width=5cm]{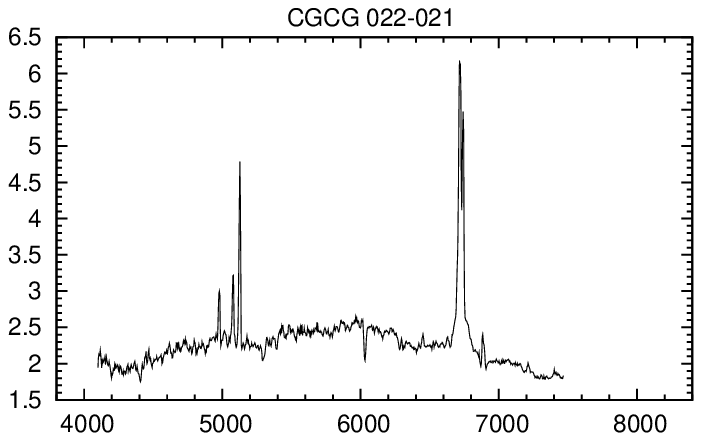}
\includegraphics[width=5cm]{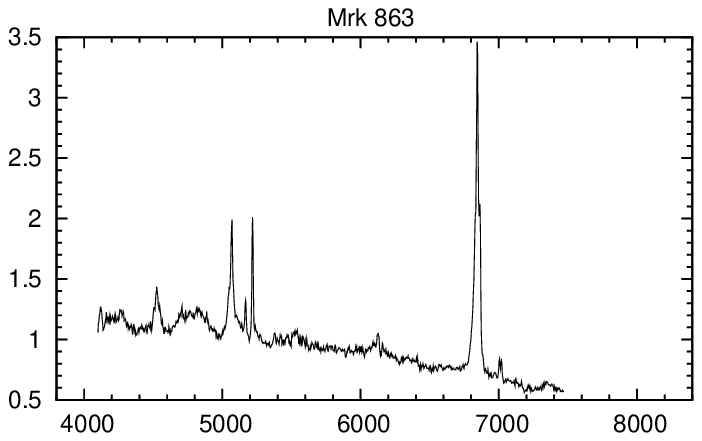}
\includegraphics[width=5cm]{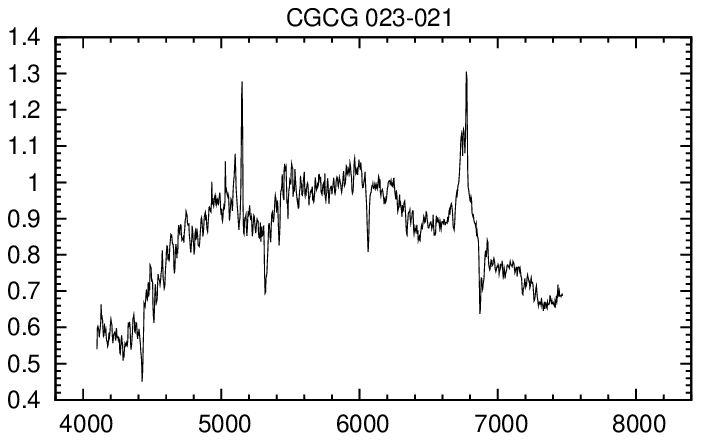}
\includegraphics[width=5cm]{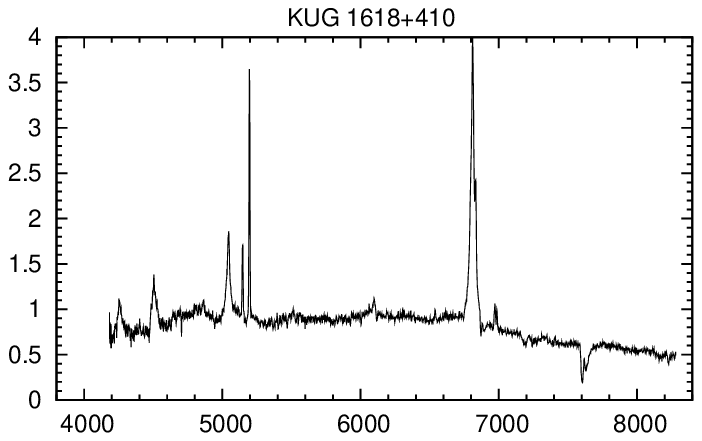}
\includegraphics[width=5cm]{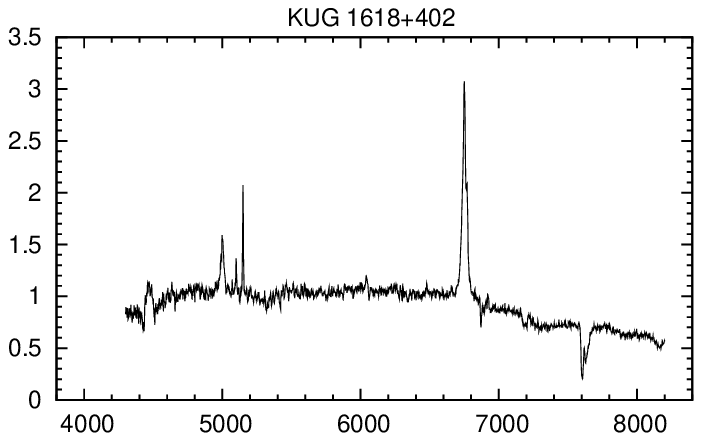}
\includegraphics[width=5cm]{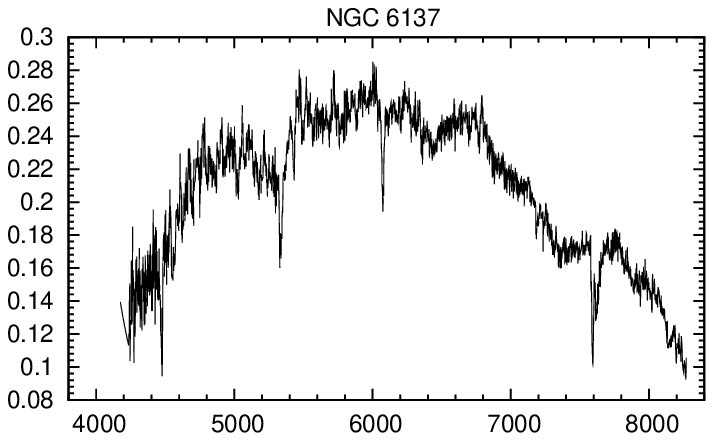}
\includegraphics[width=5cm]{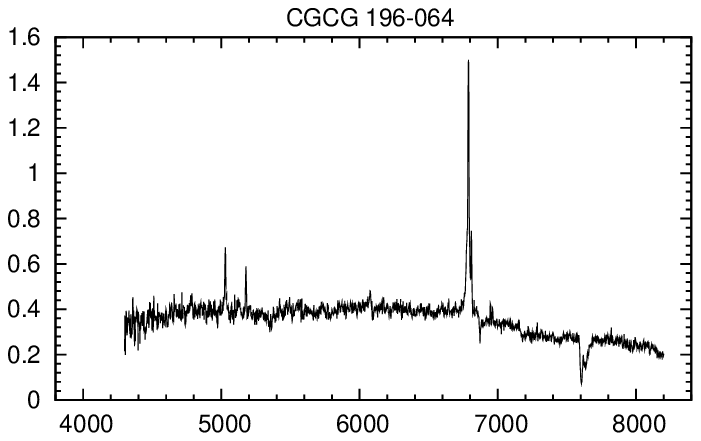}
\includegraphics[width=5cm]{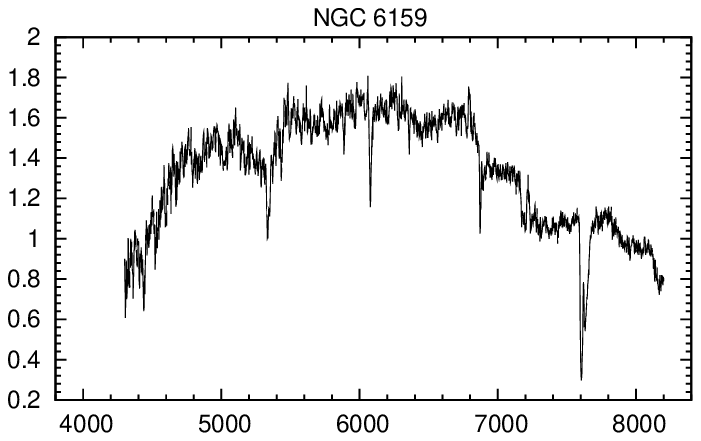}
\includegraphics[width=5cm]{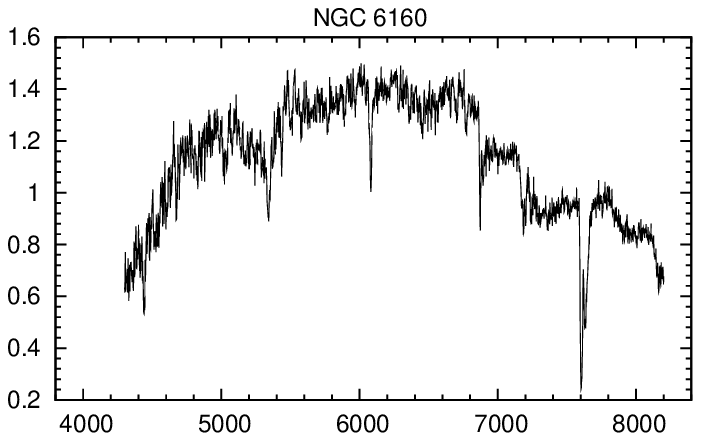}
\includegraphics[width=5cm]{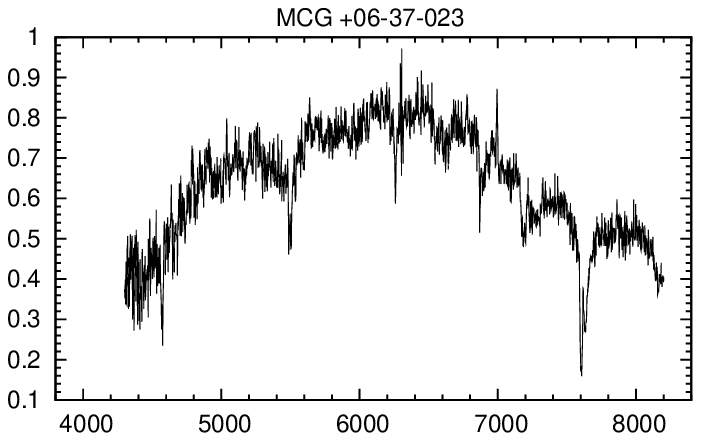}
\includegraphics[width=5cm]{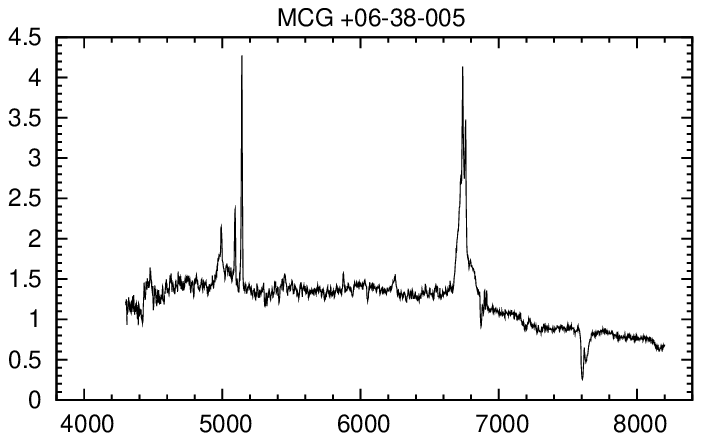}
\includegraphics[width=5cm]{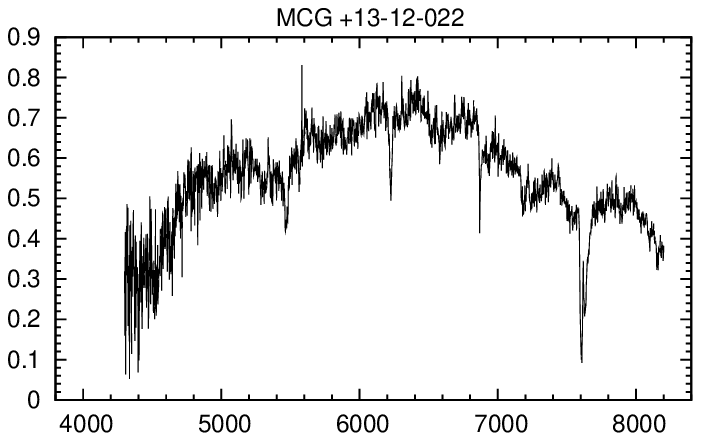}
\includegraphics[width=5cm]{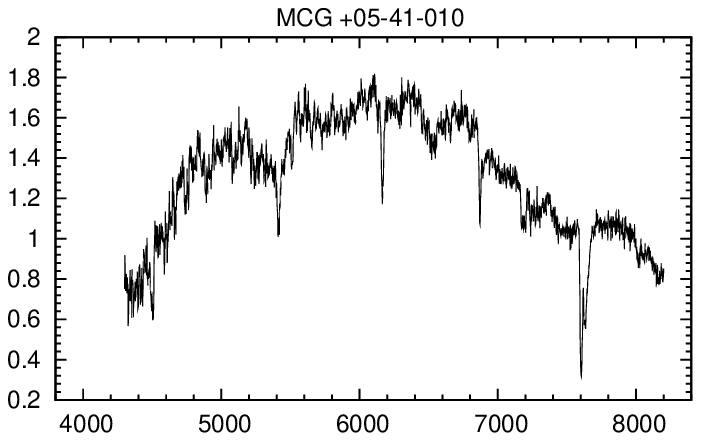}
\includegraphics[width=5cm]{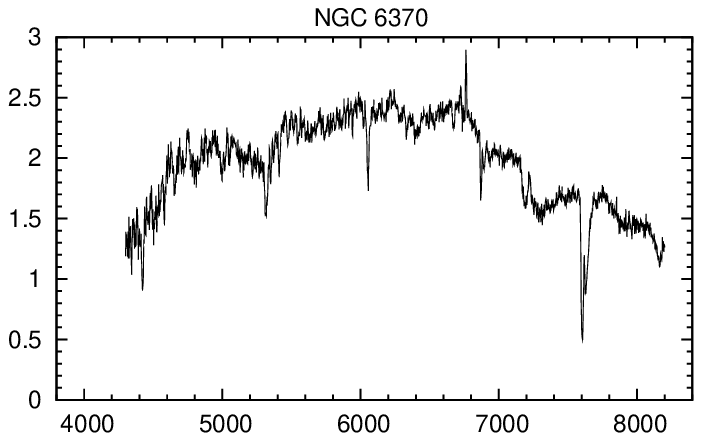}
\includegraphics[width=5cm]{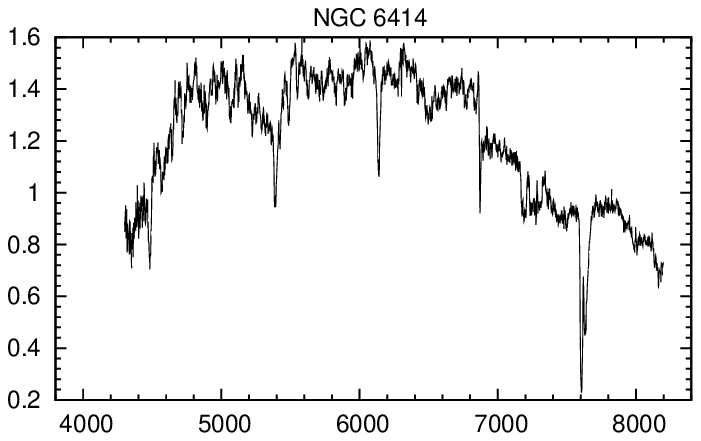}
\includegraphics[width=5cm]{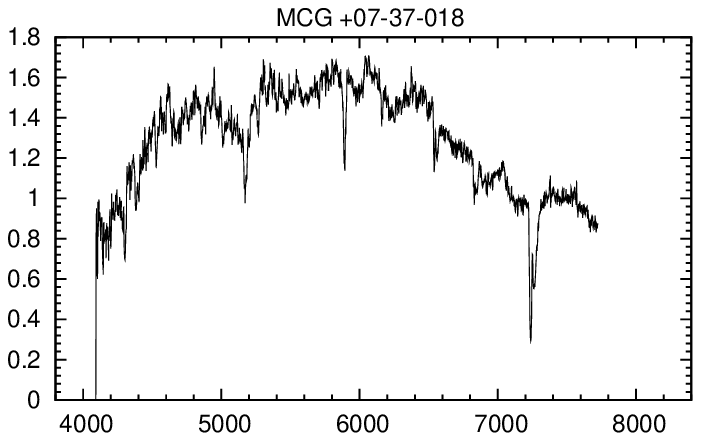}
\includegraphics[width=5cm]{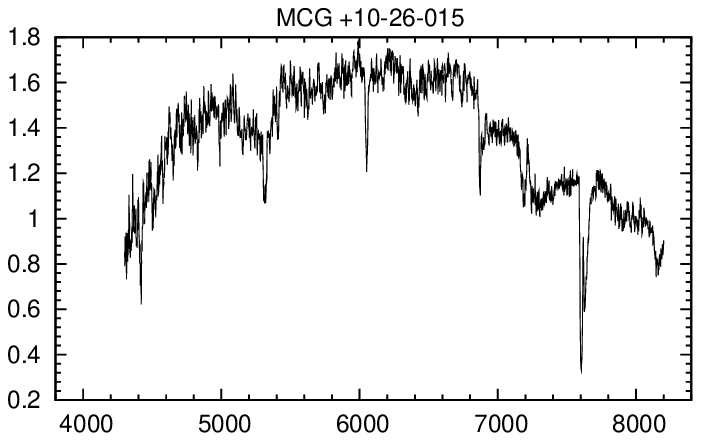}
\includegraphics[width=5cm]{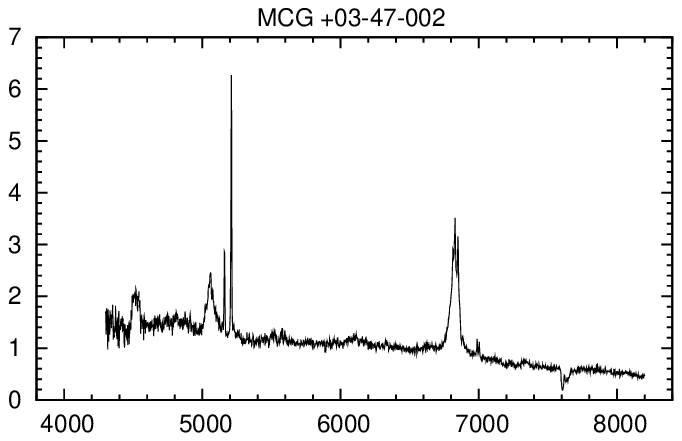}
\includegraphics[width=5cm]{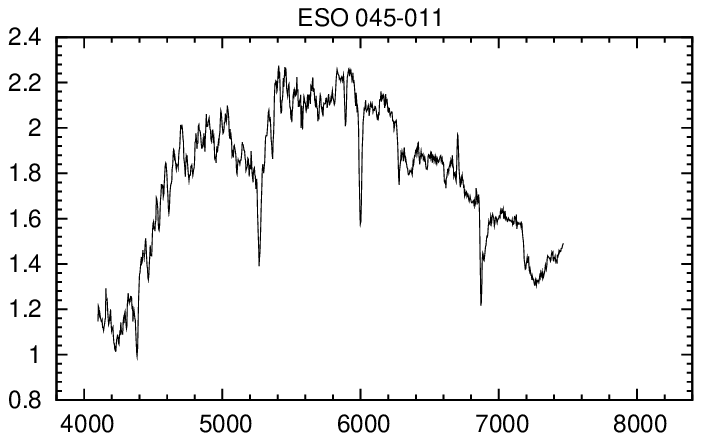}
\includegraphics[width=5cm]{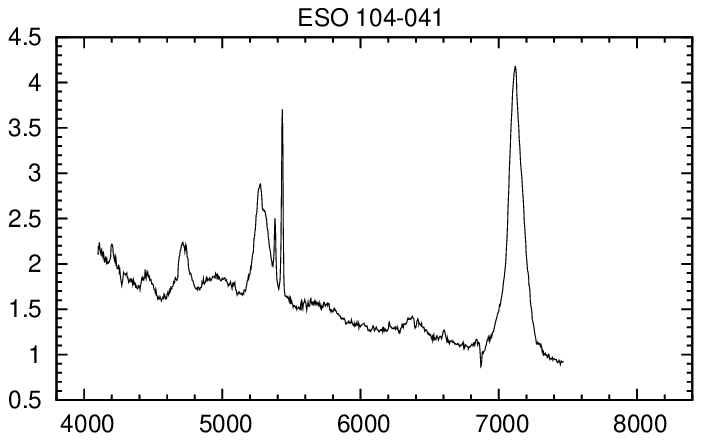}
\includegraphics[width=5cm]{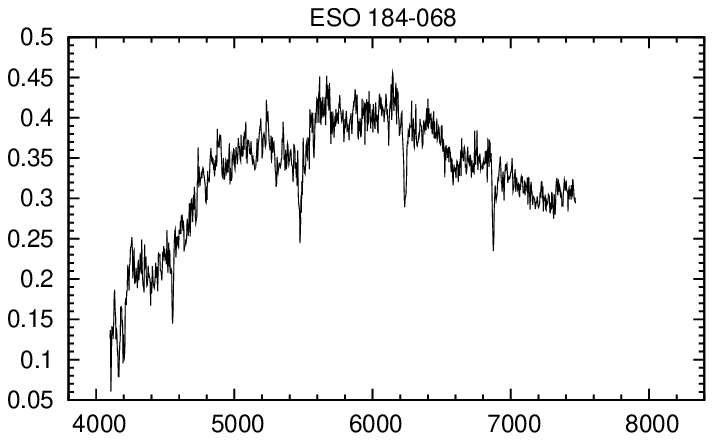}
\includegraphics[width=5cm]{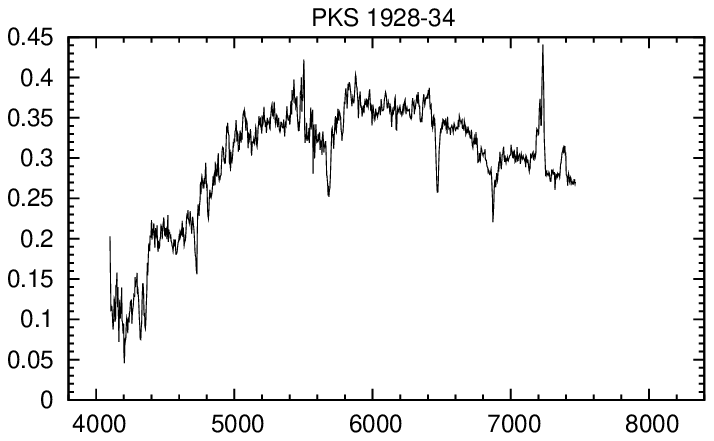}
\includegraphics[width=5cm]{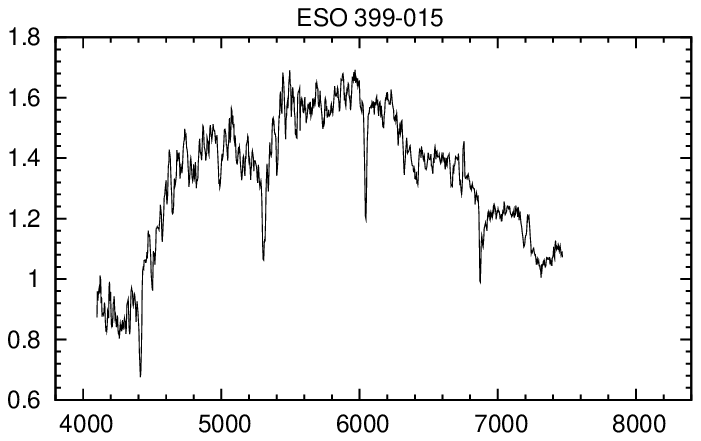}
\includegraphics[width=5cm]{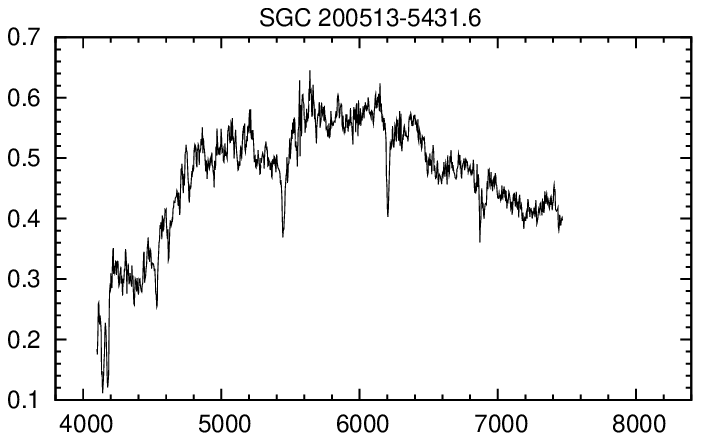}
\includegraphics[width=5cm]{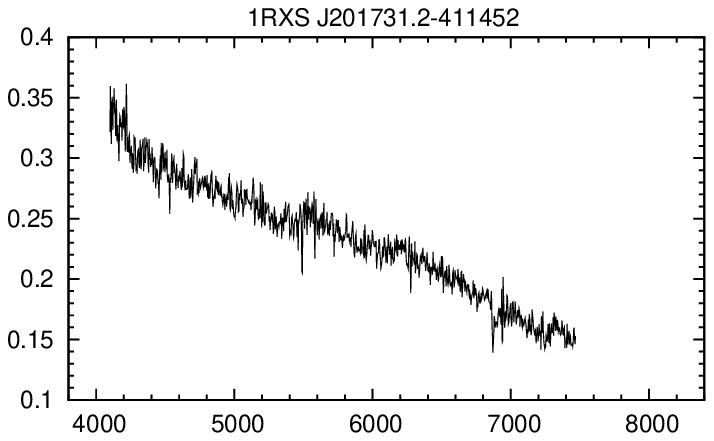}
\includegraphics[width=5cm]{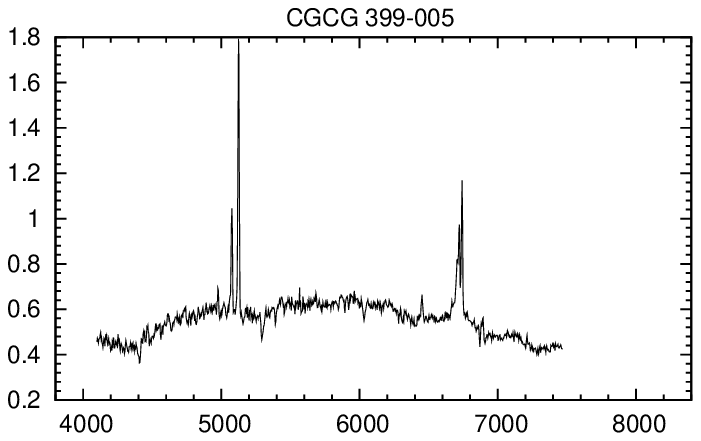}
\includegraphics[width=5cm]{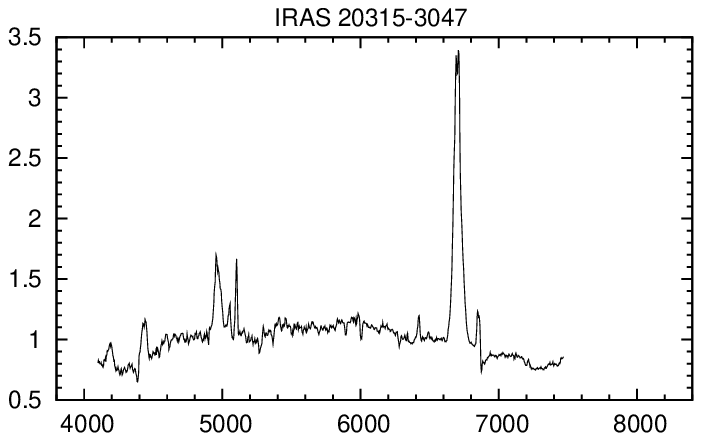}
\includegraphics[width=5cm]{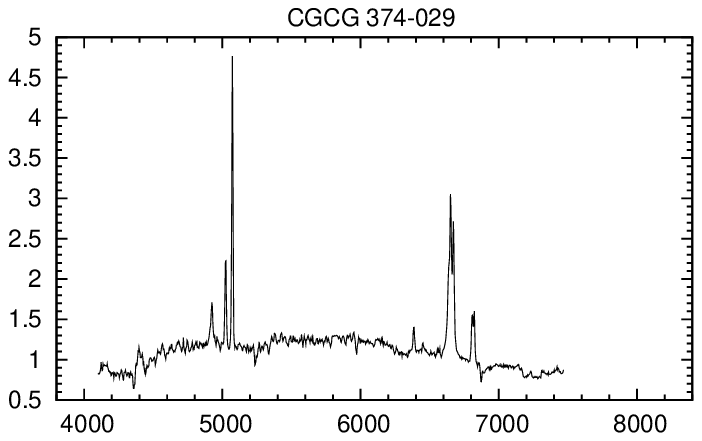}
\includegraphics[width=5cm]{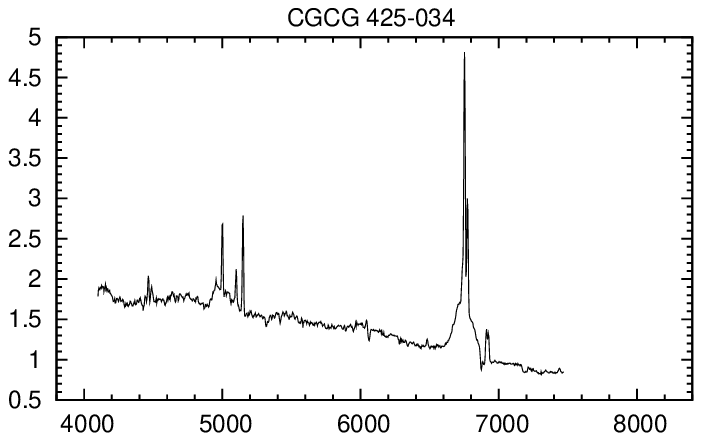}
\includegraphics[width=5cm]{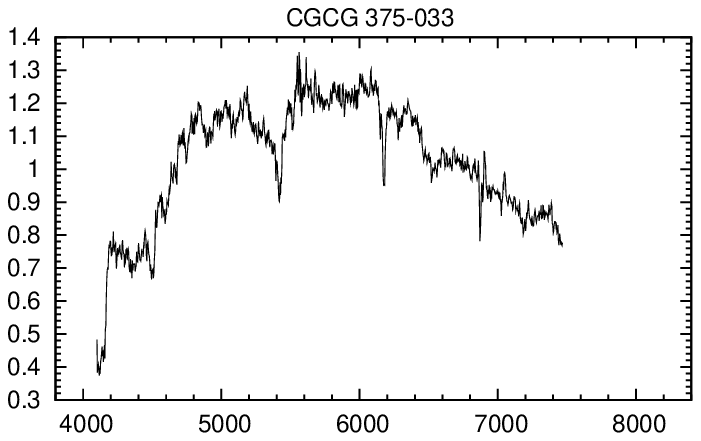}
\includegraphics[width=5cm]{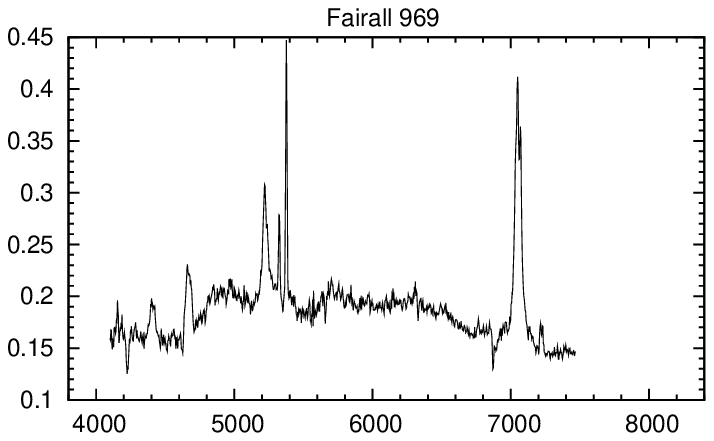}
\includegraphics[width=5cm]{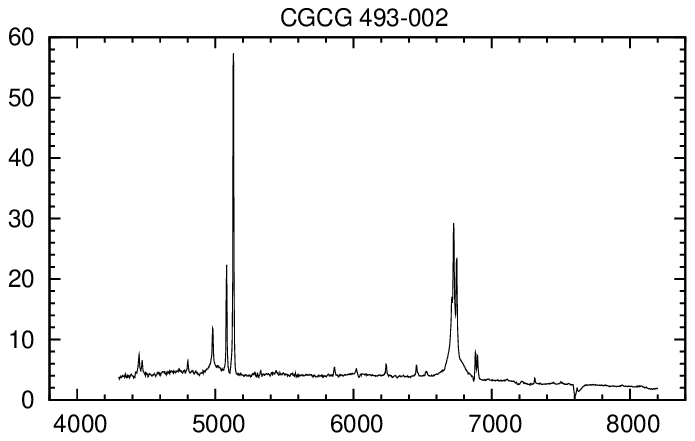}
\includegraphics[width=5cm]{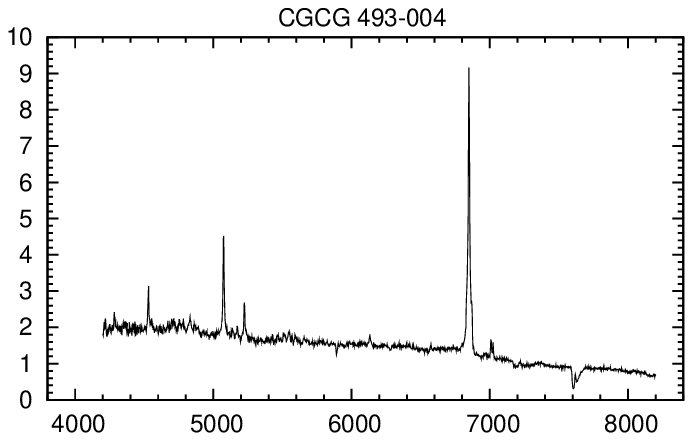}
\includegraphics[width=5cm]{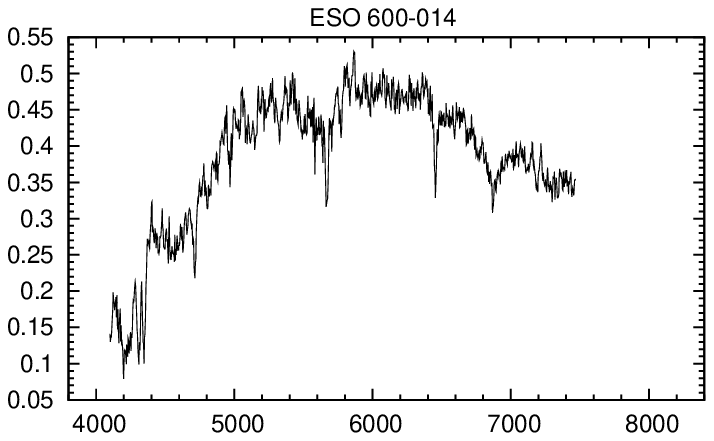}
\includegraphics[width=5cm]{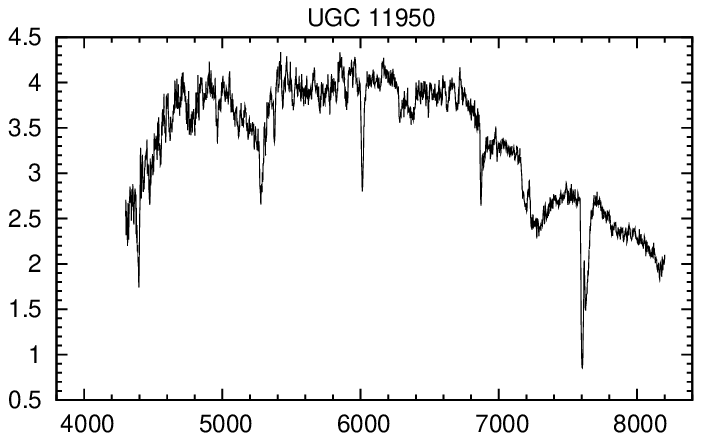}
\includegraphics[width=5cm]{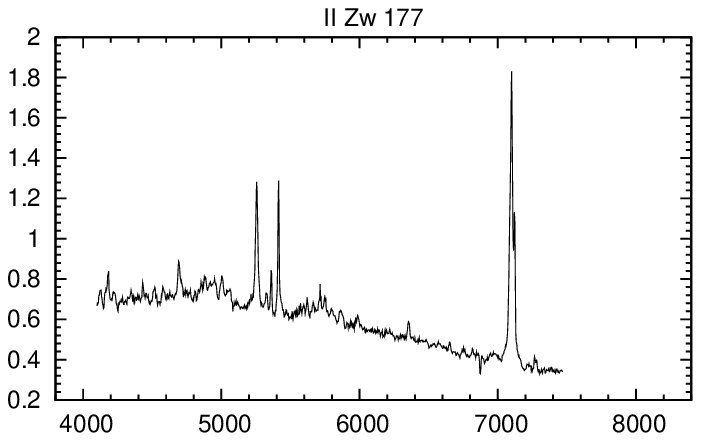}
\includegraphics[width=5cm]{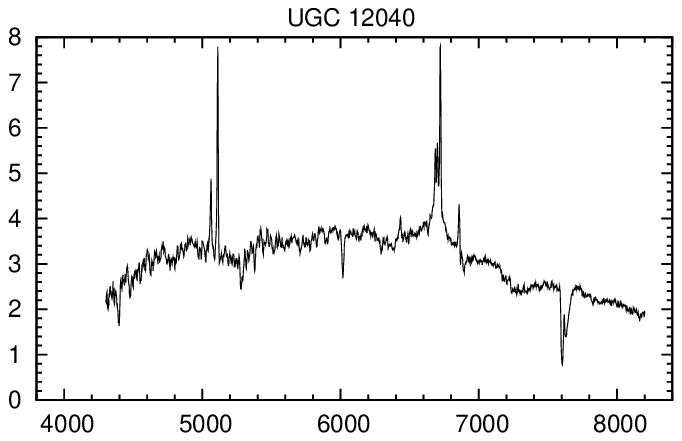}
\includegraphics[width=5cm]{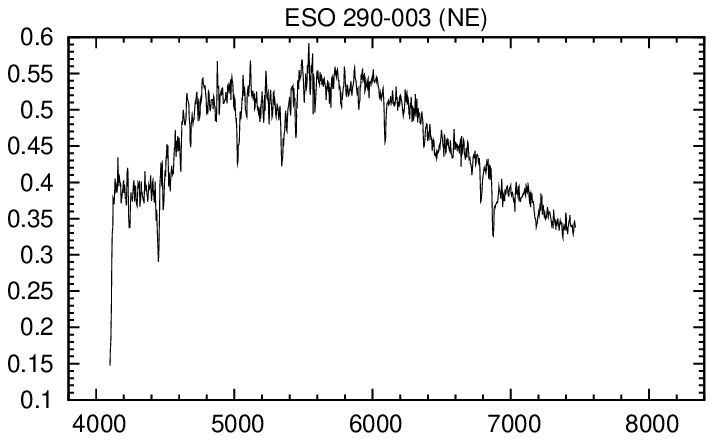}
\includegraphics[width=5cm]{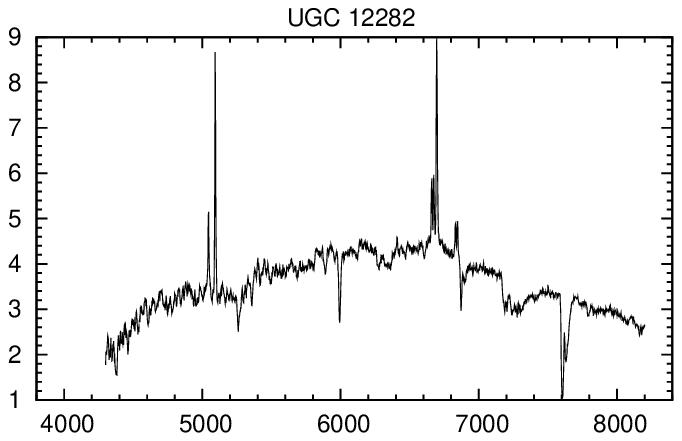}
\includegraphics[width=5cm]{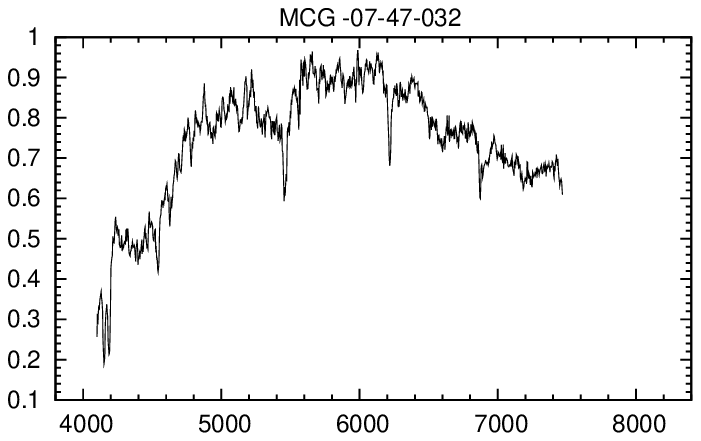}
\includegraphics[width=5cm]{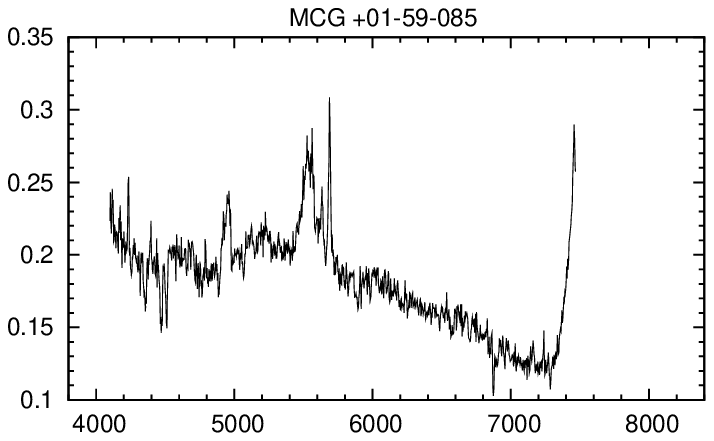}
\includegraphics[width=5cm]{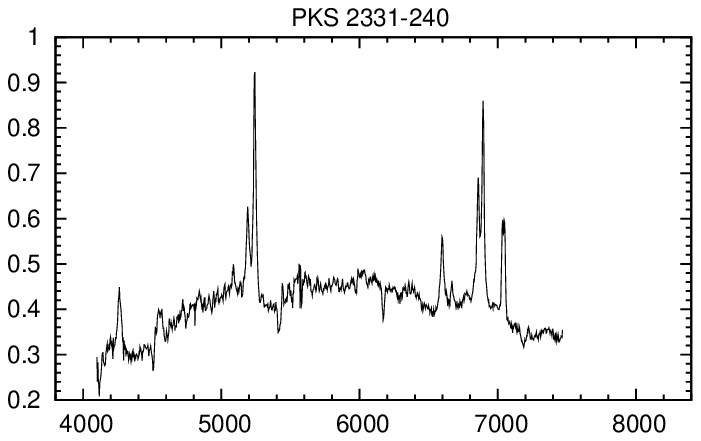}
\includegraphics[width=5cm]{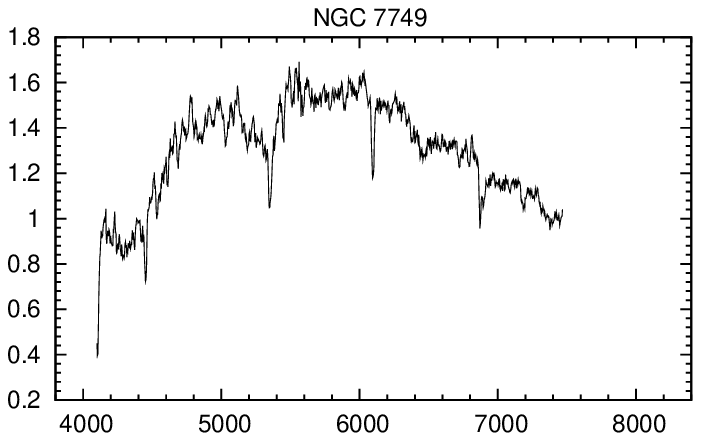}
\includegraphics[width=5cm]{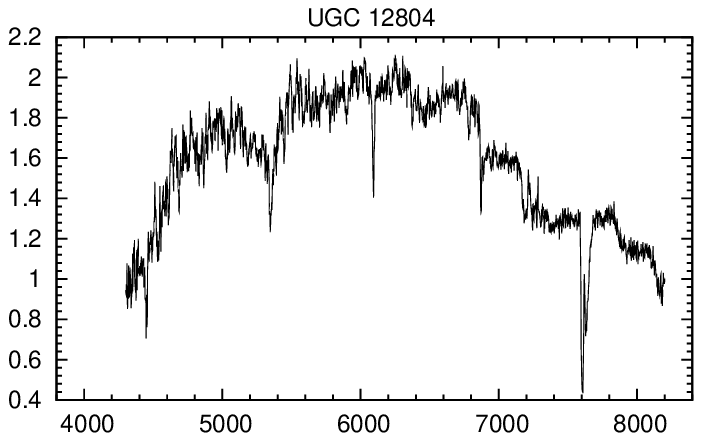}
\includegraphics[width=5cm]{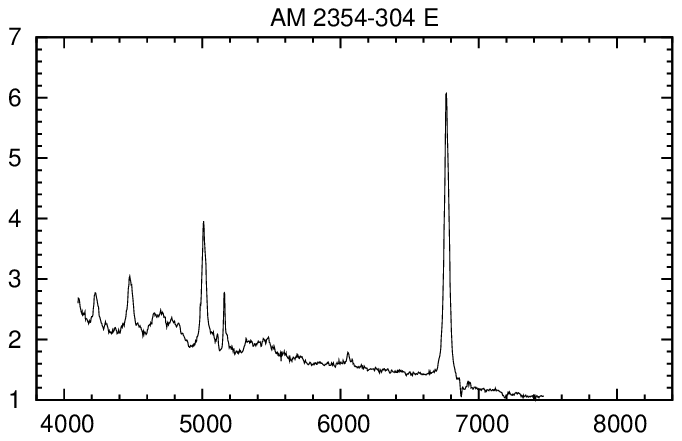}

\vfill

Companions: \medbreak
\hrule
\includegraphics[width=5cm]{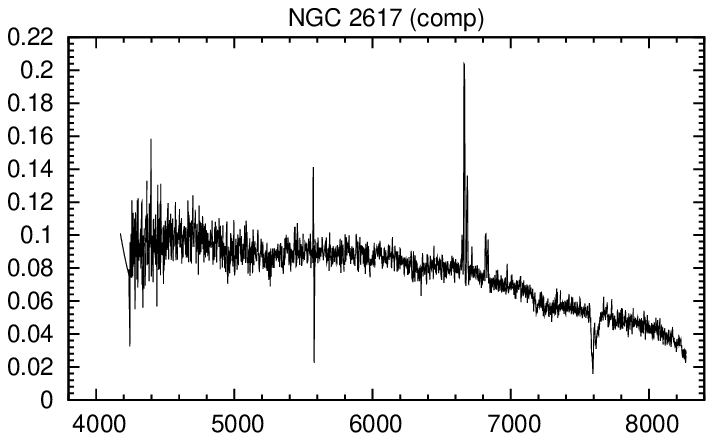}
\includegraphics[width=5cm]{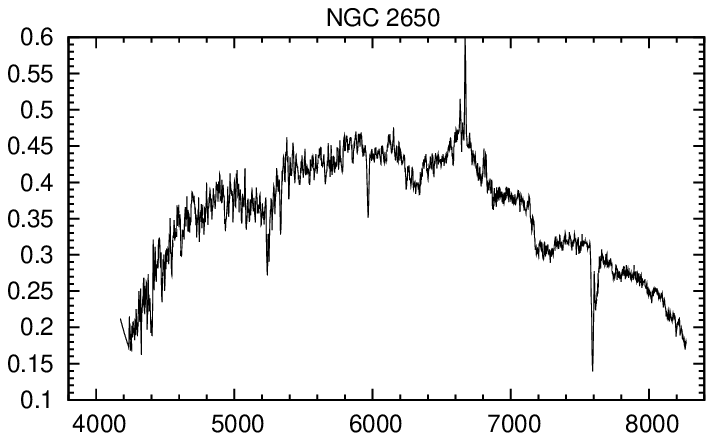}
\includegraphics[width=5cm]{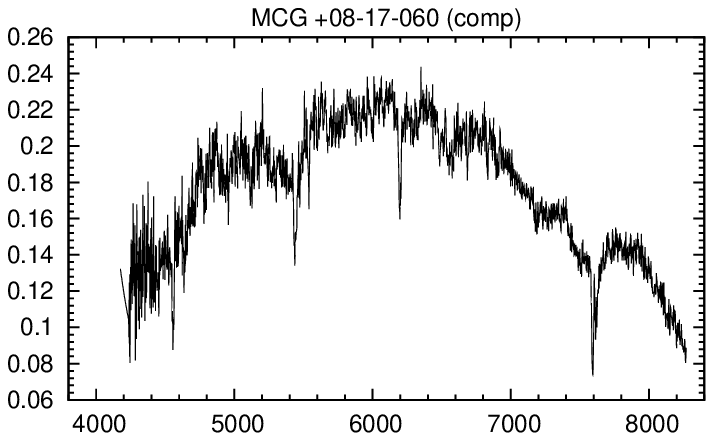}
\includegraphics[width=5cm]{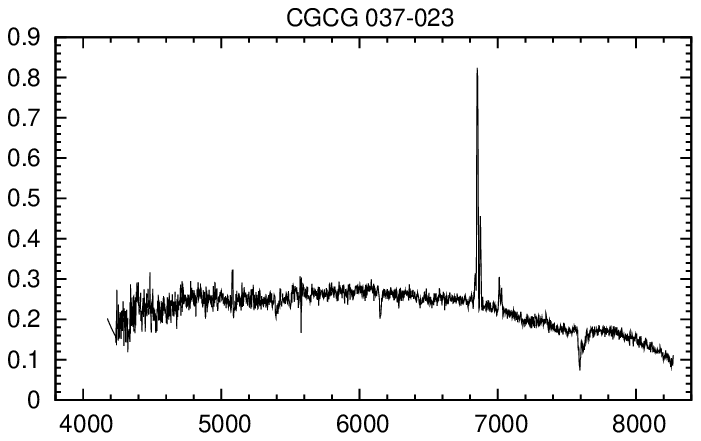}
\includegraphics[width=5cm]{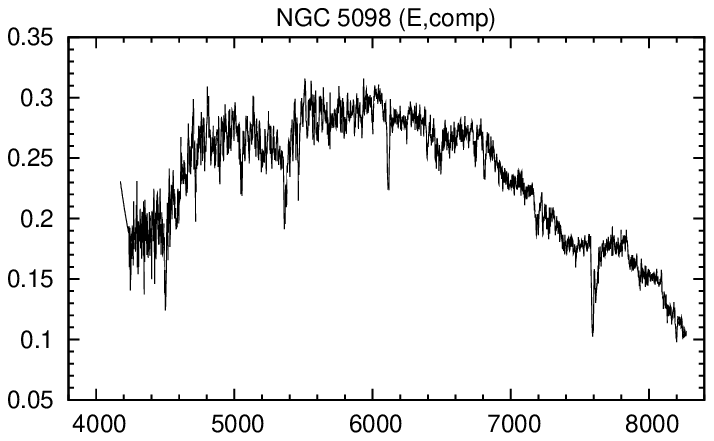}
\includegraphics[width=5cm]{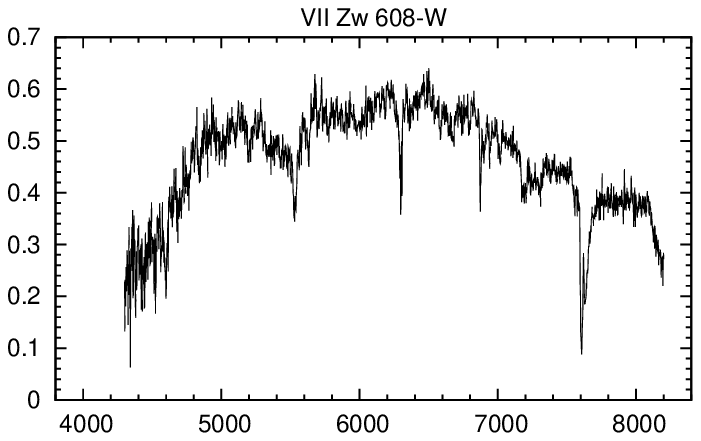}
\includegraphics[width=5cm]{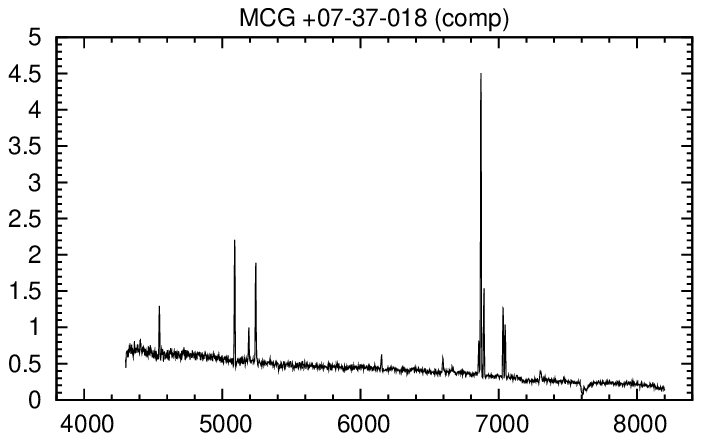}

\end{center}

\begin{figure}[b]
\caption{Spectra of all objects of our sample}
\label{fig:spectra}
\end{figure}


\begin{thebibliography}{}
         


\bibitem[2004]{Abazajian04} Abazajian, K., Adelman-McCarthy, J.~K., Ag{\"u}eros, M.~A., et al.\ 2004, AJ, 128, 502 
%
\bibitem[2006]{Adelman06} Adelman-McCarthy, J., Agueros, M.~A., Allam, S.~S.,
           et al. 2006, ApJS 162, 38
%
\bibitem[2007]{Anderson07} Anderson, S.~F., Margon, B., Voges, W et al. 2007,
     AJ 133, 313
%
\bibitem[2003]{Anderson03} Anderson, S.~F., Voges, W, Margon, B. et al. 2003,
     AJ 126, 2209
%
\bibitem[1998]{Appenzeller98} Appenzeller, I., Thiering, I., Zickgraf F.~J.
        et al. 1998, ApJS 117, 319
%
\bibitem[1998]{Bade98} Bade, N., Engels, D., Voges, W. et al. 1998,
     A\&AS 127, 145
%
\bibitem[2000]{Bauer00} Bauer, F.~E., Condon, J.~J., Thuan, T.~X., \&
  Broderick, J.~J. 2000, ApJS 129, 547
%
\bibitem[1999]{Bischoff99} Bischoff, K., Pietsch W., Boller T. et al. 1999,
         MPE Report 272, 226  
%
\bibitem[1996]{Boller96} Boller, T., Brandt, W.~N. \&  Fink, H.
            1996, A\&A 305, 53
%
\bibitem[1998]{Boller98} Boller, T., Bertoldi, F., Dennefeld, M. \&  Voges, W
            1998, A\&AS 129, 87
%
\bibitem[2000]{Brinkmann00} Brinkmann, W., Laurent-Muelheisen, S.~A.,Voges, W.
 et al. 2000, A\&A 356,445.
%
\bibitem[1990]{Dickey90} Dickey, J.~M. \& Lockman 1990, ARA\&A 28, 215 
%
\bibitem[1992]{Fabbiano92} Fabbiano, G., Kim, G.~W. \& Trinchieri, G. 
      1992, ApJS 80, 531 
%
\bibitem[1998]{Fischer98} Fischer, J.~U., Hasinger, G., Schwope, A.~D. et al.
    1998, AN 319, 347.
%
\bibitem[1999]{Grupe99} Grupe, D., Beuermann,  K., Mannheim,  K. \& Thomas,
       H.~C. 1999, A\&A 350, 805
%
\bibitem[2004]{Grupe04} Grupe, D., Wills, B.~J., Leighly,  K~.M. \& Meusinger, H.
         2004, AJ 127, 156. 
%
\bibitem[2005]{Hao05} Hao, L., Strauss, M. A.; Tremonti, C. A., et al.
      2005, AJ 129, 1783
%
\bibitem[1991]{Hasinger91} Hasinger, G., Truemper, J. \& Schmidt, M.
            1991, A\&A 246, L2
%
\bibitem[1997]{Ho97} Ho, L.C., Filippenko, A.V., Sargent, W.
      1997, ApJ 487, 568
%
\bibitem[2001]{Kewley01} Kewley, L.~J., Dopita, M.~A., Sutherland,
       R.~S. et al. 2001,  ApJ 556, 121 
%
\bibitem[1996]{Moran96} Moran, E.C., Halpern, J.P., Helfand D.J. 1996,
           ApJS 106, 341
%
\bibitem[1998]{Motch98} Motch, C., Guillout, P., Haberl, F., et al. 1998,
        A\&AS 132, 34  
%
\bibitem[1987]{Osterbrock87} Osterbrock, D.~E. 1987, Lecture Notes in Physics,
      Active Galactivc Nuclei (Springer, Heidelberg)
%
\bibitem[1985]{Osterbrock85} Osterbrock, D.~E., Pogge, R.~W. 
       1985,  ApJ 297, 166 
%
\bibitem[1989]{Paturel89} Paturel, G., Fouque, P., Bottinelli, L. \&
          Gouguenheim, L. 1989, A\&AS 80, 299
%
\bibitem[2003]{Paturel03} Paturel, G., Petit, C., Prugniel, P., et al. 2003,
  A\&A 412, 45
%
\bibitem[1998]{Pietsch98}  Pietsch, W., Bischoff, K., Boller, T., et al.
 1998,  A\&A 333, 48 (Paper 1)
%
%
%
\bibitem[1997]{Simcoe97} Simcoe, R., McLeod, K.~K., Schachter, J., Elvis M.
      1997, ApJ 489, 615
%
\bibitem[1990]{Turnshek90} Turnshek, D.A., Bohlin, R.J., Williamson, R.I.
    et al. 1990, AJ 99, 1243
%
\bibitem[2001]{Vaughan01} Vaughan, S., Edelson, R., Warwick, R. et al. 2001,
     MNRAS 327, 673.
%
\bibitem[2006]{Veroncat12ed} Veron-Cetty, M.P. \& Veron, P. 2006, VizieR
  Online Data Catalogue 7248.0
%
\bibitem[1993]{Voges93} Voges, W. 1993, AdSpR 13, 391
%
\bibitem[1999]{Voges99} Voges, W., Aschenbach, B., Boller, et al. 1999,
 A\&A 349, 389
%
\bibitem[2002]{Williams02} Williams, R.~J., Pogge, R.~W., \& Mathur, S.
           2002, AJ 124, 3042 
%
\bibitem[2001]{Xu01} Xu, D.~W., Wei J.~Y., Hu, J.~Y. 2001, Chinese J. A\&A 1,
        46
%
\bibitem[2007]{Zhou07} Zhou, H., Wang, T., Yuan, W. et al.
           2007, ApJ 658, L13 
\bibitem[2001]{Zimmermann01} Zimmermann, H.-U., Boller, T., Doebereiner, S.,
                Pietsch, W. 2001, A\&A 378, 30
%
\end{thebibliography}
\end{document}